\documentclass[openacc]{rstransa}

\usepackage{amsmath}
\usepackage{amsfonts}
\usepackage{amssymb}
\usepackage{stmaryrd}
\usepackage{bm}

\newcommand{\rf}[1]{Fig.~\ref{#1}}

\newcommand{\re}[1]{%
  \begingroup
    Eq.~\eqref{#1}%
  \endgroup
}
\newcommand{\retxt}[1]{%
  \begingroup
  (\ref{#1})%
  \endgroup
}

\allowdisplaybreaks

\begin{document}

\title{Impact of collision models on the physical properties and the stability of lattice Boltzmann methods}

\author{
C. Coreixas$^{1}$, G. Wissocq$^{2}$, B. Chopard$^{1}$ and J. Latt$^{1}$}

\address{$^{1}$Department of Computer Science, University of Geneva, 1204 Geneva, Switzerland.\\
$^{2}$CERFACS, 42 Avenue G. Coriolis, 31057 Toulouse Cedex, France.}

\subject{Statistical Physics, Computer Fluid Dynamics}


\corres{Christophe Coreixas\\
\email{christophe.coreixas@unige.ch}}

\begin{abstract}
The lattice Boltzmann method (LBM) is known to suffer from stability issues when the collision model relies on the BGK approximation, especially in the zero viscosity limit and for non-vanishing Mach numbers. To tackle this problem, two kinds of solutions were proposed in the literature. They consist in changing either the numerical discretization (finite-volume, finite-difference, spectral-element, etc) of the discrete velocity Boltzmann equation (DVBE), or the collision model. 
In this work, the latter solution is investigated in details.
More precisely, we propose a comprehensive comparison of (static relaxation time based) collision models, in terms of stability, and with preliminary results on their accuracy, for the simulation of isothermal high-Reynolds number flows in the (weakly) compressible regime.
It starts by investigating the possible impact of collision models on the macroscopic behavior of stream-and-collide based D2Q9-LBMs, which clarifies the exact physical properties of collision models on LBMs. It is followed by extensive linear and numerical stability analyses, supplemented with an accuracy study based on the transport of vortical structures over long distances. In order to draw conclusions as general as possible, the most common moment spaces (raw, central, Hermite, central Hermite and cumulant), as well as regularized approaches, are considered for the comparative studies. LBMs based on dynamic collision mechanisms (entropic collision, subgrid scale models, explicit filtering, etc) are also briefly discussed.
\end{abstract}


\begin{fmtext}

\end{fmtext}


\maketitle

\section{Introduction}

The lattice Boltzmann method (LBM) is an efficient numerical solver used to simulate various types of flows and physical phenomena~\cite{SUCCI_Book_2018}. It gives the space and time evolution of \emph{fictive} particles gathered in $V$ different groups $f_i(\bm{x},\bm{\xi}_i,t)$ with respect to their propagating speed $\bm{\xi}_i$, where $i\in\{ 0,\ldots,V-1 \}$, and which are located at point $(\bm{x},t)$. By computing statistical moments of $f_i$, one then recovers macroscopic quantities of interest for computational fluid mechanics (CFD), namely, density $\rho$, momentum $\rho \bm{u}$ and total energy $\rho E$. 

Going into more details, the evolution of these populations $f_i$ is most commonly described through two steps (propagation and collision) that are combined in the widely used stream-and-collide algorithm~\cite{KRUGER_Book_2017}. The complex collision process is usually simplified using the BGK collision operator, which assumes the relaxation of populations $f_i$ towards their local equilibrium $f_i^{eq}$ in a time $\tau$, as originally proposed by Bhatnagar, Gross and Krook~\cite{BHATNAGAR_PR_94_1954}. Despite a great success of the BGK-LBM in various fields of research, this model is well-known to encounter stability issues in the zero-viscosity limit and/or for non-vanishing Mach numbers~\cite{SUCCI_Book_2018,KRUGER_Book_2017}. 
To alleviate this problem, numerous ideas have been proposed during the past three decades. Roughly speaking, they consist in either changing the numerical discretization, the collision model, or both of them~\cite{GUO_Book_2013,COREIXAS_PRE_100_2019}.
Even if the former approach leads to more stable LBMs, the resulting numerical scheme is less efficient and accurate, as compared to the standard stream-and-collide algorithm~\cite{WANG_PRE_94_2016}. This is the reason why numerous contributions were proposed to derive more stable and accurate collision models, that would eventually be coupled with other numerical discretizations if needed be.     

Due to the various degrees of stability improvement induced by the different collision models, researchers tried to find physical interpretations that could explain the stability discrepancies between models. 
Such an investigation led to several interesting concepts. For example, by applying the collision step in a moment space, the user can control the relaxation of high-order moments -- which are not supposed to impact the physics as long as the continuum limit remains valid~\cite{MASSET_Accepted_2019} -- and to damp them when needed be, to increase the stability and accuracy of LBMs~\cite{LALLEMAND_PRE_61_2000,DELLAR_PRE_65_2002,BOSCH_ESAIM_52_2015,HOSSEINI_PRE_99_2019b,WISSOCQ_PhD_2019}. 
Interestingly, the relaxation of non-hydrodynamic modes is not restricted to the raw moment space (RM~\cite{DHUMIERE_PAA_159_1992,LALLEMAND_PRE_61_2000,DHUMIERES_TRS_360_2002}). Instead, it can be applied to several kinds of statistical quantities such as Hermite moments (HM~\cite{SHAN_IJMPC_18_2007,ADHIKARI_PRE_78_2008,CHEN_IJMPC_25_2014}), central moments (CM~\cite{GEIER_PRE_73_2006,GEIER_EPJST_171_2009,LYCETTBROWN_CMWA_67_2014,DUBOIS_CCP_17_2015}), central Hermite moments (CHM~\cite{CHEN_PATENT_Collision_2015,MATTILA_PF_29_2017,LI_PRE_100_2019,COREIXAS_PRE_100_2019}) and cumulants (K~\cite{GEIER_CMA_70_2015,GEIER_JCP_348_2017a,GEIER_JCP_348_2017b}).
Finally, the filtering of high-order contributions was also at the center of the derivation of regularization steps in both their standard~\cite{SKORDOS_PRE_49_1993,LADD_JSP_104_2001,LATT_MCS_72_2006a,CHEN_PA_362_2006} or recursive formulations (RR~\cite{MALASPINAS_ARXIV_2015,COREIXAS_PRE_96_2017,BROGI_JASA_142_2017,JACOB_JT_19_2018}).
All the resulting collision models have made it possible to successfully extend the fields of application of the LBM to the low-viscosity regime and to larger compressibility effects. The latter is perfectly illustrated, in the industrial context, by the competitive results obtained with isothermal LBMs, as compared to standard CFD solvers, for the simulation of aeroacoustic phenomena at finite Mach numbers~\cite{MANOHA_AIAA_2846_2015}.

Up to now, several criteria have been proposed in the literature to help the reader choosing the most appropriate collision model for a targeted simulation. Among them, the concept of Galilean invariance is frequently evoked as an argument of paramount importance for the choice of collision model~\cite{GEIER_PRE_73_2006,GEIER_EPJST_171_2009,PREMNATH_JSP_143_2011,GEIER_CMA_70_2015,NING_IJNMF_82_2016,GEIER_JCP_348_2017a,GEIER_CF_166_2018,GEHRKE_Chapter_2020}. Yet, this fundamental principle seems to have been deflected from its original purpose which was to guide the derivation of LBMs in order to recover the correct macroscopic behavior, and more precisely, to eliminate the velocity-dependent errors observed in the definition of transport coefficients such as kinematic viscosity and thermal diffusivity~\cite{QIAN_EPL_21_1993,CHEN_PRE_50_1994,QIAN_EPL_42_1998,LALLEMAND_PRE_61_2000,NIE_EPL_81_2008,PRASIANAKIS_PRE_79_2009,CHEN_IJMPC_25_2014,DELLAR_JCP_259_2014,SHAN_PRE_100_2019}.
Furthermore, previous efforts seem to show that as long as the correct discrete equilibrium form is adopted, then one need not to pay attention to the moment space (used for the collision step) to recover the correct macroscopic behavior whatever the targeted physics~\cite{KATAOKA_PRE_69_2004a,DELLAR_PCFD_8_2008,DELLAR_EPL_90_2010,COELHO_PRE_89_2014,COELHO_PhD_2017,GABANNA_PRE_95_2017}. 

Most of the time, the impact of collision models on the stability of LBMs is investigated through numerical simulations only, and the influence of the scaling used to compute the time step is rarely studied which further restricts the validity of these comparative studies. Consequently, the present work intends to study the stabilization properties of collision models, from a more general viewpoint, by relying on the concept of linear stability analysis (LSA) which, for example, does not depend on the time step scaling. 
Usually, the LSA is used to determine the linear stability properties of a given numerical scheme for uniform base flows and assuming a periodic simulation domain. To further include the impact of boundary conditions or grid mesh refinement techniques, one must use more complicated methods~\cite{AXELSSON_Book_1996,GOLUB_Book_3_2012}. In the CFD community, the LSA is most commonly called a von Neumann (or Fourier) analysis owing to the fact that von Neumann was among the first to study the linear stability of numerical schemes used to solve partial differential equations~\cite{HIRSCH_Book_2007}. Even if this analysis cannot quantify the impact of non-linear (de)stabilization mechanisms, it is a \emph{mandatory} step to assess to good behavior of a numerical scheme in terms of both accuracy and robustness~\cite{HIRSCH_Book_2007,BLAZEK_Book_2015}. When it comes to the LBM community, despite the tremendous number of numerical validations of collision models that have been published up to now, only a few of them rely on the LSA~\cite{STERLING_JCP_123_1996,WORTHING_PRE_56_1997,LALLEMAND_PRE_61_2000,DELLAR_PRE_65_2002,LALLEMAND_PRE_68_2003,WOLF_GLADROW_Book_2004,NIU_JSP_117_2004,DELLAR_PA_362_2006,ADHIKARI_PRE_78_2008,SIEBERT_PRE_77_2008,RICOT_JCP_228_2009,MARIE_JCP_228_2009,GINZBURG_JSP_139_2010,XU_JCP_231_2012,DUBOIS_CRM_343_2015,HOSSEINI_IJMPC_28_2017,CHAVEZMODENA_CF_172_2018,KRIVOVICHEV_IJCM__2018,COREIXAS_PhD_2018,WILDE_IJNMF_90_2019,WISSOCQ_JCP_380_2019,HOSSEINI_PRE_99_2019b,HOSSEINI_PRE_100_2019,MASSET_Accepted_2019,HOSSEINI_RSTA_378_2020}. In addition, the latter usually present results in a formalism that differs from previous studies (particular assumption on the stability criterion, reduced interval of values for the relaxation time, etc), which prevent any \emph{general} comparison of the (linear) stability of collision models.

Consequently, this paper aims at (1) clarifying the exact physical properties of a given collision model in the continuum limit, and (2) providing general results regarding the linear stability of the most common collision models. 
Since this kind of stability analysis cannot identify non-linear (de)stabilization mechanisms that might be inherent to collision models, a numerical validation is also conducted to supplement it. The latter is further used as a first quantification of the accuracy properties of the most stable collision models. To help the reader better apprehend these notions, the present work will be restricted to D2Q9-LBMs. Nevertheless, the methodology required to achieve the aforementioned goals will be presented in a systematic manner, so that one will be able to follow the same path for any LBM of interest. All studies are conducted considering isothermal and compressible LBMs in order to provide results as general as possible.

The rest of the paper reads as follows. In Section~\ref{sec:macroBehavior}, the impact of the equilibrium on the discrete velocity Boltzmann equation (DVBE) is investigated. The latter allows us to explain the observations that were reported in the literature regarding the effect of collision models on the physical properties of the DVBE in the continuum limit.
After recalls on the linear behavior of the Navier-Stokes equations, the influence of the moment space is further investigated through the LSA of the DVBE, where a lattice- and equilibrium-dependent critical Mach number is pointed out (Section~\ref{sec:LSAcontinu}). It is then proposed in Section~\ref{sec:LSAdiscret} to rely on the LSA to quantify the impact of the numerical discretization on the LBM, and on top of that, the impact of the collision model on numerical errors. This is further investigated through the vortex transport by a uniform flow in Section~\ref{sec:Further}.
Eventually, conclusions are drawn in Section~\ref{sec:Conclu}.

For the sake of completeness, the main steps of the LSA are recalled in Appendix~\ref{sec:AppLSA}, and technicalities related to collision models of interest are compiled in Appendix~\ref{sec:CollModels}. The latter briefly discuss how moment-space-based collision models
can be build on top of the simplest one, namely, the RM-LBM. It is further explained how to obtain the RR approach by recursively computing high-order non-equilibrium HMs in the context of the HM-LBM. Eventually, an example of the spectral sampling induced by coarse grid meshes is proposed in Appendix~\ref{sec:filtering}.

\section{Discrete equilibrium and macroscopic behavior \label{sec:macroBehavior}}

This work lies in the context of isothermal fluid flow simulations at finite Mach numbers. The macroscopic behavior of that type of flow is described by the isothermal and compressible Navier-Stokes (NS) equations:
\begin{equation}
\begin{array}{c}
\partial_t \rho + \partial_{\alpha}(\rho u_{\alpha})=0,\\[0.1cm]
\partial_t (\rho u_{\alpha}) + \partial_{\beta}(\rho u_{\alpha}u_{\beta} + p \delta_{\alpha\beta})=\partial_{\beta}(\Pi_{\alpha\beta}),
\end{array}
\label{eq:NS}
\end{equation}  
where external accelerations were not accounted for. The viscous stress tensor $\Pi_{\alpha\beta}$ is defined by 
\begin{equation}
\Pi_{\alpha\beta} = \rho \nu (S_{\alpha\beta}-\tfrac{2}{D}\partial_{\chi}u_{\chi}\delta_{\alpha\beta}) + \rho \nu_b\partial_{\chi}u_{\chi}\delta_{\alpha\beta},
\label{eq:PiNeq}
\end{equation}
with $S_{\alpha\beta}=\partial_{\alpha}u_{\beta}+\partial_{\beta}u_{\alpha}$, while $\nu$ and $\nu_b$ are transport coefficients named kinematic and bulk viscosities respectively. $D$ is the number of physical dimensions, and Greek symbols stand for the physical coordinates $x$ and $y$. Eventually, the equation of state for ideal gases leads to the thermodynamic relationship between temperature $T_0$ (which is constant) and pressure $p=\rho T_0$, hence, closing the set of equations (\ref{eq:NS}).

Interestingly, this kind of fluid flow can also be described through the (force-free) DVBE. The latter is a set of $V$ equations, which directly flows from the velocity discretization of the Boltzmann equation, and that governs the evolution of $V$ groups of particles $f_i$:
\begin{equation}
\forall i\in \{ 0,\ldots,V-1 \},\quad \partial_t f_i + \xi_{i,\alpha}\partial_{\alpha} f_i = \Omega_{f_i^{neq}}.
\label{eq:DVBE}
\end{equation}
The latter translates the balance between the transport at speed $\xi_{i,\alpha}$ and the collision of $V$ populations $f_i$, respectively represented by the left- and right-hand-side terms in Eq.~(\ref{eq:DVBE}). Following basic rules of the kinetic theory of gases~\cite{HUANG_Book_2nd_1987}, collisions must drive $f_i$ towards its local (discrete) equilibrium $f_i^{eq}$. This is done here by assuming $\Omega_{f_i^{neq}}$ explicitly depends on the departure from equilibrium $f_i^{neq}=f_i-f_i^{eq}$. 
In addition, the above methodology is not restricted to the standard BGK operator (named after Bhatnagar, Gross and Krook~\cite{BHATNAGAR_PR_94_1954}), but instead, it can also account for more complex collision models~\cite{SEEGER_PhD_2003,FOX_QBMM_2014}. The best way to achieve such a goal is to rely only on moments of $f_i$, or more precisely, moments of $f^{eq}_i$.

Contrarily to the Boltzmann equation, the DVBE assumes that particles can only propagates in particular directions. The number, norm and orientation of these velocities $\bm{\xi}_i$ must follow particular rules that depend on the macroscopic behavior of interest~\cite{SUCCI_Book_2018,KRUGER_Book_2017}. Similarly, a number of constraints must be satisfied by the discrete equilibria $f_i^{eq}$, where the latter is the discrete counterpart of the Maxwell-Boltzmann distribution (or Maxwellian)
\begin{equation}
f^{eq}=\frac{\rho}{(2\pi T_0)^{D/2}}\exp\bigg[-\frac{(\bm{\xi}-\bm{u})^2}{2 T_0}\bigg],
\label{eq:eqMB}
\end{equation}
with $(p, q)\in \mathbb{N}^2$, and $D=2$ in the present work. More specifically, the moments of $f_i^{eq}$ must equal those of its continuous counterpart $f^{eq}$, 
\begin{equation}
\sum_i f^{eq}_i \xi_{i,x}^p\xi_{i,y}^q = M_{pq}^{eq} = M^{\mathrm{MB}}_{pq} =\displaystyle{\int \xi_x^p\xi_y^q f^{eq}\,d\bm{\xi}},
\label{eq:MatchingCondition}
\end{equation}
at least, up to a given order $p+q=N\in \mathbb{N}$ that depends on the physics of interest~\cite{SHAN_JFM_550_2006}. To recover the isothermal behavior of the NS equations, up to third-order moments of the Maxwellian must be recovered by the discrete equilibrium. This is explained by the fact that the NS equations (\ref{eq:NS}) can be rewritten as
\begin{equation}
\begin{array}{l @{\,=\,}l}
\partial_t (M^{\mathrm{MB}}_{00}) + \partial_{x}(M^{\mathrm{MB}}_{10}) + \partial_{y}(M^{\mathrm{MB}}_{01}) & 0,\\
\partial_t (M^{\mathrm{MB}}_{10}) + \partial_{x}(M^{\mathrm{MB}}_{20})+\partial_{y}(M^{\mathrm{MB}}_{11}) & \partial_{x}(\Pi_{20})+\partial_{y}(\Pi_{11}),\\
\partial_t (M^{\mathrm{MB}}_{01}) + \partial_{x}(M^{\mathrm{MB}}_{11})+\partial_{y}(M^{\mathrm{MB}}_{02}) & \partial_{x}(\Pi_{11})+\partial_{y}(\Pi_{02}),
\end{array}
\label{eq:NSMomentConstraints}
\end{equation}
where the diffusive term is further related to equilibrium moments through the Chapman-Enskog expansion~\cite{CHAPMAN_Book_3rd_1970}
\begin{equation}
\begin{array}{l @{\,\propto\,}l}
 \Pi_{20} & \partial_t (M^{\mathrm{MB}}_{20}) + \partial_{x} (M^{\mathrm{MB}}_{30})+ \partial_{y} (M^{\mathrm{MB}}_{21}),\\
 \Pi_{11} & \partial_t (M^{\mathrm{MB}}_{11}) + \partial_{x} (M^{\mathrm{MB}}_{21})+ \partial_{y} (M^{\mathrm{MB}}_{12}),\\
  \Pi_{02} & \partial_t (M^{\mathrm{MB}}_{02}) + \partial_{x} (M^{\mathrm{MB}}_{12})+ \partial_{y} (M^{\mathrm{MB}}_{03}).
\end{array}
 \label{eq:DiffusiveFluxesMomentConstraints}
 \end{equation}
and the equilibrium moments are  
\begin{equation}\label{eq:EqMomentsTensor}
\begin{array}{c}
M^{\mathrm{MB}}_{00}=\rho, \: M^{\mathrm{MB}}_{10}=\rho u_x, \: M^{\mathrm{MB}}_{20}=\rho(u_x^2+T_0), \: M^{\mathrm{MB}}_{11}=\rho(u_x u_y),\\
M^{\mathrm{MB}}_{30}=\rho(u_x^3+3 T_0 u_x),\: M^{\mathrm{MB}}_{21}=\rho(u_x^2+T_0)u_y.
\end{array}
\end{equation}
$M_{01}^{\mathrm{MB}}$, $M_{02}^{\mathrm{MB}}$, $M_{12}^{\mathrm{MB}}$ and $M_{03}^{\mathrm{MB}}$ are obtained by index permutations. 

For standard lattices, discrete velocities are defined as $\xi_{i,\alpha} \in \{0, \pm 1\}$ ($\alpha=x$ or $y$) which leads to the aliasing defect $\xi_{i,\alpha}^3=\xi_{i,\alpha}$~\cite{KRUGER_Book_2017}. In theory, this decreases the accuracy of such velocity discretization to $N=2$, which corresponds to the isothermal and weakly compressible regime~\cite{SHAN_JFM_550_2006}. Nevertheless, it can be shown that for lattices built through tensor-products of D1Q3 lattices in each space direction (D2Q9 and D3Q27), high-order terms can be accounted for in the definition of the discrete equilibrium. In the context of the D2Q9 velocity discretization, the discrete equilibrium reads~\cite{WAGNER_PRE_59_1999,PRASIANAKIS_PRE_76_2007,KARLIN_PA_389_2010}
\begin{equation}
\begin{array}{l}
f_{(0,0)}^{eq} = \tfrac{1}{\rho}\left(M_{00}^{eq}-M_{20}^{eq}\right)\left(M_{00}^{eq}-M_{02}^{eq}\right)= M_{00}^{eq}-(M_{20}^{eq}+M_{02}^{eq}) + M_{22}^{eq},\\
f_{(\sigma,0)}^{eq} = \tfrac{1}{2\rho}\left(\sigma M_{10}^{eq} + M_{20}^{eq}\right) \left(M_{00}^{eq}-M_{02}^{eq}\right) = \tfrac{1}{2}\left(\sigma M_{10}^{eq} + M_{20}^{eq} - \sigma M_{12}^{eq} - M_{22}^{eq}\right), \\
f_{(0,\lambda)}^{eq} = \tfrac{1}{2\rho}\left(M_{00}^{eq}-M_{20}^{eq}\right)\left(\lambda M_{01}^{eq} + M_{02}^{eq}\right) = \tfrac{1}{2}\left(\lambda M_{01}^{eq} + M_{02}^{eq} - \lambda M_{21}^{eq} - M_{22}^{eq}\right),\\
f_{(\sigma,\lambda)}^{eq} = \tfrac{1}{4\rho}\left(\sigma M_{10}^{eq} + M_{20}^{eq}\right)\left(\lambda M_{01}^{eq} + M_{02}^{eq}\right) = \tfrac{1}{4}\left(\sigma\lambda M_{11}^{eq} + \sigma M_{12}^{eq}+ \lambda M_{21}^{eq} + M_{22}^{eq}\right).
\end{array}
\label{eq:extendedFeq}
\end{equation}
where $(\sigma,\lambda)\in \{\pm1 \}^2$, and $M_{22}^{eq}=M_{20}^{eq}M_{02}^{eq}$. By following this procedure, one can show that the resulting equilibrium populations will satisfy \emph{nine} matching conditions, i.e., for all $p,q\leq 2$. Hence, the above equilibrium states do recover some of the third- and fourth-order moments of the Maxwell-Boltzmann distribution. While fourth-order equilibrium moments do not have any impact on the isothermal macroscopic behavior (but only on the numerical stability~\cite{COREIXAS_PhD_2018,HOSSEINI_PRE_99_2019b,HOSSEINI_PRE_100_2019,WISSOCQ_PhD_2019}), third-order ones are required to recover the correct viscous stress tensor [see Eq.~(\ref{eq:PiNeq})]. Indeed, considering the standard second-order equilibrium [by imposing $M_{pq}^{eq}=0$ for $p+q\geq3$ in Eq.~(\ref{eq:extendedFeq})], macroscopic errors related to the viscous stress tensor $\Delta\Pi_{pq}$ are
\begin{equation}\label{eq:RemainingErrors}
\begin{array}{l @{\,\propto\,}l}
 \Delta\Pi_{20} & \partial_{x} (M^{eq}_{30})+ \partial_{y} (M^{eq}_{21}),\\
 \Delta\Pi_{11} & \partial_{x} (M^{eq}_{21})+ \partial_{y} (M^{eq}_{12}),\\
  \Delta\Pi_{02} & \partial_{x} (M^{eq}_{12})+ \partial_{y} (M^{eq}_{03}).
\end{array}
\end{equation}
whereas accounting for third-order terms ($M^{eq}_{21}$ and $M^{eq}_{12}$) corrects all non-diagonal terms (which are related to shear phenoma): 
\begin{equation}
\Delta\Pi_{20} \propto \partial_{x} (M^{eq}_{30}),\: \Delta\Pi_{02} \propto \partial_{y} (M^{eq}_{03}),
\end{equation}
and $\Delta\Pi_{11} = 0$. 

Interestingly, the extended equilibrium populations~(\ref{eq:extendedFeq}) are naturally incorporated in the most sophisticated collision models (cascaded, cumulant, recursive regularized) while the common BGK operator is usually based on the second-order formulation. This seems to be the reason why the better macroscopic behavior of these LBMs was attributed to the use of more ``physically sound'' collision models, i.e., that better fits, in the asymptotic limit, the isothermal NS equations. Nevertheless, the extended equilibrium~\retxt{eq:extendedFeq} is not restricted to the latter collision models, but on the contrary, it can be included in the definition of standard (originally second-order) collision models for them to benefit from the inclusion of higher-order velocity terms~\cite{COREIXAS_PRE_100_2019}.

Consequently, collision models are not expected to have any impact on the macroscopic behavior of the resulting LBMs, as long as, (1) they are based on the very same equilibrium state, (2) relaxation parameters satisfy the continuum limit required to perform the Chapman-Enskog expansion. If one wants to improve the asymptotic behavior of LBMs in an a priori manner, one must adapt either the discrete equilibrium state or the velocity discretization. This statement obviously extends to LBMs based on collision models with dynamic relaxation frequencies (entropic~\cite{KARLIN_PRL_81_1998,BOGHOSIAN_PRSLA_457_2001a,ATIF_PRL_119_2017}, KBC~\cite{BOSCH_ESAIM_52_2015}, subgrid scale models~\cite{SAGAUT_CMA_59_2010}, space filtering~\cite{RICOT_JCP_228_2009,MARIE_JCP_333_2017}, etc). This is explained by the fact that they either lead to (1) a space- and time-dependent kinematic viscosity $\nu(\bm{x},t)$, or (2) a dynamic relaxation of third- and higher-order \emph{non-equilibrium} moments that should not have any impact on the \emph{isothermal} macroscopic behavior. It is finally worth noting that one could rely on approaches based on velocity dependent relaxation times~\cite{DELLAR_JCP_259_2014}, or even correction terms~\cite{PRASIANAKIS_PRE_76_2007,PRASIANAKIS_PRE_78_2008,GEIER_CMA_70_2015,FENG_JCP_394_2019,RENARD_ARXIV_2020_03644,TAYYAB_CF_211_2020,HOSSEINI_RSTA_378_2020}, to alleviate the remaining deficiencies of standard velocity discretizations. 
\newpage
\section{Linear stability analyses of the DVBE \label{sec:LSAcontinu}}

\subsection{Motivation\label{sec:motivation}}

In the previous section, the impact of the equilibrium on the macroscopic behavior was highlighted. Nevertheless, the latter behavior is only one of the properties of the DVBE and its corresponding LBM. Hence, changing the collision model might influence other properties, such as, the (linear) stability of the DVBE. In addition, LBMs based on different collision models do not share similar numerical properties, and some of them turn out to be more robust and accurate than others under given flow conditions~\cite{GEIER_CMA_70_2015,GEHRKE_CF_156_2017,HAUSSMANN_IJMPC_30_2019}. In order to explain these discrepancies --which cannot be evidenced by only looking at the asymptotic behavior of either the DVBE or its numerical discretization-- it is proposed to perform their LSA.

As a reminder, the LSA has been extensively used to investigate the spectral properties of different sets of partial differential equations~\cite{STRUCHTRUP_PoF_15_2003,TORRILHON_ARFM_48_2016,MASSET_Accepted_2019}, as well as, the robustness and accuracy of their numerical discretizations~\cite{VONNEUMANN_Book_1966,HIRSCH_Book_2007,VANHAREN_JCP_337_2019}. 
This kind of stability analysis consists in studying the evolution of perturbations ($f'_i$ in the lattice Boltzmann context) that are injected, as monochromatic plane waves, in the set of linearized equations of interest. In the case of a temporal analysis, this technique leads to a time-advance matrix that contains all the properties of the linearized equations. As an example, looking at the propagation speed (dispersion) and growth rate (dissipation) of these waves, it is then possible to confirm if they follow rules obtained with the linearized NS equations, for a given set of parameters ($\mathrm{Ma}$, $\nu$, $\nu_b$, etc.).

In the context of the linearized DVBE, and as proposed by several authors~\cite{LALLEMAND_PRE_61_2000,DELLAR_PRE_65_2002,ADHIKARI_PRE_78_2008,MARIE_JCP_228_2009,WISSOCQ_JCP_380_2019,MASSET_Accepted_2019}, this powerful tool can be used as an extension to standard asymptotic expansions (Chapman-Enskog, Grad, Hilbert, etc), which are only valid in the vanishing wave number limit ($k \rightarrow 0$). It can further be used to check the validity of choices that have been made during the derivation of the DVBE: (1) the velocity discretization, (2) the form of the equilibrium state, and (3) the type of collision model. When it comes to linearized LBMs, Sterling and Chen~\cite{STERLING_JCP_123_1996} were among the firsts to apply such analyses to the LBM after linearizing the equilibrium distribution function, as the sole nonlinear term involved in the BGK collision model. Later, several authors performed LSA of the LBM, which significantly improved the understanding of its numerical properties and stability limits~\cite{WORTHING_PRE_56_1997,LALLEMAND_PRE_61_2000,DELLAR_PRE_65_2002,SIEBERT_IJMPC_18_2007,ADHIKARI_PRE_78_2008, MARIE_JCP_228_2009, GINZBURG_JSP_139_2010,WISSOCQ_JCP_380_2019}. In addition, a number of works investigated the behavior of collision models operating in a particular moment space (raw~\cite{LALLEMAND_PRE_61_2000,LALLEMAND_PRE_68_2003,DUBOIS_CRM_343_2015}, Hermite~\cite{DELLAR_PRE_65_2002,DELLAR_PA_362_2006,ADHIKARI_PRE_78_2008}, central~\cite{DUBOIS_CRM_343_2015,CHAVEZMODENA_CF_172_2018} and central Hermite~\cite{HOSSEINI_RSTA_378_2020}) or based on a regularization step~\cite{COREIXAS_PhD_2018,HOSSEINI_PRE_99_2019b,HOSSEINI_PRE_100_2019,HOSSEINI_RSTA_378_2020}. These studies were usually presented in their own framework, and using different methodologies, which prevents any general comparison of the results obtained for each collision model.  

Concretely, this work focuses on the temporal LSA in the context of \textit{isothermal} but \textit{compressible} fluid flows. The aim here is to investigate both the physical properties of the DVBE and the numerical properties of every collision model of interest. The latter being re-written in a common framework, a fair comparison is indeed made possible using the extended equilibrium~(\ref{eq:extendedFeq}), so that any effect of a given model can be isolated. Furthermore, analyzing the linear behavior of the DVBE will make it possible to clearly exhibit the effects of the time/space discretization. The exact methodology followed in this work is introduced in Appendix~\ref{sec:AppLSA}. All of this is done by studying the time evolution of perturbations superimposed to a uniform mean base flow. At first glance, the uniform base flow assumption might be a bit confusing because such a macroscopic flow is pretty far from realistic ones. Nonetheless, it totally makes sense considering that any realistic flow can actually be decomposed into a mean and a fluctuating part, without any hypothesis. The only assumption arises when neglecting second-order effects in this decomposition, i.e., the linear assumption [Eqs.~(\ref{eq:Perturbed_VDF}) and~(\ref{eq:Perturbed_Collision})]. This assumption does not prevent the exhibited instabilities to arise in more realistic cases, if nonlinear phenomena (related to high-order terms) do not take the upper hand. This is illustrated, in Section~\ref{sec:Further}, with a very simple testcase: the transport of vortical structures.

Before moving to the results obtained through the LSA of both the DVBE (Section~\ref{sec:LSAcontinu}\ref{sec:LSADVBE}) and its numerical discretization (Section~\ref{sec:LSAdiscret}), let us recall some basic features about the linearized \emph{isothermal} NS equations, which will be used as references.

\subsection{Hydrodynamic waves}

In the linear context, where perturbations have small amplitudes with respect to mean flow quantities, the macroscopic behavior of the \emph{isothermal} and \textit{compressible} NS equations can be fully described through \emph{two} types of characteristic waves~\cite{LEKKERKERKER_PRA_10_1974,COREIXAS_PhD_2018,WISSOCQ_JCP_380_2019}. They are related to the propagation [$\mathrm{Re}(\omega)$] and the dissipation [$\mathrm{Im}(\omega)$] of: (1) shear ($\omega_{S}$) and (2) acoustic ($\omega_{\pm}$) perturbations. Their analytical formulas read as~\cite{LEKKERKERKER_PRA_10_1974,WISSOCQ_JCP_380_2019,SHAN_PRE_100_2019}
\begin{equation}
\mathrm{Re}(\omega_S) = {u}k,\quad\mathrm{Re}(\omega_{\pm}) = ({u} \pm c)k,
\end{equation}
and
\begin{equation}
\mathrm{Im}(\omega_S) = -\nu k^2,\quad
\mathrm{Im}(\omega_{\pm}) = - \frac{1}{2}\left[\left(\frac{2(D-1)}{D} \nu + \nu_b\right)\right]k^2,
\end{equation}
where $c$ is the mean (Newtonian) speed of sound, and can further be identified as the lattice constant $c_s$ in the present context. 
It is worth noting that while the above relations are exact for the shear wave, those for acoustic waves are obtained through a truncated Taylor expansion~\cite{SHAN_PRE_100_2019}.
In addition, it is assumed here that perturbations propagate in the direction of the mean flow (i.e, $\bm{{u}}\cdot\bm{k} = {u}k$) to ease the description of the LSA basic concepts, whereas this assumption will be dropped for the computation of linear stability domains [Section~\ref{sec:LSAdiscret}\ref{sec:linStabDomain}].

The \emph{isothermal} NS equations is a set of $(D+1)$ conservation equations. Hence $(D+1)$ characteristic waves fully describe the linear behavior of this set of equations. In the one-dimensional case, these hydrodynamic modes are divided into two acoustic waves. The shear wave, which is linked to transverse velocity perturbations, has no physical meaning in this particular case. For bidimensional flows, one shear wave can further be identified as a solution of the eigenvalue problem. Eventually, three-dimensional flows have their linear behavior fully described by four characteristic waves: two shear waves and two acoustic waves. Regarding now the properties of these waves, the shear wave propagates at the mean flow speed ${u}$, whereas forward ($+$) and backward ($-$) acoustic waves propagate at ${u} \pm c$. Furthermore, the attenuation of the shear wave is directly controlled by its diffusivity coefficient $\nu$. Regarding acoustic waves, the attenuation process is divided into two parts: dissipation induced by (i) shear ($\nu$), and (ii) compression/dilation ($\nu_b$).

In the present work, we are interested in 2D \emph{isothermal} and \textit{compressible} LBMs. As a consequence, three characteristic waves (one shear, and two acoustic waves) compose hydrodynamic modes, whereas incompressible LBMs should not recover the two modes related to acoustics. Furthermore, the bulk viscosity is linked to its kinematic counterpart via $\nu_b = (2/D)\nu = \nu$~\cite{DELLAR_PRE_64_2001}. Thus, shear and acoustic waves will have the very same attenuation rate $\mathrm{Im}(\omega_{\pm})=\mathrm{Im}(\omega_S) = -\nu k^2$.     

\begin{figure}[bp!]
\begin{tabular}{c @{\hspace{0.1cm}} c @{\hspace{0.1cm}} c}
 \includegraphics[totalheight=0.17\textheight]{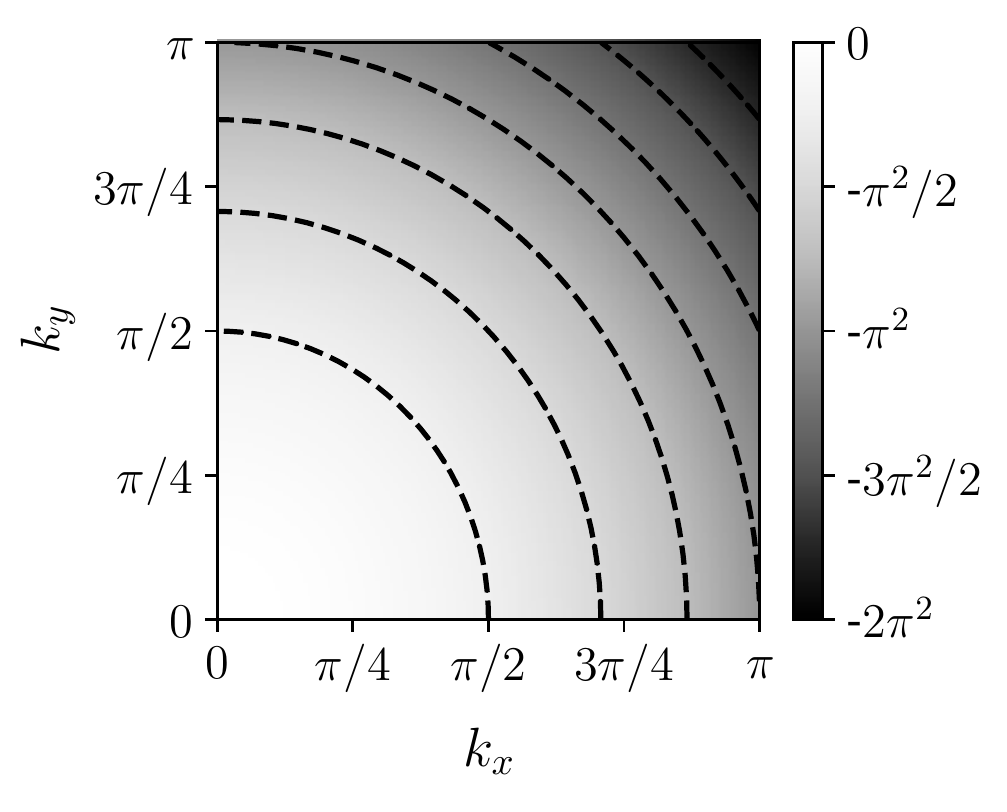} &
  \includegraphics[totalheight=0.17\textheight]{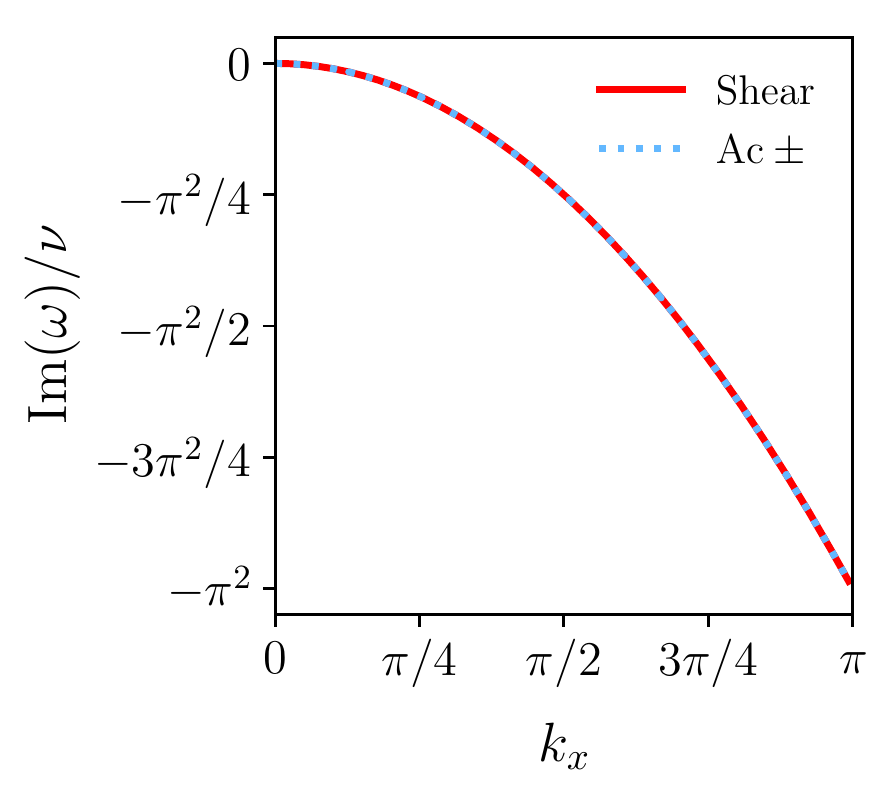} &
\includegraphics[totalheight=0.17\textheight]{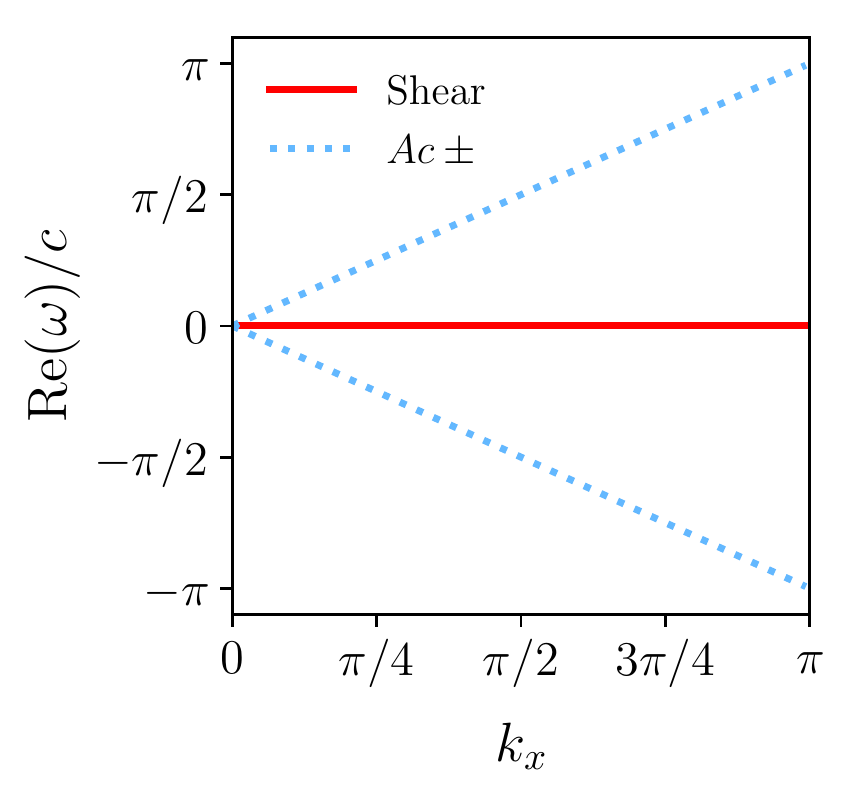}
\end{tabular}
\caption{LSA of the isothermal NS equations. From left to right, maximal growth rate $\mathrm{max}[\mathrm{Im}(\omega)/\nu]$, dissipation, and dispersion properties ($k_y=0$) of a 2D gas flow at rest, with the same characteristics as for isothermal LBMs ($\nu_b = \nu$). For the 2D map, dashed isolines range from $-\pi^2/4$ to $-7\pi^2/4$ with a step of $\pi^2/4$. Expressions of theoretical curves are~\cite{LEKKERKERKER_PRA_10_1974,COREIXAS_PhD_2018,WISSOCQ_JCP_380_2019}: $\mathrm{Im}(\omega_S)/\nu = \mathrm{Im}(\omega_{\pm})/\nu= -k^2$, $\mathrm{Re}(\omega_S)/c =\mathrm{Ma}k$, and $\mathrm{Re}(\omega_{\pm})/c = (\mathrm{Ma} \pm 1)k$, where $\mathrm{Ma}$ is the mean Mach number.}
\label{fig:LSA_NS_2D_isothermal_LBM}
\end{figure}

To help the reader understand the behavior of shear and acoustic waves in the Fourier space, the spectral properties of a 2D isothermal gas flow typical of LBMs of interest ($\nu_b = \nu$) are plotted in~\rf{fig:LSA_NS_2D_isothermal_LBM}. The 2D map of the maximal growth rate highlights their isotropic behavior in the Fourier space, meaning that dissipation rates are independent of the direction along which waves propagate. This is also true for the group velocity of each wave. In addition, one can notice that normalized dispersion curves [$\mathrm{Re}(\omega)/c$] are linear with respect to the wave number $\bm{k}$. Knowing that the propagation speed, or group velocity, is the slope of dispersion curves, one can conclude that hydrodynamic waves propagate at constant speed whatever the value of $k$, at least, in the limit of vanishing viscosity~\cite{SHAN_PRE_100_2019}. When it comes to dissipation rates, they fit a quadratic trend with respect to the wave number [$\mathrm{Im}(\omega)/\nu\propto k^2$] and they are always negative. This means that the dissipation of waves is even more important when their spatial frequency $k$ is high. This translates the fact that viscosity impacts more small vortical structures than bigger ones. 

\begin{figure}[bp!]
\centering
\begin{tabular}{c c c}
\hspace*{1.1cm}$\mathrm{Ma} = 0.0$ & \hspace*{1.1cm}$\mathrm{Ma} = 0.4$ & \hspace*{1.1cm}$\mathrm{Ma} = 0.8$\\
\hspace*{0.15cm} \includegraphics[totalheight=0.17\textheight]{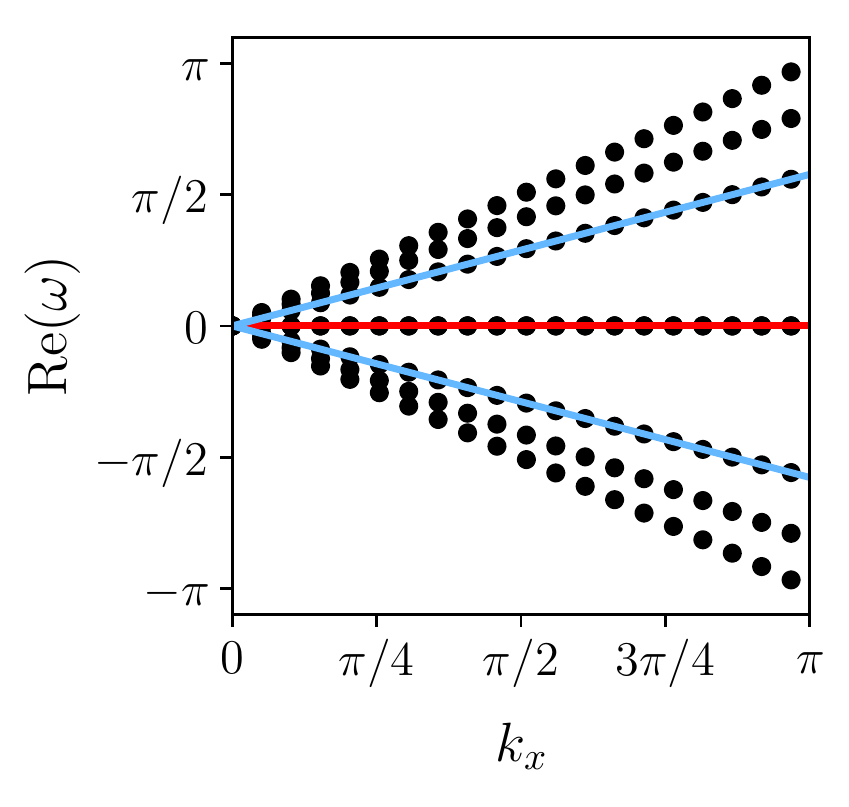} & 
\hspace*{0.15cm}\includegraphics[totalheight=0.17\textheight]{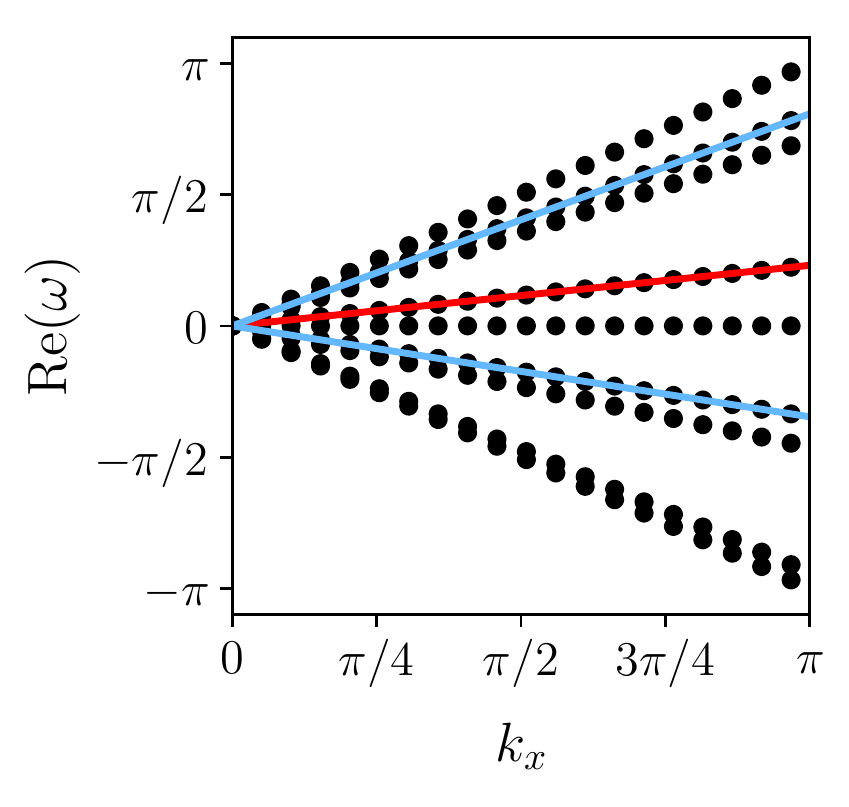} & 
\hspace*{0.15cm}\includegraphics[totalheight=0.17\textheight]{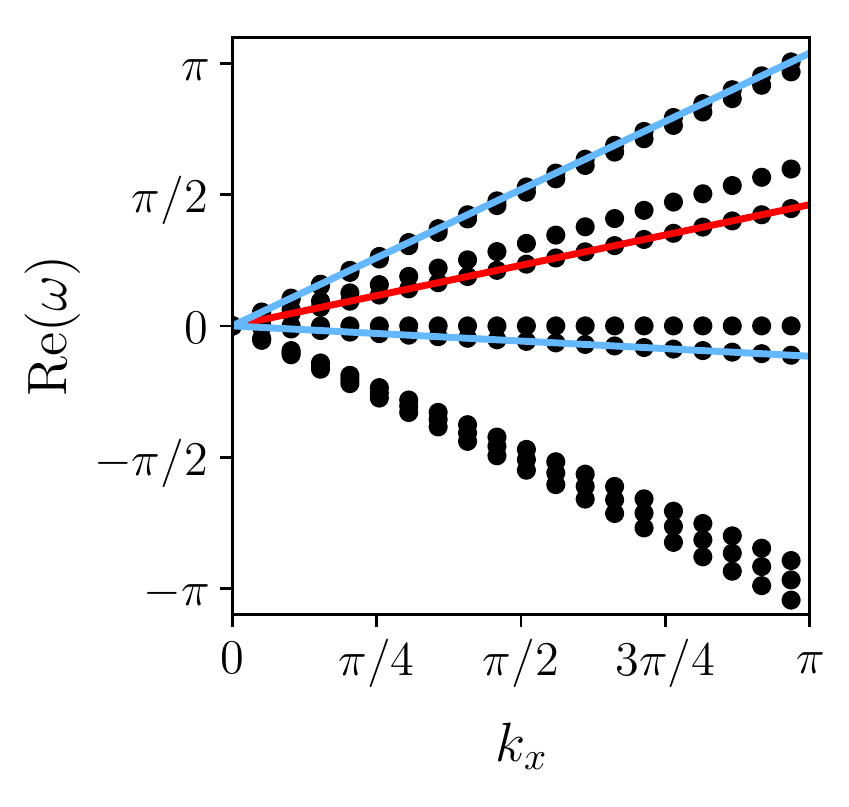}\\
\includegraphics[totalheight=0.17\textheight]{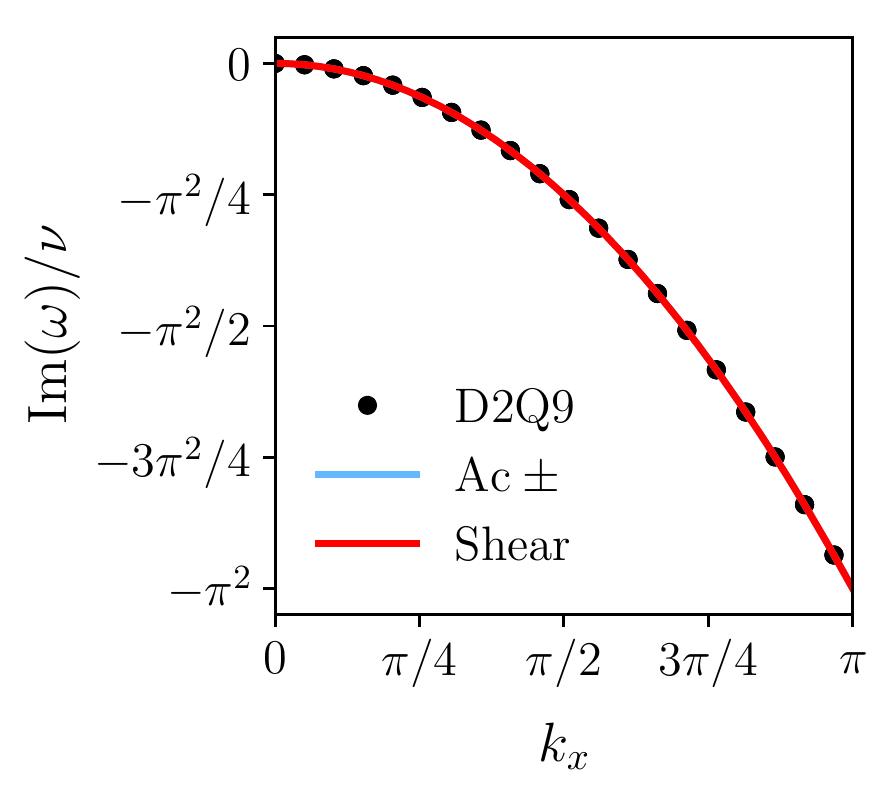} & 
\includegraphics[totalheight=0.17\textheight]{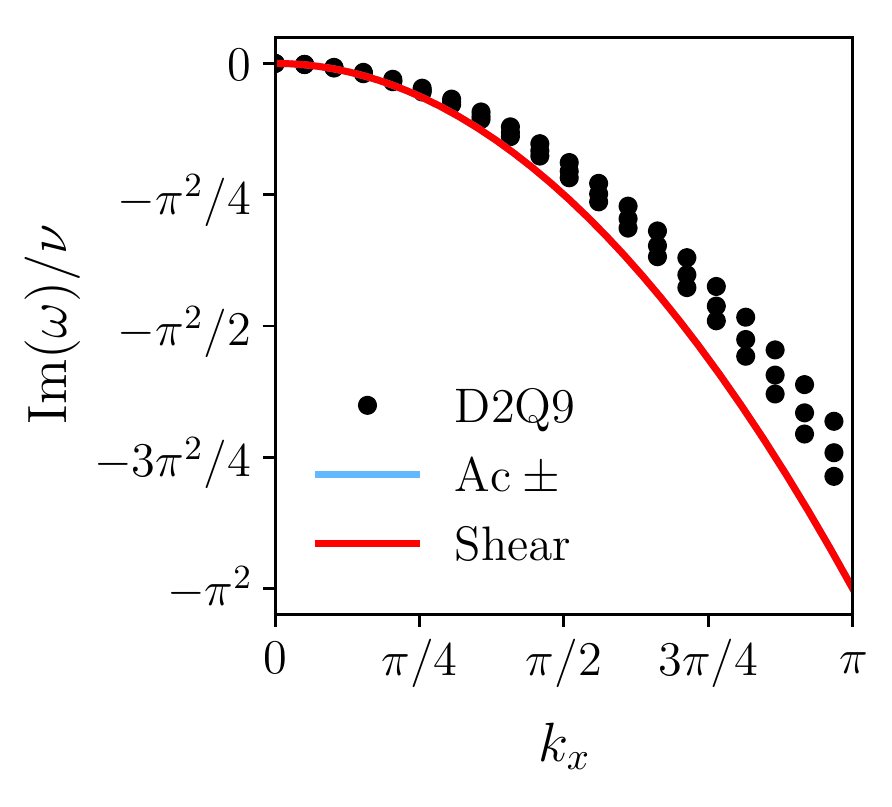} & 
\includegraphics[totalheight=0.17\textheight]{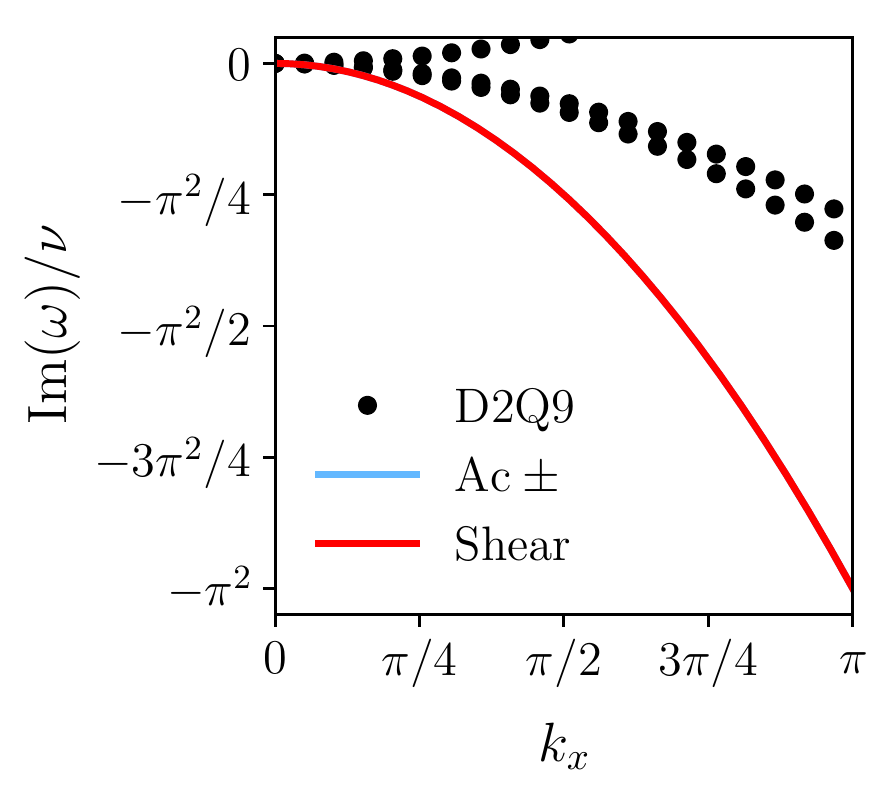}
\end{tabular}
\caption{LSA of the D2Q9-BGK-DVBE with a second-order equilibrium $(N=2)$. Impact of the Mach number on dissipation rates (top) and dispersion properties (bottom) for $\nu = 10^{-6}$ and $k_y = 0$.}
\label{fig:LSA_BGK_N2_MachImpact}
\end{figure}

\begin{figure}[btp!]
\centering
\begin{tabular}{c c c}
\hspace*{1.1cm}$\mathrm{BGK}\,(N=2)$ & \hspace*{1.1cm}$\mathrm{CM/CHM/K}$ & \hspace*{1.1cm}$\mathrm{BGK}\,(N=4)$\\
\includegraphics[totalheight=0.17\textheight]{./spectrum_dissipation_D2Q9_continuous_BGK_fEqO2_M0_8_Theta00_nu0_000001_deg0.pdf} & 
\includegraphics[totalheight=0.17\textheight]{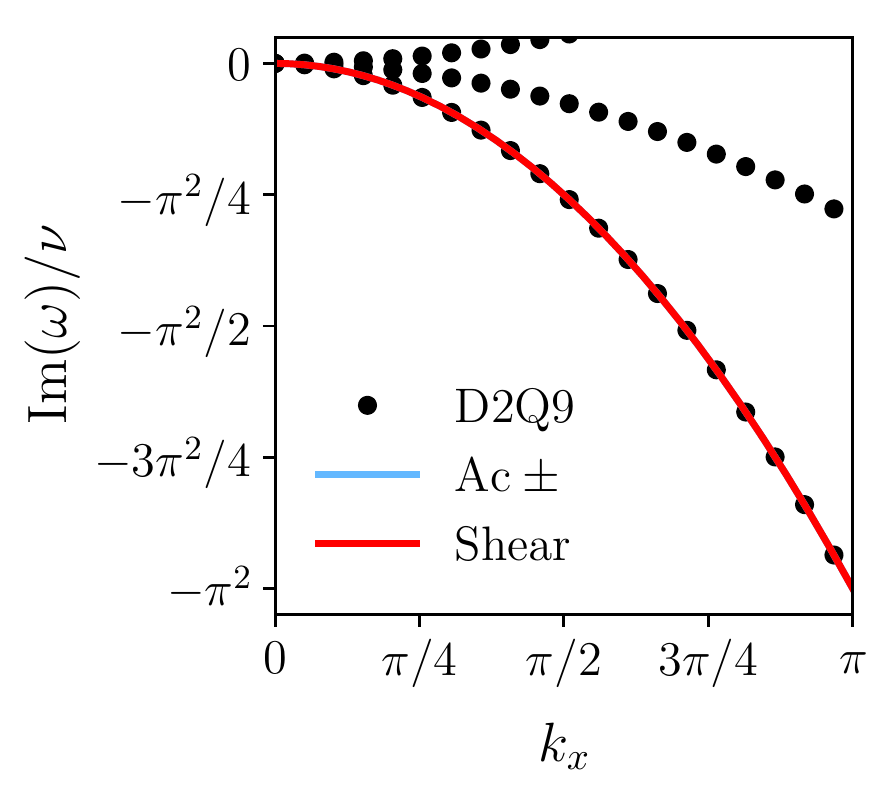} & 
\includegraphics[totalheight=0.17\textheight]{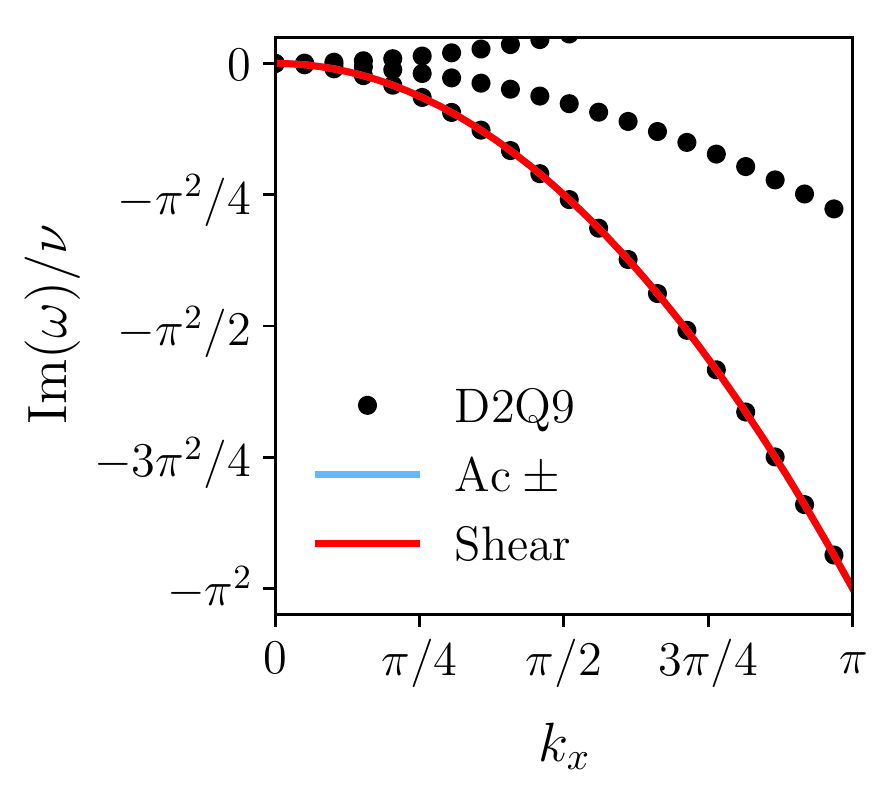}
\end{tabular}
\caption{Impact of the moment space on the dissipation rates obtained from the LSA of the D2Q9-DVBE with $\nu = 10^{-6}$, $\mathrm{Ma}=0.8$ and $k_y = 0$. From left to right, BGK with the standard equilibrium, collision models based on different moment spaces (but a single relaxation time), BGK with extended equilibrium. CM, CHM and K stand for central, central Hermite moment and cumulant based LBMs.}
\label{fig:LSA_Comp_CollModel_Ma08}
\end{figure}

\begin{figure}[btp!]
\centering
\includegraphics[width=0.95\textwidth]{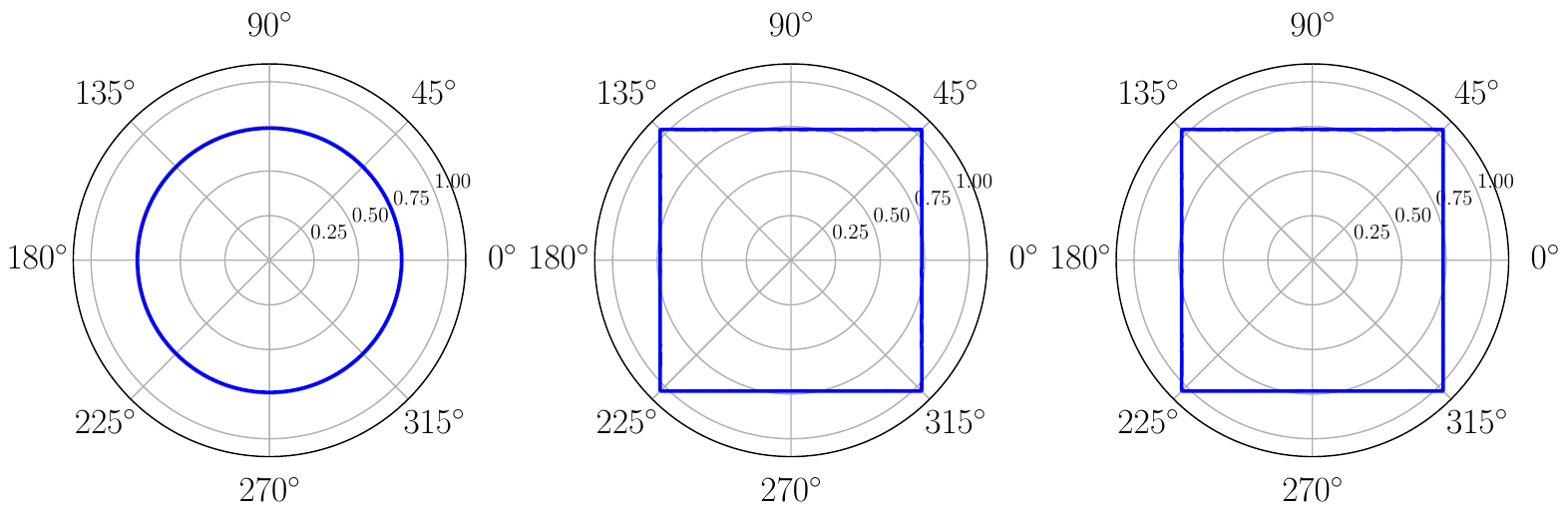}
\caption{Mach number ($\mathrm{Ma}$) corresponding to the linear stability limit of the D2Q9-BGK-DVBE, for various angles $\phi$ of the mean flow and $\nu=10^{-6}$. Impact of the truncation order of the equilibrium state~\retxt{eq:extendedFeq}: standard second-order, partial third-order and partial fourth-order (from left to right). The critical Mach number is defined as $\mathrm{Ma_c}=\min_{\phi}(\mathrm{Ma})$. For the D2Q9-DVBE based on a polynomial equilibrium, one obtains $\mathrm{Ma_c}=\sqrt{3} - 1\approx 0.73$~\cite{WISSOCQ_PhD_2019,MASSET_Accepted_2019}.}
\label{fig:Anisotropic_Mach_D2Q9_BGK_LBE}
\end{figure}

\subsection{Linear behavior of the D2Q9-DVBE\label{sec:LSADVBE}}

In what follows, the linear behavior of the D2Q9-DVBE is investigated by solving the eigenvalue problem related to the perturbed equation~\retxt{eq:ContinuousPerturbedEq}. Hence, it is important to understand that $V$ waves will be evolving in the Fourier space, instead of the $D+1$ waves encountered with the isothermal NS equations. This is because the DVBE, and a fortiori the LBM, is a set of $V$ equations whose linear behavior is entirely linked to $V$ waves. Among them, $D+1$ modes linked to hydrodynamics will be encountered, while $V-(D+1)$ non-hydrodynamic waves will also be present in the Fourier space. By comparing modes obtained via the linearization of the DVBE with those resulting from the linearized macroscopic equations, one is then able to investigate potential defects of the DVBE in terms of macroscopic behavior \emph{independently} from the numerical discretization that further leads to the LBM. Hence the LSA of the DVBE aims at providing general results that do not depend on the numerical discretization adopted (stream-and-collide, finite-volume, etc), and which are not restricted to a particular time step scaling. 
Hereafter, we will only focus on hydrodynamic modes, but the interested reader is referred to Refs.~\cite{DELLAR_PRE_65_2002,DELLAR_PA_362_2006,WISSOCQ_JCP_380_2019,WISSOCQ_PhD_2019,ASTOUL_ARXIV_2020_11863} for a detailed investigation of the properties of non-hydrodynamic modes. 

Let us start with the D2Q9-BGK-DVBE based on the standard second-order equilibrium, i.e., which is recovered imposing $M_{pq}^{eq}=0$ if $p+q>2$ in \re{eq:extendedFeq}. Its dispersion and dissipation curves are plotted in \rf{fig:LSA_BGK_N2_MachImpact}. From them, it is clear that this model is able to recover the correct macroscopic behavior, in terms of propagation speed and dissipation of shear and acoustic waves, for low values of the Mach number. Nevertheless, deviations in the dissipation rate start appearing for higher values. The latter deviations are related to the famous $\mathcal{O}(\mathrm{Ma}^3)$ errors that are obtained in the definition of the viscous stress tensor~\cite{QIAN_EPL_21_1993}. Hence, the LSA of the DVBE is able to capture the weakly compressible deficiency of the D2Q9 velocity discretization, as previously reported for the D3Q19 lattice~\cite{MARIE_JCP_228_2009}. In fact, previous studies showed that by adopting higher-order velocity discretizations, the LSA further leads to shear and acoustic waves that do not suffer anymore from compressibility errors~\cite{COREIXAS_PhD_2018,WISSOCQ_PhD_2019}.  

If we now consider results obtained with methods that are supposed to restore the Galilean invariance of standard lattices, such as the cascaded and the cumulant based LBMs~\cite{GEIER_PRE_73_2006,GEIER_EPJST_171_2009,GEIER_CMA_70_2015}, it seems clear from \rf{fig:LSA_Comp_CollModel_Ma08} that these methods do correct the errors usually observed on the dissipation of shear waves. Nonetheless, this has nothing to do with the moment space used for the collision step, but instead, this improvement is only due to the extended equilibrium~\retxt{eq:extendedFeq} that is naturally used with these methods, as proven by the results obtained from the LSA of the BGK-DVBE based on the same extended equilibrium. This misinterpretation results from a common mistake encountered in numerous comparisons of collision models in the literature, where two parameters are changed at the same time: equilibrium and moment space. Consequently, the LSA of the DVBE confirms that moment spaces do not have any impact on the macroscopic behavior of this set of equations, and that only the equilibrium does have an influence in the continuum limit.

It is also worth noting that even with the extended equilibrium \retxt{eq:extendedFeq}, there are still velocity-dependent errors remaining on the diagonal part of the viscous stress tensor [see Eq.~(\ref{eq:RemainingErrors})]. While these errors obviously impact the dissipation of acoustic waves, in a less intuitive way, they also impact the attenuation of shear waves that are not propagating along coordinate axes. Hence, to \emph{correctly assess} the Galilean invariance properties of transport coefficients, one must study the impact of the mean flow orientation (with respect to the grid mesh) on these coefficients~\cite{DELLAR_JCP_259_2014}.

Eventually, since the linearized D2Q9-DVBE shows compressibility errors that act as anti-diffusion, there is a (critical) Mach number $\mathrm{Ma_c}$ above which this set of equations becomes linearly unstable. Such an upper limit can be obtained by checking the value of $\mathrm{Ma}$ for which at least one of the nine waves has a positive growth rate ($\mathrm{Im}(\omega)>0$) for a given wavenumber $k$~\cite{HIRSCH_Book_2007}. For the D2Q9-BGK-DVBE, one ends up with $\mathrm{Ma_c}=\sqrt{3}-1\approx 0.73$ (for values of $\nu$ satisfying the continuum limit assumption), as illustrated in \rf{fig:Anisotropic_Mach_D2Q9_BGK_LBE} and further demonstrated in Refs.~\cite{WISSOCQ_PhD_2019,MASSET_Accepted_2019}. Interestingly, this upper stability limit can be shown to depend on both the lattice and the type of equilibrium (polynomial, entropic, etc). In addition, it does not seem to be impacted by either the collision model~\cite{HOSSEINI_PRE_99_2019b,HOSSEINI_PRE_100_2019,WISSOCQ_PhD_2019}, or the numerical discretization~\cite{WILDE_IJNMF_90_2019}.

In what follows, we will further investigate the linear stability domain of LBMs based on collision models considered in our previous work~\cite{COREIXAS_PRE_100_2019}, and further recalled in Appendix~\ref{sec:CollModels}. 
This will allow us to: (1) quantify the impact of collision model on errors introduced during the numerical discretization of the DVBE, and (2) see if the critical Mach number of the DVBE is also an upper limit for LBMs whatever the collision model considered.

\section{Sophisticated collision models as a way to improve the linear stability of LBMs? \label{sec:LSAdiscret}}

\subsection{Numerical discretization}

In previous sections, results regarding the asymptotic limit and the linear behavior of the DVBE have been provided before its numerical discretization. Hence, their validity holds whatever the numerical scheme used to solve that set of partial differential equations (provided it is consistent~\cite{HIRSCH_Book_2007}), e.g., stream-and-collide~\cite{KRUGER_Book_2017,SUCCI_Book_2018}, finite-difference~\cite{CAO_PRE_55_1997,GUO_PRE_67_2003,KATAOKA_PRE_69_2004a,WATARI_PA_382_2007,KEFAYATI_JFS_83_2018}, finite-volume~\cite{NANNELLI_JSP_68_1992,XI_PRE_60_1999,UBERTINI_PRE_68_2003,SBRAGAGLIA_PRE_82_2010,GUO_PRE_88_2013,GUO_PRE_91_2015,FENG_CF_131_2016,YANG_PF_30_2018}, spectral-element discontinuous Galerkin~\cite{SHI_IJNMF_42_2003,MIN_JCP_230_2011,PATEL_IJNMF_82_2016,SHAO_CMWA_75_2018}. 
For the sake of fairness, however, it is also important to understand that numerical properties (i.e., stability and accuracy) of LBMs do depend on the collision model.
This statement is even more true in the context of the stream-and-collide discretization for which numerical errors appear due to the numerical discretization of the collision term in the DVBE~(\ref{eq:DVBE}).
To understand why, let us start from the numerical discretization of the DVBE (based on this algorithm), and referred to as LBM:
\begin{equation}
\forall i\in \{ 0,\ldots,V-1 \},\quad f_i(\bm{x} +\bm{\xi}, t+1) - f_i(\bm{x}, t) = - \Omega_{ij} f_j^{neq}(\bm{x}, t). \label{eq:StreamCollide}
\end{equation}
The left-hand-side (convective) term has been discretized using the method of characteristics. In the context of standard lattices, composed of \emph{constant} discrete velocities whose components are \emph{integers} (in lattice units), this numerical discretization is \emph{exact}, and as a consequence, does not introduce any numerical error. On the contrary, the right-hand-side (collision) term is discretized following the trapezium rule which leads to second-order accuracy~\cite{HE_JCP_146_1998,DELLAR_CMA_65_2013}. The latter discretization is then the only one introducing numerical errors in the derivation of the LBM from the DVBE. This is the reason why, one can show that the \emph{numerical} behavior of LBMs is drastically impacted by the choice of both the moment space and the relaxation parameters~\cite{DHUMIERE_CMA_58_2009,DUBOIS_CRM_343_2015,GEIER_CMA_70_2015,GEIER_JCP_348_2017a,GEIER_CF_166_2018,CHAVEZMODENA_CF_172_2018,HOSSEINI_PRE_99_2019b}.

In what follows, we will then further quantify errors that emerge from the numerical discretization of the DVBE by using the linearized formulation of the stream-and-collide algorithm~\retxt{eq:DiscretePerturbedEq}. This will allow us to identify the potential impact of collision models on the linear stability of corresponding LBMs.

\subsection{Linear behavior of D2Q9-LBMs \label{sec:linStabDomain}}

\begin{figure}[bp!]
\begin{tabular}{c @{\hspace{0.1cm}} c @{\hspace{0.1cm}} c}
\includegraphics[totalheight=0.17\textheight]{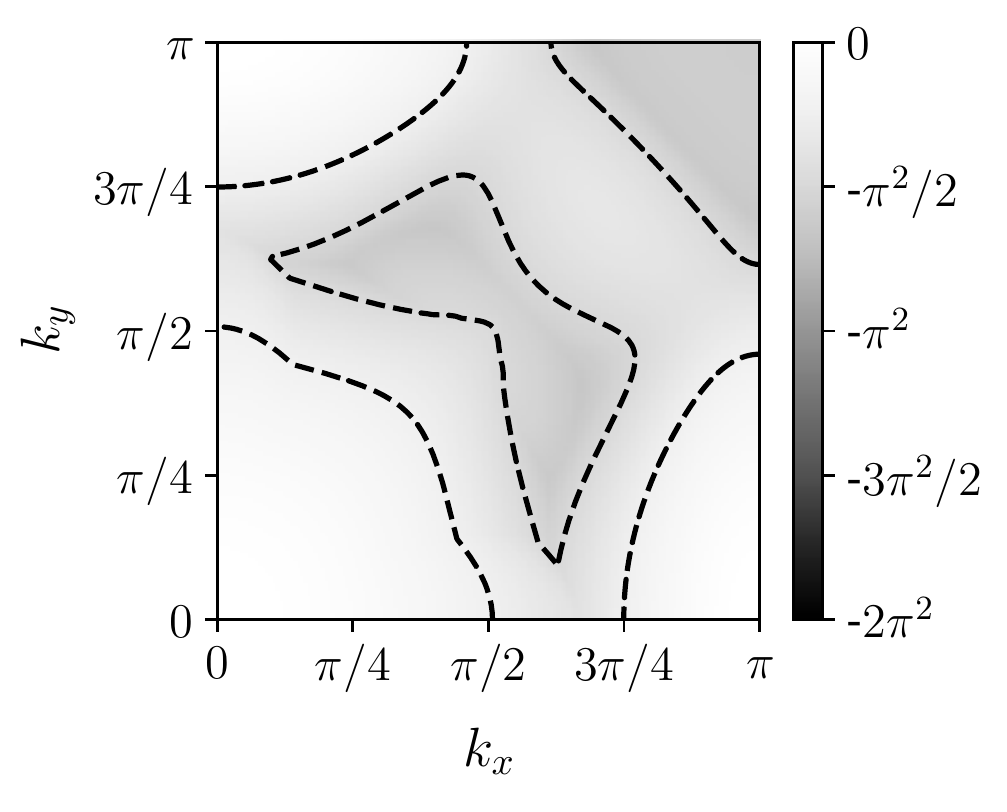} &
  \includegraphics[totalheight=0.17\textheight]{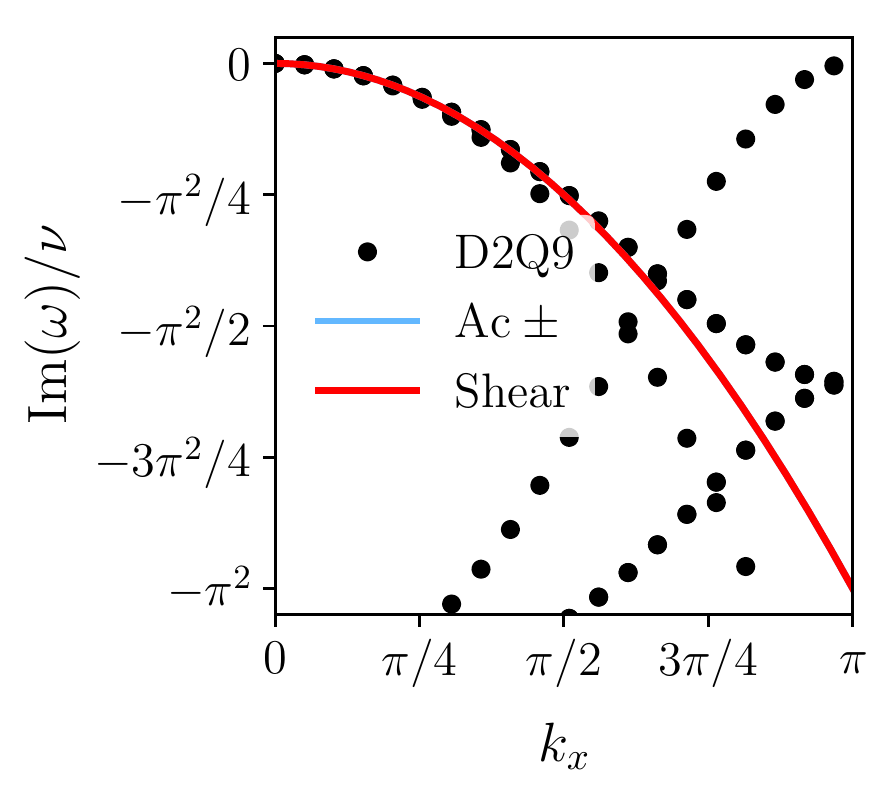} &
\includegraphics[totalheight=0.17\textheight]{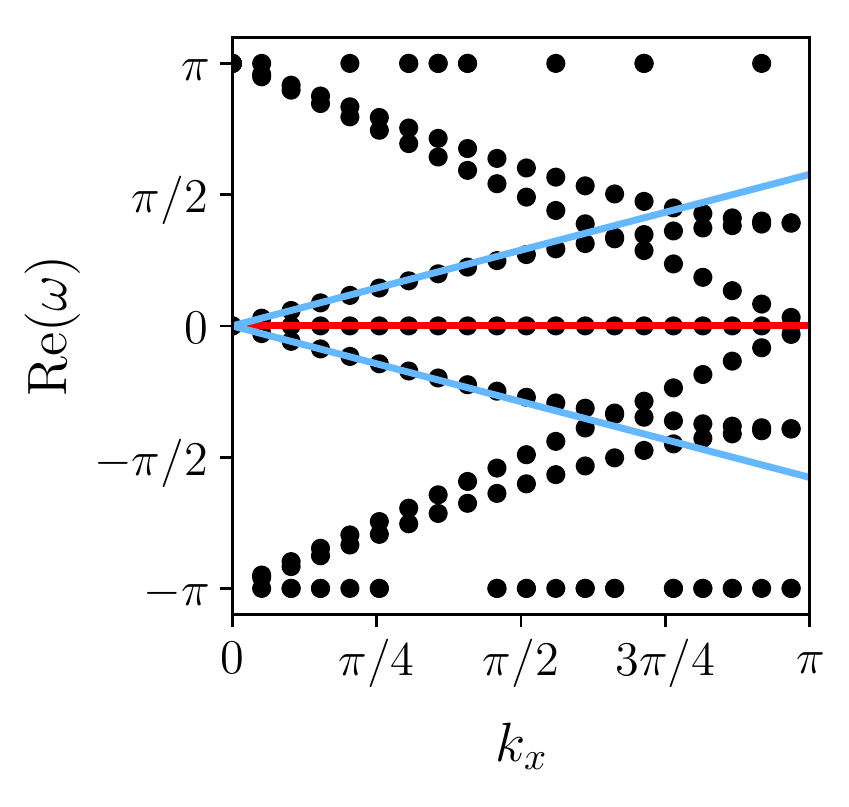}
\end{tabular}
\caption{Spectral properties of the D2Q9-BGK-LBM for $\mathrm{Ma} =0$ and  $\nu = 10^{-6}$. Map of dimensionless maximal growth rate, dissipation rates and dispersion properties (from left to right). Regarding the 2D map, dashed isolines range from $-\pi^2/4$ to $-7\pi^2/4$ with a step of $\pi^2/4$. For 1D plots, symbols and solid lines correspond to LSA results and theoretical curves, where the latter read as: $\mathrm{Im}(\omega_S) = \mathrm{Im}(\omega_\pm) =-\nu k_x^2$, $\mathrm{Re}(\omega_S) = k_x c_s \mathrm{Ma}$ and  $\mathrm{Re}(\omega_{\pm}) = k_x c_s(\mathrm{Ma} \pm 1)$  in lattice units.}
\label{fig:LSA_LBM_ContinuousVSDiscrete}
\end{figure}

As a start, the configuration of a flow at rest ($\mathrm{Ma}=0$), with a viscosity of $\nu=10^{-6}$, is considered. Corresponding results are compiled in \rf{fig:LSA_LBM_ContinuousVSDiscrete}. The most striking result that should be pointed out is the strong anisotropic behavior of the maximal growth rate, as compared to the theoretical one (\rf{fig:LSA_NS_2D_isothermal_LBM}). More precisely, the numerical discretization introduces both dissipation and dispersion errors. In other words, when hydrodynamic waves are not sufficiently discretized (e.g., due to a coarse space step) they will neither propagate at the correct speed nor be dissipated correctly. Another key thing to notice is that non-hydrodynamic behaviors might override hydrodynamic modes, due to their very low dissipation rate, for very high wave numbers ($k\sim \pi$). The latter might then be the source of several spurious behaviors that have been reported in the literature for a flow at rest~\cite{SHAN_PRE_73_2006,LEE_PRE_74_2006,CHEN_IJHMT_76_2014}.

In addition, it should be noted that for a mean flow at rest no positive growth rates were found, meaning that the D2Q9-LBM is linearly stable for $\mathrm{Ma}=0$ no matter how low the value of $\nu$ is. This is true whatever the equilibrium and the collision model considered, with the exception of the \emph{cumulant} based LBM. This seems to be in contradiction with previous studies~\cite{GEIER_CMA_70_2015,GEHRKE_CF_156_2017,GEHRKE_Chapter_2020}, but in fact, this defect might be related to the linear stability analysis itself, which cannot account for non-linear mechanism without increasing the truncation order of the Taylor expansion~(\ref{eq:Perturbed_VDF}).

Eventually, the numerical discretization leads to spurious couplings between modes, that were not observed for the DVBE. These couplings induce energy transfers between non-hydrodynamic and hydrodynamic ones, eventually leading to positive growth rates, i.e., linear instabilities. Consequently, one can expect the critical Mach number, that was obtained with the D2Q9-DVBE, to be an upper limit of linear stability domains obtained after the numerical discretization. This point will be further investigated hereafter.

\subsection{Linear stability domain}
 
\begin{figure}[b!]
\centering
\begin{tabular}{c c}
\hspace*{.65cm}$\mathrm{SRT}$ & \hspace*{-1.6cm} $\mathrm{REG}$\\
\includegraphics[totalheight=0.22\textheight]{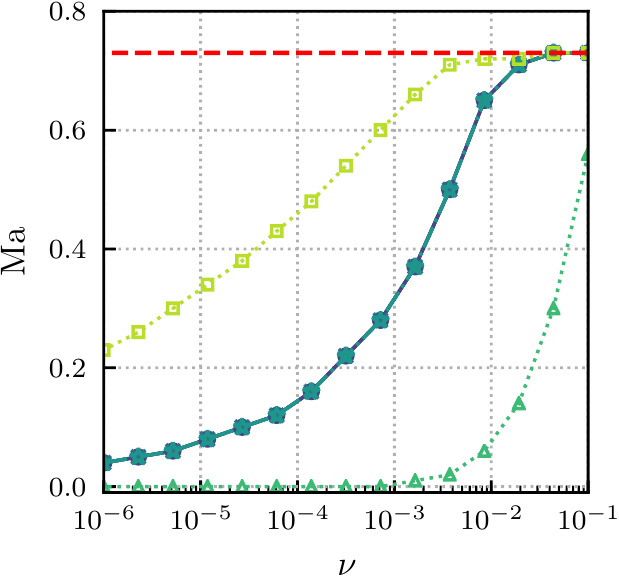} &
\includegraphics[totalheight=0.217\textheight]{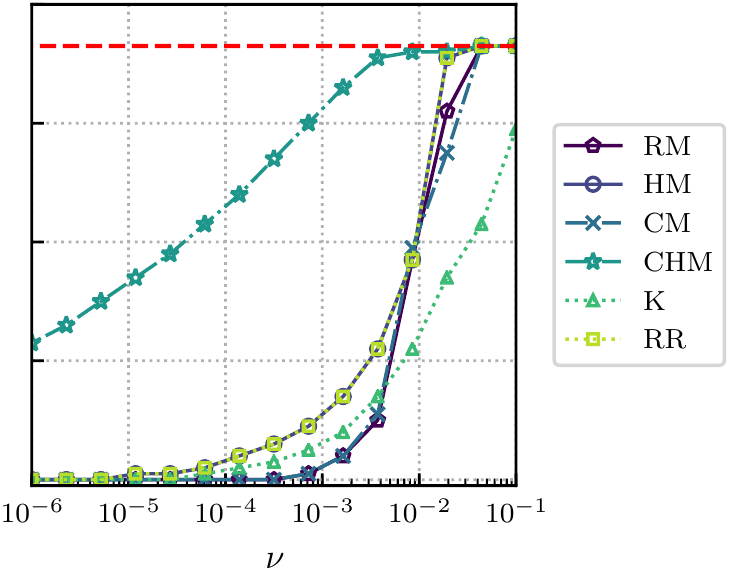}
\end{tabular}
\caption{Linear stability domain of the D2Q9-LBMs. Impact of the collision model using a single relaxation time (left) and equilibrating high-order moments (right). The red dotted line corresponds to the critical Mach number of the D2Q9-DVBE: $\mathrm{Ma}_c=\sqrt{3}-1\approx 0.73$~\cite{WISSOCQ_PhD_2019,MASSET_Accepted_2019}. In the SRT configuration, RM, HM, CM and CHM models recover the same behavior (BGK). For regularized approaches, the CHM also recovers the behavior of the SRT-RR. Both HM and RR further lead to the standard regularized approach as predicted in our previous work~\cite{COREIXAS_PRE_100_2019}.}
\label{fig:LSA_StabDomain_Coll}
\end{figure}

A set of partial differential equations, either continuous or discrete in both space and time, is considered to be linearly stable if and only if the growth rate of perturbations $\mathrm{Im}(\omega)$ remains negative whatever the value of the wavenumber $k$~\cite{CRANK_MPCPS_43_1947,VONNEUMANN_Book_1966,HIRSCH_Book_2007}. Put simply, it translates the fact that perturbations should not be amplified over time by the system, but instead, they should decay following dissipative laws, imposed by both the physics and the numerical discretization, and that depend on their wavenumber $k$.

Hereafter, we will use this stability criterion to compute the maximal Mach number for $\nu\in[10^{-6},10^{-1}]$, where the lower value is typical of airflow simulations in under-resolved conditions while the upper value is chosen so that the continuum limit assumption remains valid. In addition, contrary to previous analyses assuming a critical stability for waves propagating in the direction of the mean flow~\cite{LALLEMAND_PRE_61_2000,SIEBERT_PRE_77_2008}, every combination have been considered here. The same goes for the orientation of the mean flow, meaning that $\bm{u}=(u \cos{\phi}, u \sin{\phi})$ with $\phi\in[0,2\pi]$ in the most general case. The above methodology drastically increases the time required for the computation of the stability domains. Hence, it is worth noting that lattice symmetry properties can be invoked to reduce the number of possibilities for $\phi$, and consequently, speed up the computation of linear stability curves~\cite{WISSOCQ_PhD_2019,HOSSEINI_PRE_99_2019b}.

The comparative study will be restricted to the extended equilibrium \retxt{eq:extendedFeq} so that the impact of each collision model can be isolated. Two configurations frequently encountered in the literature are further considered: (1) the single-relaxation-time (SRT) approach, and (2) the equilibration of high-order moments, the latter corresponding to the standard regularization (REG) procedure based on the extended equilibrium expressed in the moment space of interest~\cite{COREIXAS_PRE_100_2019}. Noting $\omega_n$ the relaxation frequency of the $n$th statistical quantity of interest (RM, HM, CM, CHM and K), the SRT approach is obtained by fixing $\omega_3=\omega_4=\omega_{\nu}$, with $\omega_{\nu}=1/(\nu/c_s^2 + 1/2)$, whereas the REG approach discards high-order non-equilibrium contributions through $\omega_3=\omega_4=1$.

Corresponding results are compiled in \rf{fig:LSA_StabDomain_Coll}, and from them, several remarks can be drawn. 
Qualitatively speaking, it is clear that the collision model does have an impact on the linear stability of LBMs. Interestingly, changing the collision model does not allow us to overcome the DVBE upper limit $\mathrm{Ma}_c=\sqrt{3} -1 \approx 0.73$ that is recovered for very high values of the kinematic viscosity $\nu$~\cite{HOSSEINI_PRE_99_2019b,WISSOCQ_PhD_2019,MASSET_Accepted_2019}. 
From a more quantitative point of view, all moment-based LBMs (RM, HM, CM and CHM) lead to the same behavior (BGK) when (1) a SRT is employed, and (2) the same (extended) equilibrium~\retxt{eq:extendedFeq} is used for all collision models.
Furthermore, it is confirmed that SRT-RR- and SRT-K-LBMs do not recover the behavior of the BGK-LBM. The SRT-RR collision model leads to the widest linear stability range whereas the SRT-K-LBM shows an unexpected linear behavior. 
Mathematically speaking, its main difference with other collision models is the non-linear relaxation related to the fourth-order cumulant.  
Hence, by equilibrating high-order cumulants, one would expect that such a defect would disappear, but surprisingly, the linear stability of the regularized (REG-)K-LBM is not improved. The same goes for other collision models, with the exception of the REG-CHM that recovers the results obtained by the SRT-RR, as mathematically anticipated in our previous work~\cite{COREIXAS_PRE_100_2019}. 

In the end, the LSA provides precious information, notably, regarding the impact of the equilibrium and the collision model on the linear stability of both DVBEs and LBMs. It leads to a first ranking of collision models in terms of linear stability, and confirmes that none of them can help D2Q9-LBMs (based on polynomial equilibria) exceeding the Mach number upper limit ($\mathrm{Ma_c\approx 0.73}$).
Yet, one may wonder if both results only hold in the linear regime, or if they remain valid when running simulations -- for which non-linear phenomena may be encountered. A partial answer can be found in the literature. Indeed, a critical Mach number of approximately $0.73$ was observed by Wilde et al.~\cite{WILDE_IJNMF_90_2019}, by adopting several numerical discretizations of the BGK-DVBE. Hereafter, we will check if this result can be extended to the opposite viewpoint, i.e., for which simulations will be run using a sole numerical discretization (stream-and-collide algorithm), and several collision models will be compared.

\section{Further investigations\label{sec:Further}}

Hereafter, the transport of a vortex by a mean flow is investigated to quantify the numerical stability domain of collision models, which will be compared to their linear one. This may allow us to identify (de)stabilization mechanisms that could not be observed in the linear context. All the simulations are performed using Palabos software~\cite{LATT_CMA_CorrectedProof_2020}, for which, collision models were implemented following instructions provided in our previous work~\cite{COREIXAS_PRE_100_2019}, and recalled in Appendix~\ref{sec:CollModels}. 

\subsection{From linear to numerical instabilities}

\begin{figure}[t!]
\centering
\begin{tabular}{c @{\,} c @{\,} c}
\includegraphics[trim={0.25cm 0 1.78cm 0},clip,totalheight=0.175\textheight]{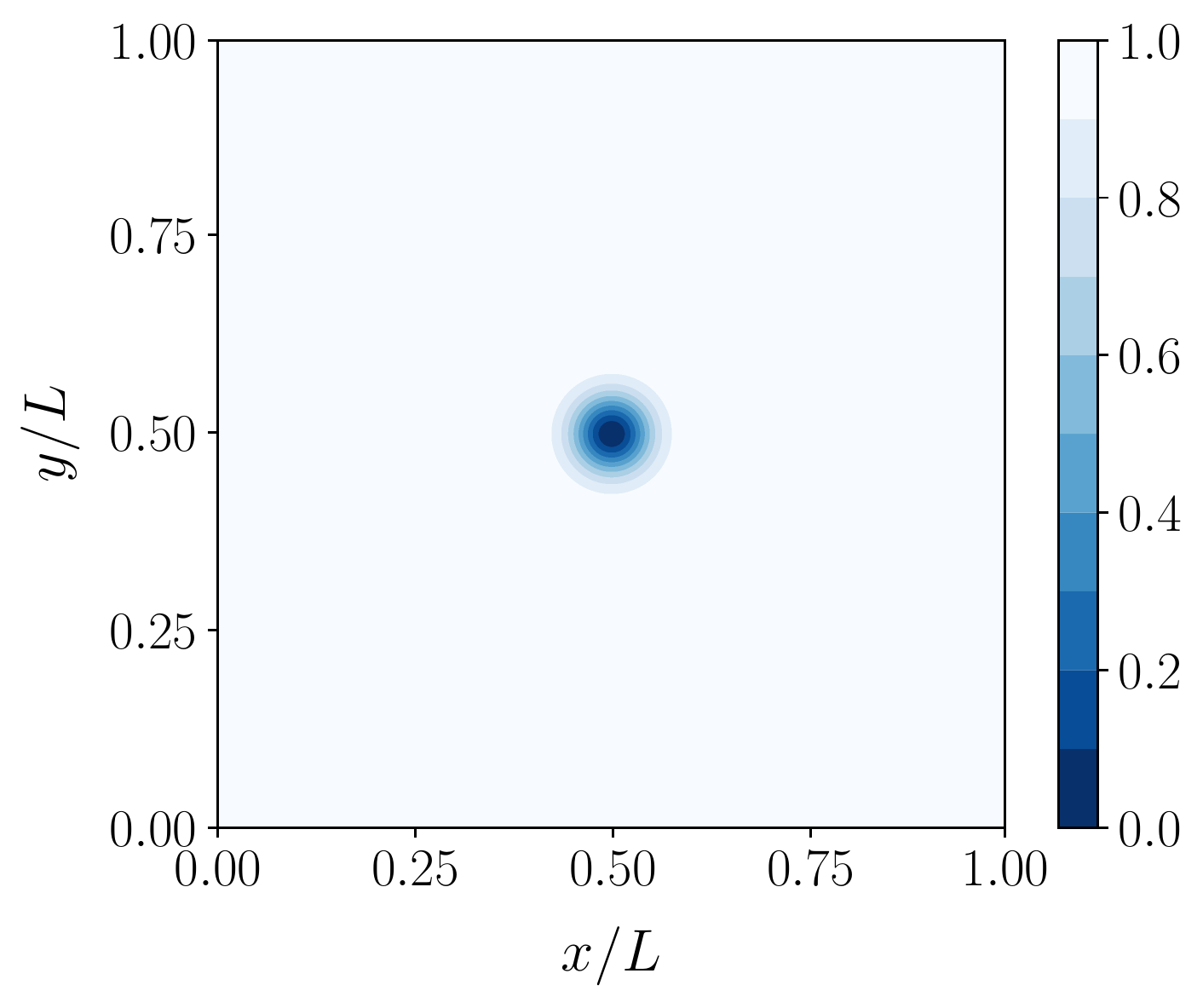} &
\includegraphics[trim={1.5cm 0 0 0},clip,totalheight=0.175\textheight]{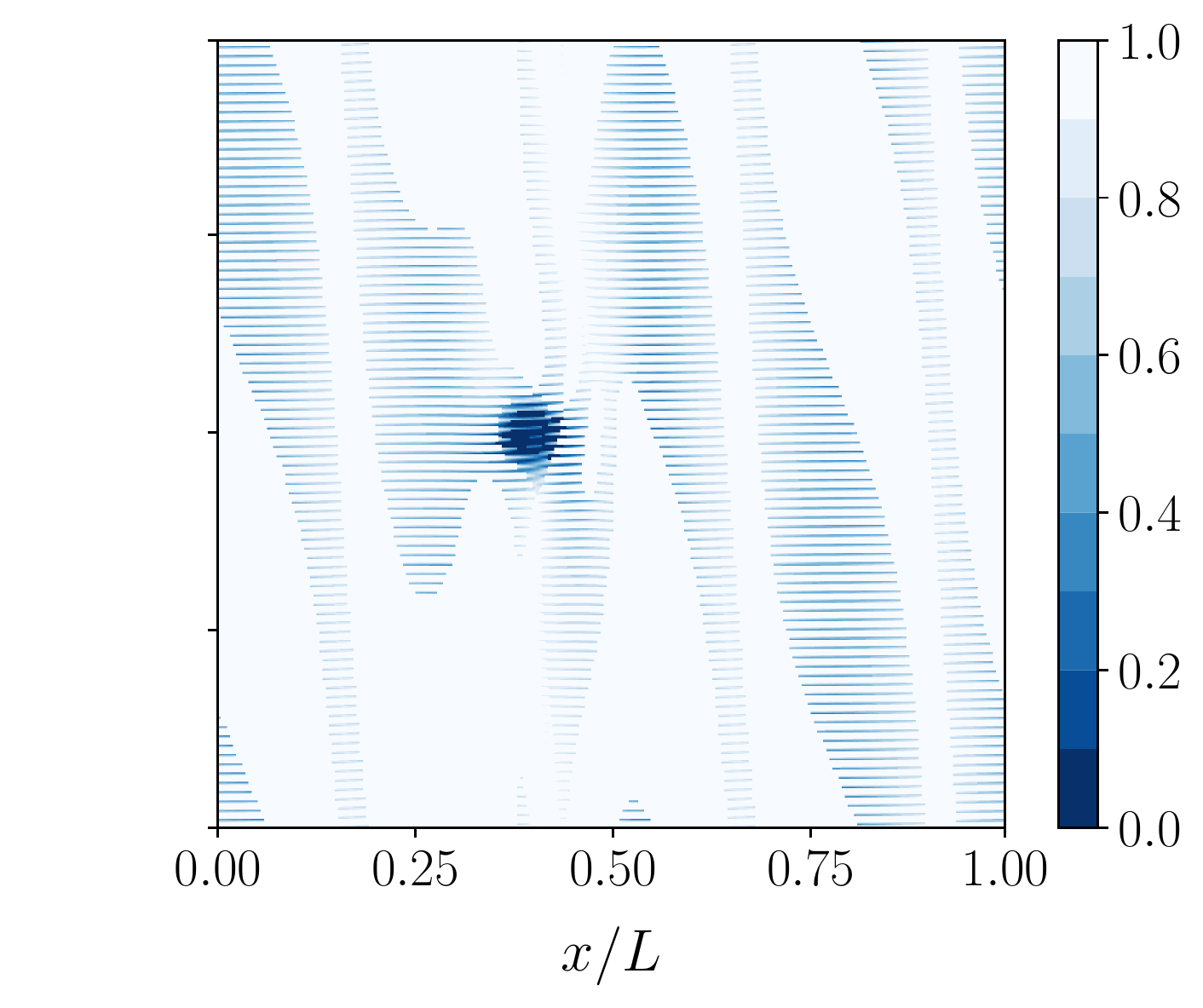} &
\includegraphics[totalheight=0.175\textheight]{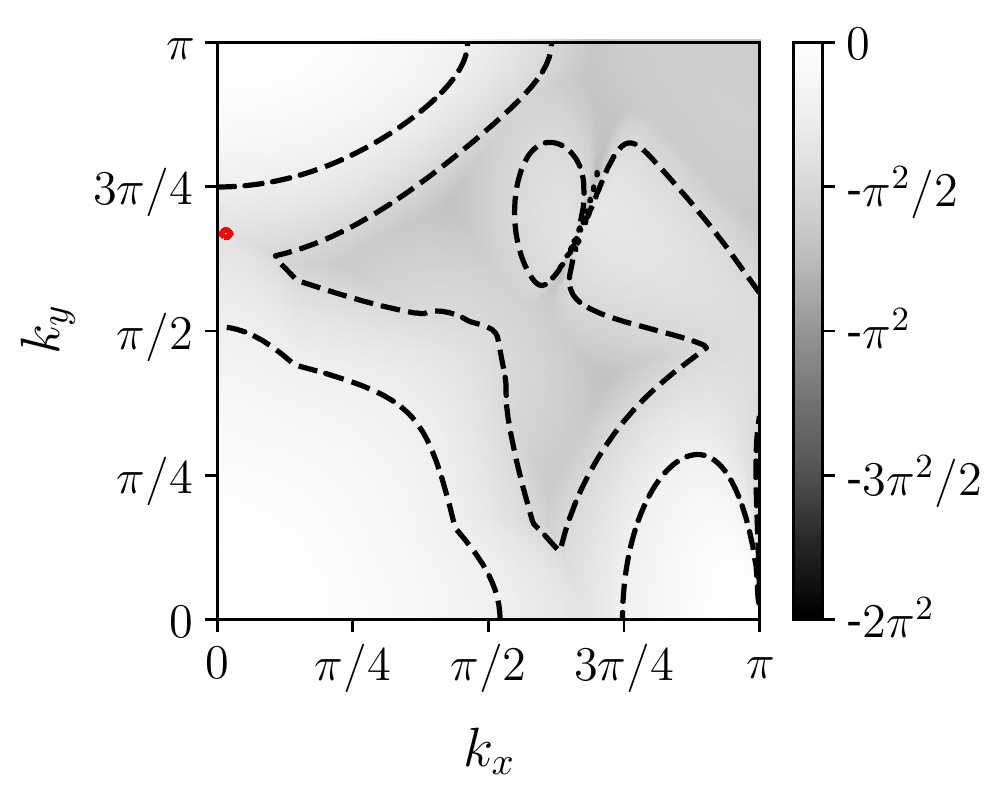}
\end{tabular}
\caption{Isothermal convected vortex simulated with the D2Q9-BGK-LBM based on the second-order equilibrium: $\mathrm{Ma}=0.1$, $\nu=10^{-5}$ and $L=256$. From left to right, normalized density field $(\rho - \rho^{\mathrm{min}}_{t=0})/(\rho^{\mathrm{max}}_{t=0} - \rho^{\mathrm{min}}_{t=0})$ at $t=0$, $t=4.9 t_c$, and 2D map of the maximal growth rate for the same parameters. In the latter, an instability bubble arises close to $k_x\approx 0.1$ and $k_y\approx 2.1$.}
\label{fig:instabilityCoVo}
\end{figure}

This testcase consists in the superposition of a uniform flow (density $\rho_0$, velocity $\bm{u}=(u_x,u_y)$ and temperature $T_0$) and a vortical structure (characteristic radius $R$ and strength $\beta$) located at point $(x_c,y_c)$. 
This test is of particular interest since it complies with the linear assumptions -- small perturbations with respect to the mean flow and periodic simulation domain -- which allows us to directly compare the present (numerical) stability domains with those computed using the LSA. 

In the context of isothermal LBMs, the isentropic initialization of the vortex (as proposed in Ref.~\cite{WANG_IJNMF_72_2013}) cannot be applied as it is. Gendre \emph{et al.}~\cite{GENDRE_PRE_96_2017} proposed to derive its isothermal counterpart by applying a Taylor expansion about the lattice temperature $T_0=c_s^2$. The first-order approximation leads to the initial state $(t=0)$
\begin{equation}
\rho =\rho_0\left[1-\tfrac{\epsilon^2}{2c_s^2}\exp{(-r^2)}\right], \:u_x = u_0-\tfrac{\epsilon (y-y_c)}{R}\exp{(-r^2/2)},\: u_y = \tfrac{\epsilon (x-x_c)}{R}\exp{(-r^2/2)},
\end{equation}
with $\epsilon=\beta u_0$, $\beta=0.5$, $R=1/20$, $r=\sqrt{(x-x_c)^2 + (y-y_c)^2}/R$ and $u_0 = \mathrm{Ma}\, c_s$. 

Before moving to the evaluation of the stability domain of each collision model, let us have a look at the kind of instabilities that emerge during simulations. For example, simulations based on the BGK-LBM with the second-order equilibrium show the emergence of unstable waves in the low-viscosity regime ($\nu=10^{-5}$) for a relatively small Mach number ($\mathrm{Ma=0.1}$). This is illustrated in Fig.~\ref{fig:instabilityCoVo}, where after several characteristic times ($t_c=L/u_0$ with $L$ the number of points per direction), a spurious wave emerges and eventually leads to the blow-up of the simulation. This spurious phenomenon can be explained by looking at the 2D map of the maximal growth rate where the coupling between two modes leads to a small instability bubble close to $k_x\approx 0.1$ and $k_y\approx 2.1$. The latter wavenumbers correspond to space frequencies that are observed in the above simulation (low-frequency in the $x$-direction and high-frequency in the $y$-direction). This shows that the LSA -- which in our case relies on a uniform base flow -- seems to be able to predict instabilities arising during numerical simulations, \emph{if non-linear phenomena do not come into play}.

\subsection{Methodology}

As for stability domains obtained with the LSA, two configurations of relaxation parameters will be investigated hereafter, namely, SRT ($\omega_{\nu}=\omega_3=\omega_4$) and REG ($\omega_3=\omega_4=1$). The stability criterion is based on the normalized kinetic energy averaged over the whole simulation domain $\langle u^2 \rangle / u_0^2$, that is supposed to monotonically decrease over time for this testcase. Hence, for a given kinematic viscosity $\nu$ and Mach number $\mathrm{Ma}$, if it becomes larger than one (before a certain number $M$ of characteristic time $t_c$) then the simulation is very likely to be unstable in the next few time iterations. Hence the stability criterion reads $\langle u^2 \rangle / u_0^2<1$ for $t\leq M t_c$, with $M$ a parameter to be defined, and $L$ being the number of points used to discretized the $L\times L$ simulation domain.

Before comparing the stability domains obtained for all collision models of interest, the impact of the parameter $M$ has been investigated. Results obtained for the most sensitive collision model (which was identified as the BGK operator) are plotted for different mesh sizes in \rf{fig:CoVo_ImpactTc_RM_SRT}. At first glance, the fact that the stability of LBMs depends on the number of iterations (required to run the simulation) might be surprising. Nonetheless, one must remember that in the context of LSA, even if a wave is proven to lead to a linear instability, it will take some time before it emerges from the uniform background of the flow. Especially, if its growth rate is extremely low, let us say $\mathrm{Im}(\omega)=10^{-5}$, then its amplitude would be increased by a factor $\exp[\mathrm{Im}(\omega)]\approx 1 + 10^{-5}$ at each iteration. Consequently, it would require several tens or hundreds of thousands of iterations before it leads to the blow-up of the simulation. In any case, it seems safe to assume that $M=50$ is sufficient to accurately evaluate the numerical stability domains of LBMs. In addition, the stability domains obtained with the D2Q9-BGK clearly show a dependence on the grid mesh. This translates the fact that the grid mesh acts as a space filter which will automatically prevent the growth of (spurious) waves with a particular space frequency. In other words, even if the LSA shows that a particular type of LBM should encounter stability issues due to the growth of spurious waves for a given set of parameters ($\mathrm{Ma}$, $\nu$, etc), they might not emerge during the simulation because of the numerical sampling of the discrete wavenumbers, and the simulation might then remain stable. This was pointed out, for example, in Refs.~\cite{COREIXAS_PhD_2018,WISSOCQ_PhD_2019}, and it is further illustrated in Appendix~\ref{sec:filtering} for the standard regularized collision model using the double shear layer testcase~\cite{BROWN_JCP_122_1995,MINION_JCP_138_1997}. 

\begin{figure}[t!]
\centering
\begin{tabular}{c c}
\hspace*{.65cm}$L=32$ & \hspace*{-1.5cm} $L=256$\\
\includegraphics[totalheight=0.22\textheight]{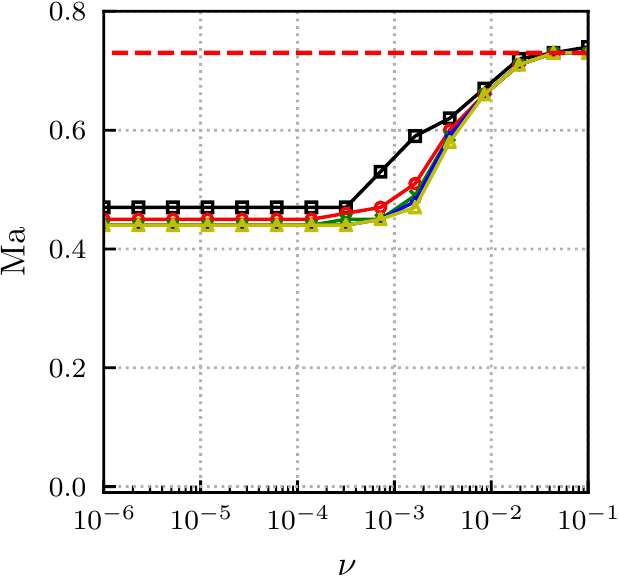} &
\includegraphics[totalheight=0.217\textheight]{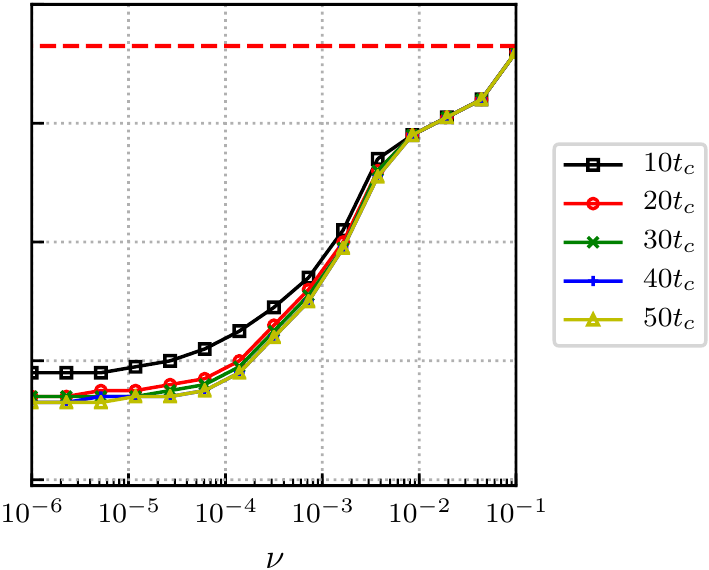}
\end{tabular}
\caption{Isothermal convected vortex. Impact of the number of characteristic times $M$ on the stability domain of the D2Q9-LBM, with the BGK operator based on the extended equilibrium, and using $L=32$ and $256$ points (from left to right). The red dotted line corresponds to the critical Mach number of the D2Q9-DVBE: $\mathrm{Ma}_c=\sqrt{3}-1\approx 0.73$~\cite{WISSOCQ_PhD_2019,MASSET_Accepted_2019}.}
\label{fig:CoVo_ImpactTc_RM_SRT}
\end{figure}

\subsection{Numerical stability domains}

Stability curves obtained for all collision models of interest (RM, HM, CM, CHM, K and RR) are compiled in \rf{fig:CoVo_StabDomain_Coll} for $L=256$, $M=50$ and considering the two configurations: SRT and REG. 

From them, several mathematical results obtained in our previous work~\cite{COREIXAS_PRE_100_2019} can further be confirmed. 
First, when only one relaxation time is used (SRT configuration), all moment based LBMs recover the same behavior (BGK) with the exception of the cumulant approach. The RR-LBM further diverges from the BGK-LBM due to the reconstruction of the non-equilibrium part of populations through recursive formulas. When high-order moments are equilibrated, the REG-CHM model recovers the behavior of the SRT-RR approach. The REG-HM and REG-RR further leads to the standard regularized approach. The latter is explained by the fact that this configuration (REG) amounts to discarding non-equilibrium part of populations that are computed using recursive formulas.

In addition, it is confirmed that none of the collision model is able to remove the Mach number limitation, that was obtained in the linear case, and which is imposed by the lattice and the equilibrium. In fact, by adopting a different type of equilibrium, e.g., the entropic formulation obtained by minimizing the H-functional under the constraint of density and momentum conservation~\cite{ANSUMALI_EPL_63_2003}, this limitation could be alleviated but at the cost of part of the physics~\cite{HOSSEINI_PRE_100_2019}. 
To improve both the physical and numerical properties of LBMs, one could include correction terms for the velocity-dependent errors that remain in the definition of the viscous stress tensor~\cite{PRASIANAKIS_PRE_76_2007,PRASIANAKIS_PRE_78_2008,DELLAR_JCP_259_2014,GEIER_CMA_70_2015,FENG_JCP_394_2019,RENARD_ARXIV_2020_03644,TAYYAB_CF_211_2020,HOSSEINI_RSTA_378_2020}. 

Going into more details, the SRT-RR approach clearly outperforms other collision models based on a SRT. Even though the SRT-K model leads to better stability domains than in the linear case, it remains pretty similar to those obtained with the BGK operator while being far more complicated. Interestingly, the tendency is reversed for the REG configuration where it leads to better stability curves than standard moment space approaches (RM, HM and CM), which confirms tendencies observed in previous works~\cite{GEIER_CMA_70_2015,GEHRKE_CF_156_2017,GEHRKE_Chapter_2020}. Regarding the only moment space that was not investigated by the latter authors (CHM), it also leads to very satisfying results in terms of stability. 

In addition, while all collision models lead to similar stability properties for high values of the kinematic viscosity $\nu$, the SRT-RR, REG-CHM and REG-K seem to be the best candidates for the simulation of isothermal flows in the zero-viscosity limit. It is also worth noting that despite its simplicity, the BGK operator based on the extended equilibrium~(\ref{eq:extendedFeq}) also leads to rather satisfying results, as already pointed out by Geier et al.~\cite{GEIER_CMA_70_2015}. 

\begin{figure}[t!]
\centering
\begin{tabular}{c c}
\hspace*{.65cm}$\mathrm{SRT}$ & \hspace*{-1.6cm} $\mathrm{REG}$\\
\includegraphics[totalheight=0.22\textheight]{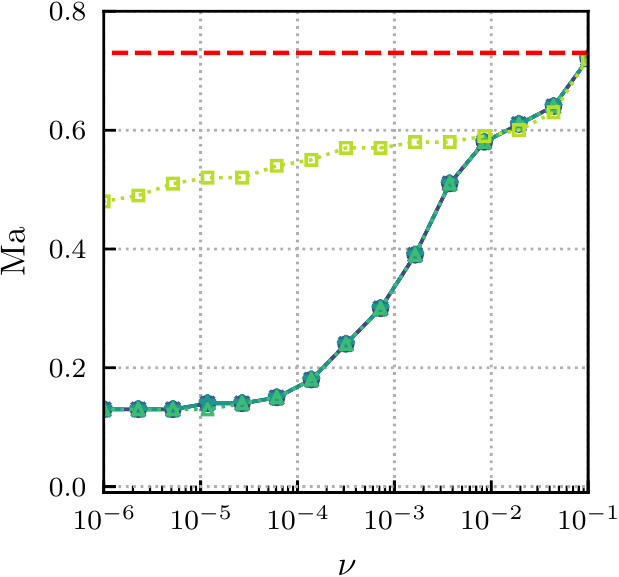} &
\includegraphics[totalheight=0.217\textheight]{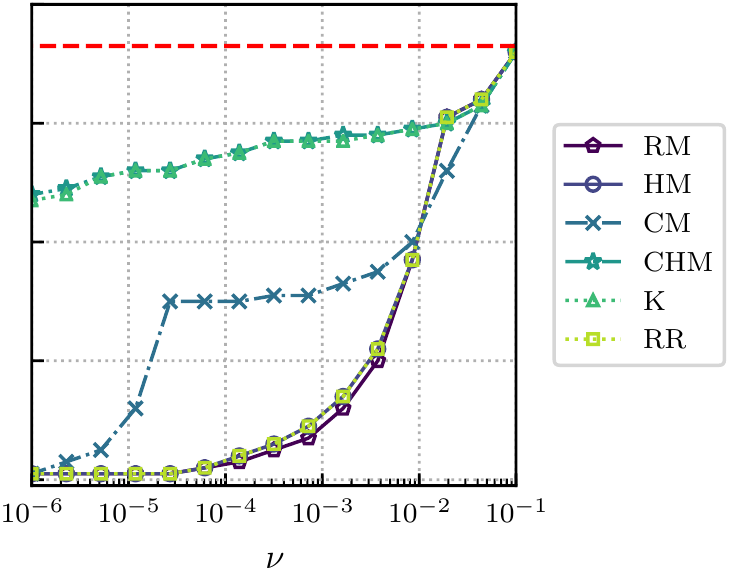}
\end{tabular}
\caption{Isothermal convected vortex with $L=256$. Impact of the collision model on the stability domain of the D2Q9-LBMs: using a single relaxation time $\omega_{\nu}=\omega_3=\omega_4$ (left), and equilibrating high-order moments $\omega_3=\omega_4=1$ (right). The red dotted line corresponds to the critical Mach number of the D2Q9-DVBE: $\mathrm{Ma}_c=\sqrt{3}-1\approx 0.73$~\cite{WISSOCQ_PhD_2019,MASSET_Accepted_2019}. In the SRT configuration, RM, HM, CM and CHM models recover the same behavior (BGK). For regularized approaches, the CHM also recovers the behavior of the SRT-RR. Both HM and RR further lead to the standard regularized approach as predicted in our previous work~\cite{COREIXAS_PRE_100_2019}.}
\label{fig:CoVo_StabDomain_Coll}
\end{figure}

\subsection{Impact on the accuracy}

\begin{figure}[t!]
\centering
\begin{tabular}{c @{\,} c @{\,} c}
\includegraphics[trim={0 1.4cm 1.4cm 0},clip,totalheight=0.166\textheight]{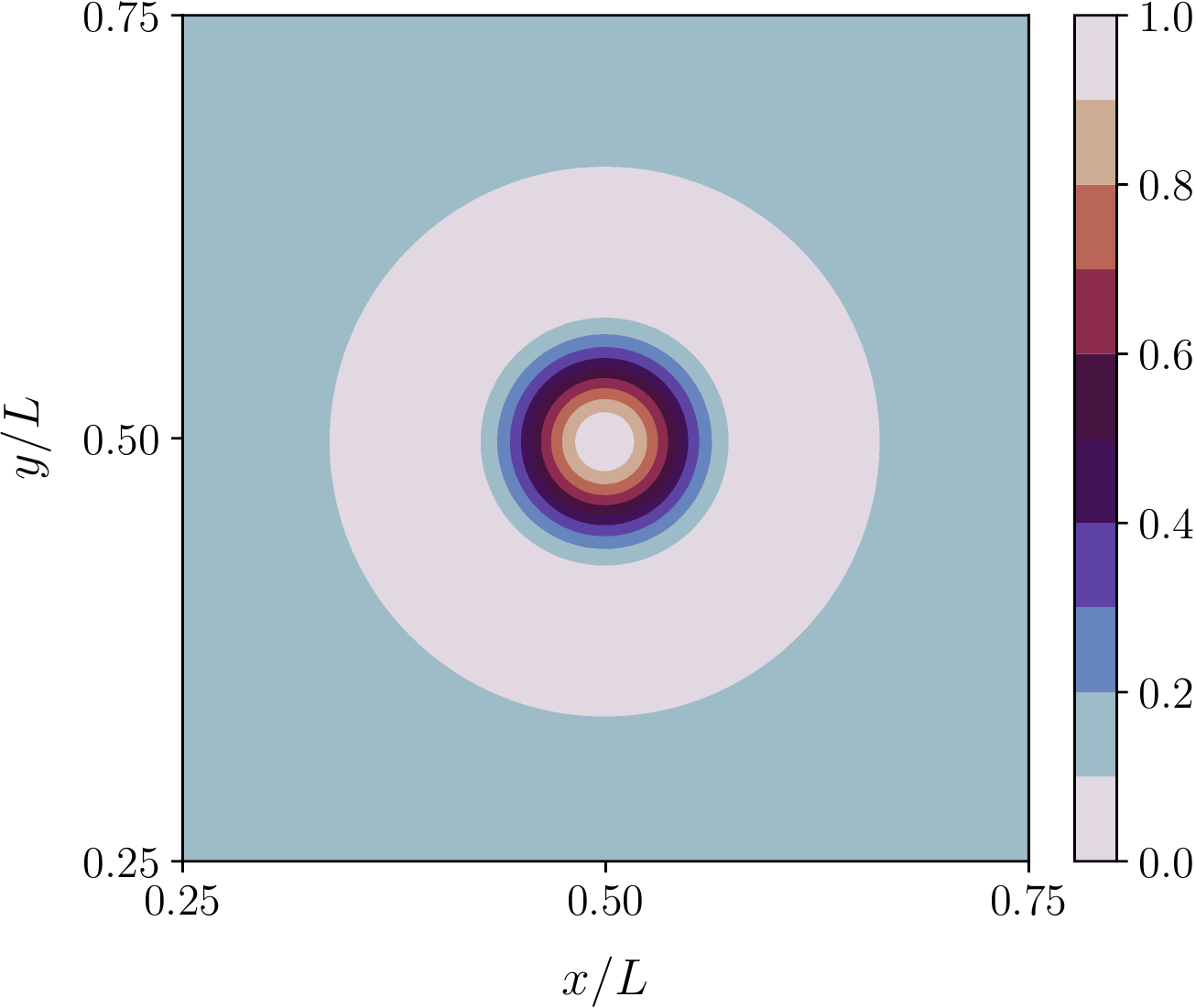} &
\includegraphics[trim={1.25cm 1.4cm 1.4cm 0},clip,totalheight=0.166\textheight]{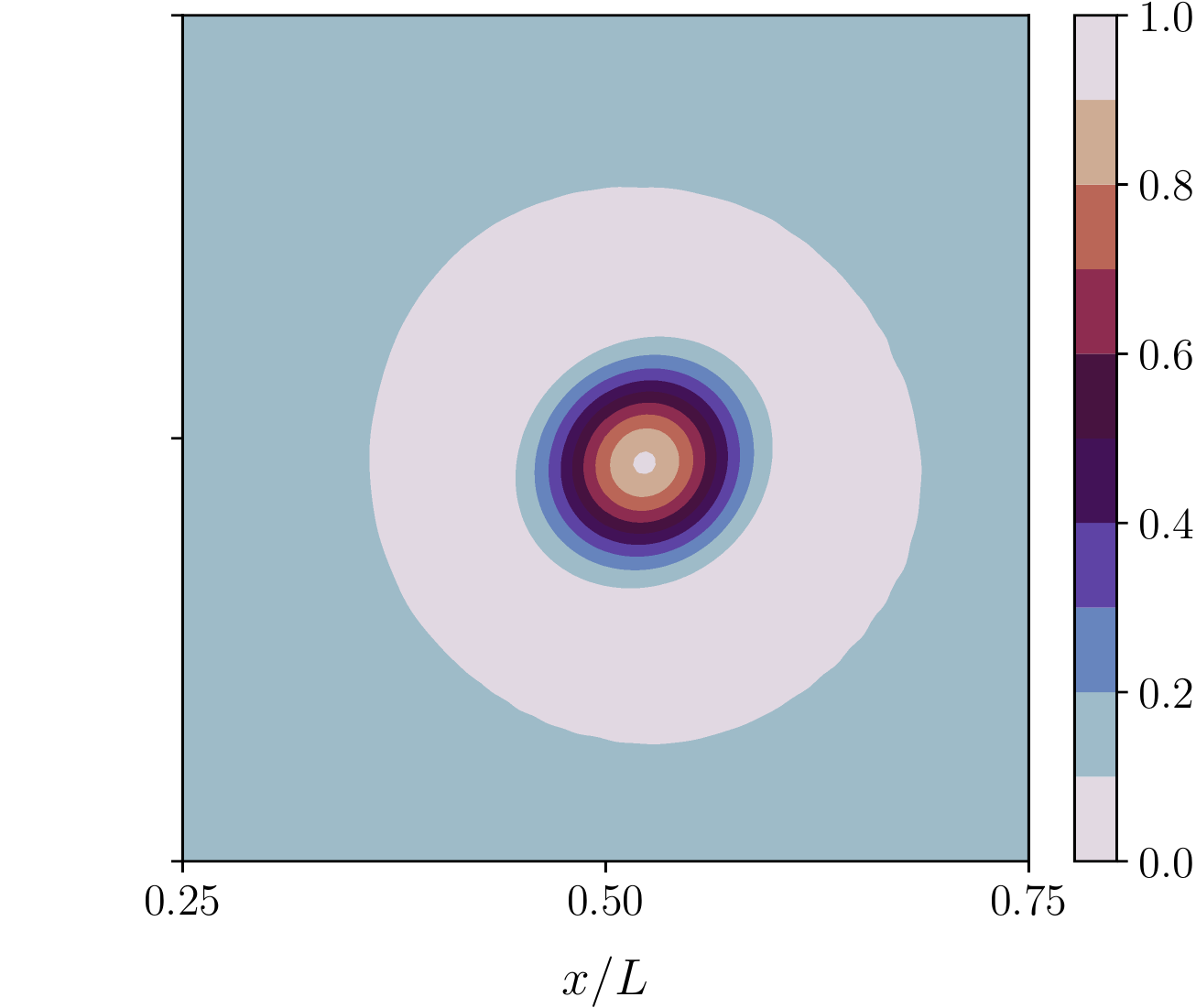} &
\includegraphics[trim={1.25cm 1.4cm 0 0},clip,totalheight=0.166\textheight]{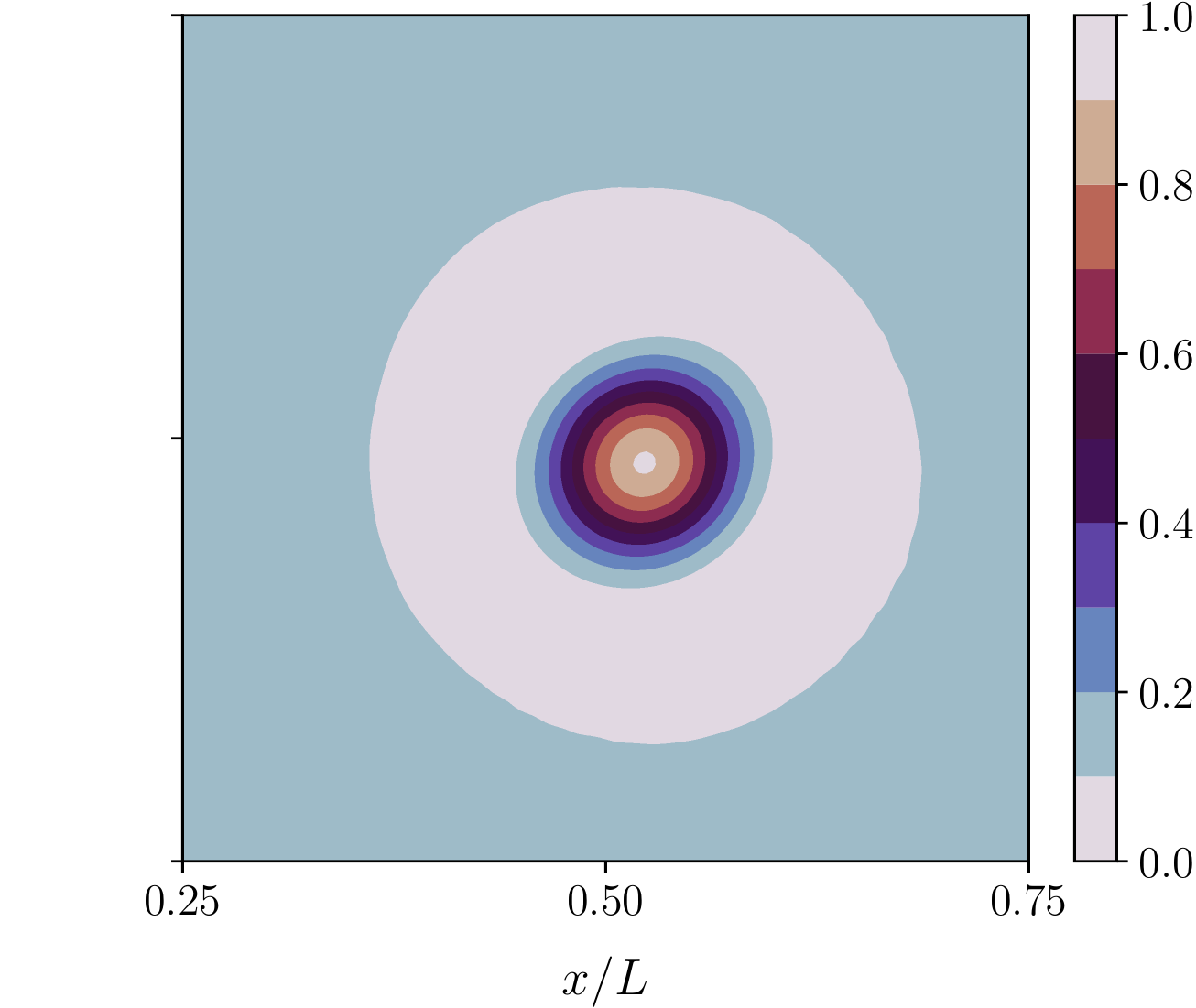}\\
\includegraphics[trim={0 0 1.4cm 0},clip,totalheight=0.19\textheight]{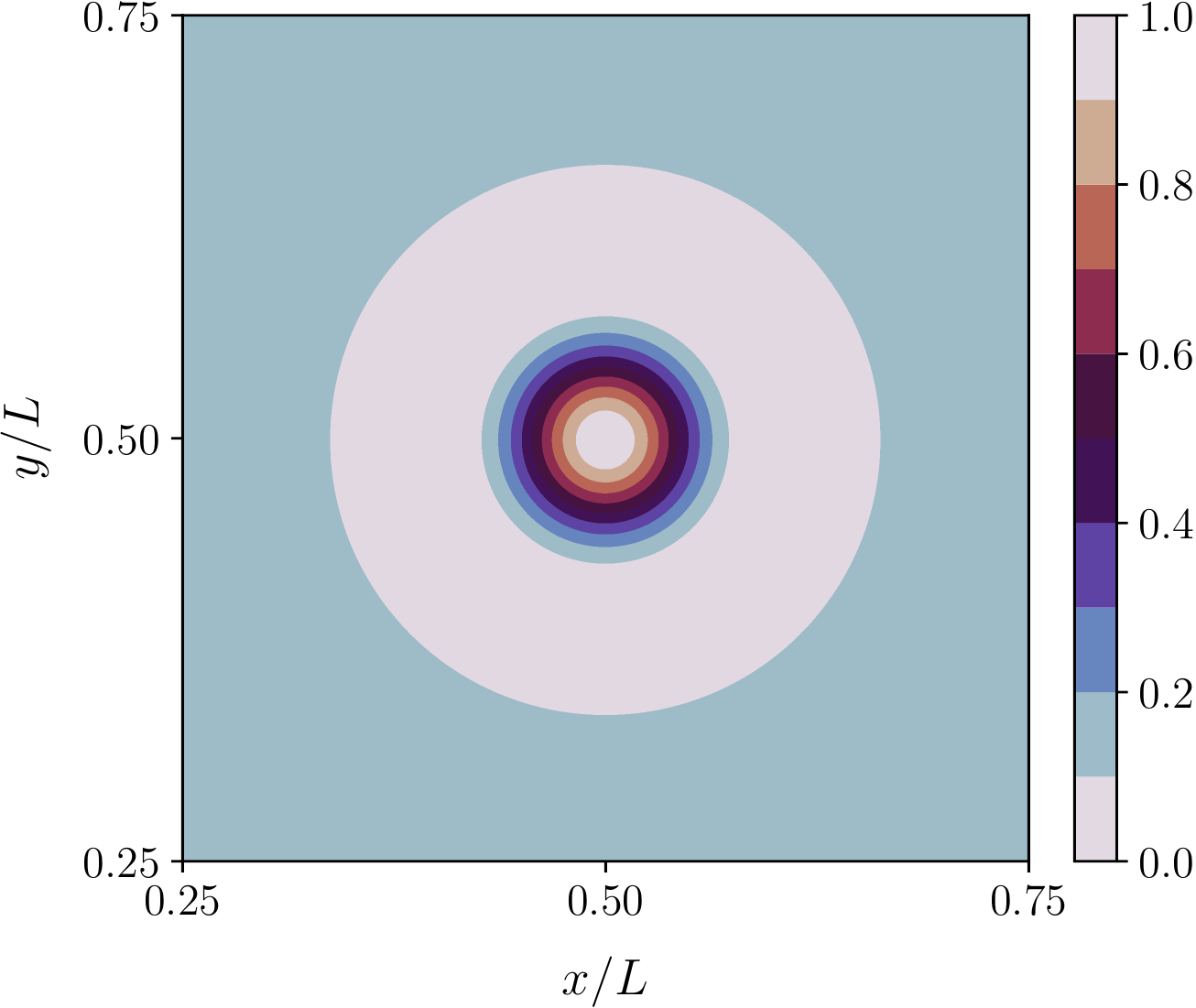} &
\includegraphics[trim={1.25cm 0 1.4cm 0},clip,totalheight=0.19\textheight]{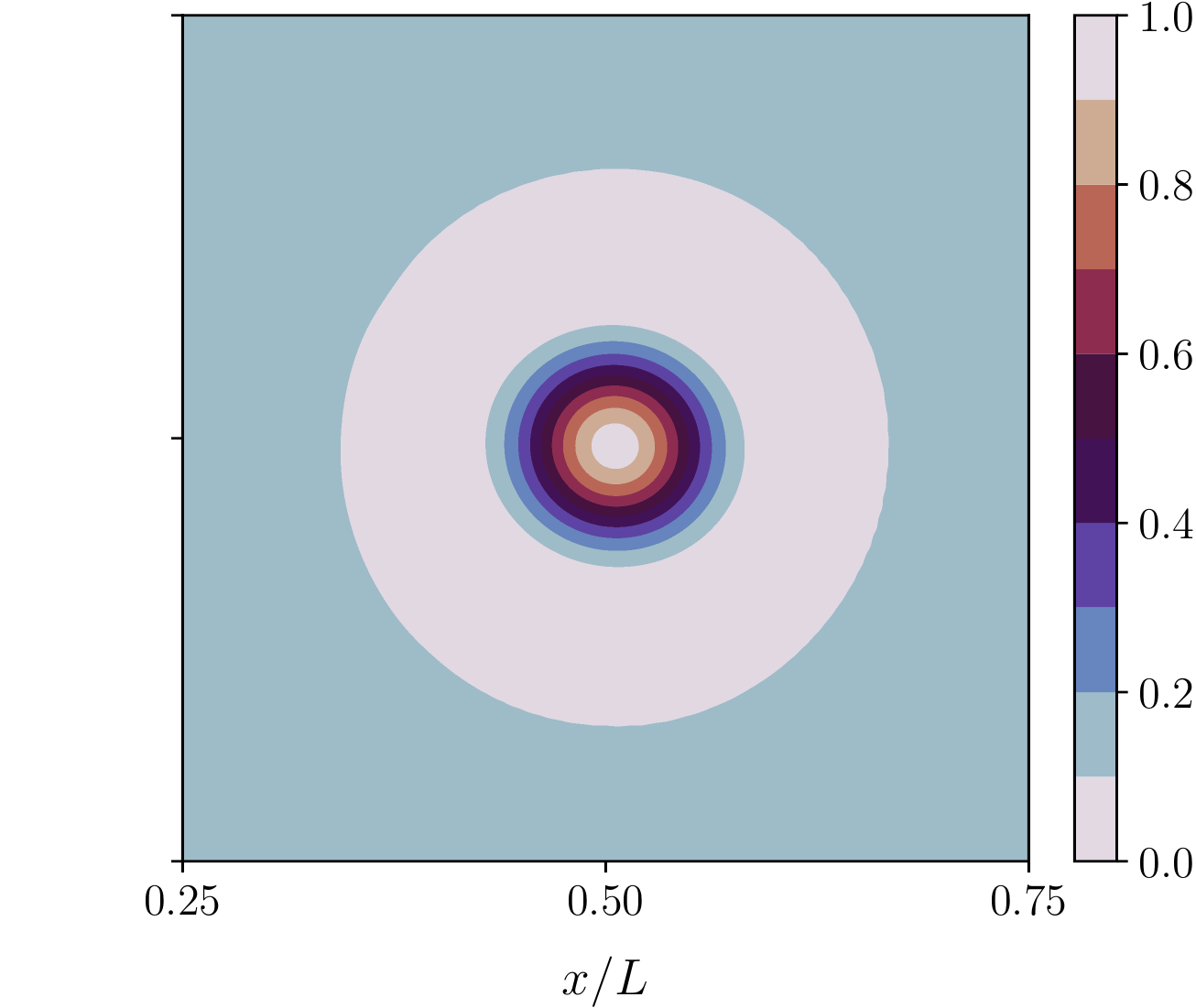} &
\includegraphics[trim={1.25cm 0 0 0},clip,totalheight=0.19\textheight]{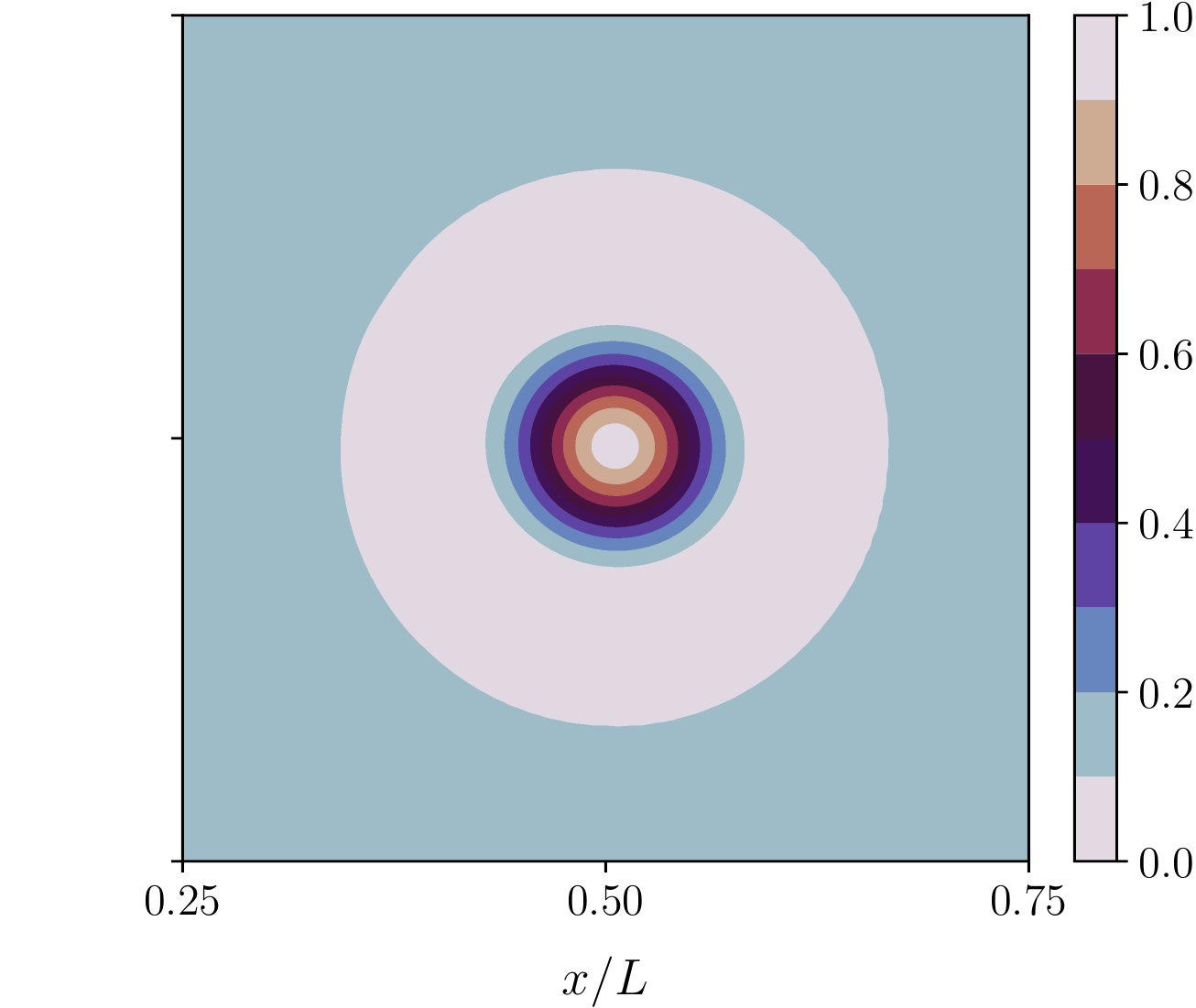}
\end{tabular}
\caption{Normalized vorticity field $(\omega_z - \omega^{\mathrm{min}}_{z,t=0})/(\omega^{\mathrm{max}}_{z,t=0} - \omega^{\mathrm{min}}_{z,t=0})$ of the isothermal convected vortex simulated with $\mathrm{Ma}=0.5$, $\nu=10^{-4}$, $L=256$ (top) and $L=512$ (bottom). From left to right: initial state ($t=0$), REG-K and SRT-RR ($t=50 t_c$). The REG-CHM collision model leads to results identical with those obtained with the SRT-RR.}
\label{fig:stabCollModel}
\end{figure}

Nevertheless, when it comes to numerics, it is not sufficient to only focus on the impact of collision models on the robustness of LBMs without also considering the impact on their accuracy. The latter point is investigated for the most stable collision models (SRT-RR, REG-CHM and REG-K) considering a configuration which is close to their stability limit, i.e., $\mathrm{Ma}=0.5$ and $\nu=10^{-4}$. Corresponding results are plotted in Fig.~\ref{fig:stabCollModel}. The visualization of the normalized vorticity fields at $t=50 t_c$ is very satisfying for all collision models --whose accuracy is almost identical. More precisely, using a moderate resolution of the simulation domain ($L=256$ points per direction), the vortex is properly convected over a long distance, and shows only little dispersion and dissipation errors. 
The dispersion error is identified by looking at the location and the shape of the vortex, whereas the dissipation error is quantified focusing on isocontours of the normalized vorticity. By increasing the resolution of the simulation domain, e.g., considering $L=512$, these errors are further reduced but are still observable. The remaining dissipation error can be explained, in part, because $\nu\neq 0$. Interestingly, the fact that LBMs introduce more dispersion than dissipation errors, as shown by Mari\'{e} et al.~\cite{MARIE_JCP_228_2009}, is further confirmed here by the non-circular shape of the vortex for the finest resolution.

In the end, taking one step back and looking at the big picture, it seems clear that these collision models (SRT-RR, REG-CHM and REG-K) not only lead to stable simulations for non-negligible Mach numbers and low-viscosity values, but they also allow the resulting LBMs to accurately transport information over \emph{long} distances, even if one must keep in mind dispersion issues. Of course, more numerical validations are required to extensively quantify the accuracy of these models, for the simulation of finite Mach number flows in the low-viscosity regime. Nevertheless, these first results are very promising.

\section{Conclusion \label{sec:Conclu}}

The quest for always more sophisticated collision models has led many researchers to propose new approaches with various degrees of success.
In this context, our work further confirmed that only the equilibrium state does impact the macroscopic behavior of LBMs, as long as, the continuum limit remains valid.
The equilibrium is then the key element to improve at the same time the physical and numerical properties of LBMs. 

Another very powerful, known feature of some LBMs was further highlighted by this quest: their numerical properties can be drastically improved by simply changing the collision model. In the context of CFD based on NS solvers, such an achievement cannot be reached without adopting complex numerical schemes which usually require extended stencils~\cite{HIRSCH_Book_2007}. The latter eventually deteriorates the parallel performances of the corresponding solvers, whereas the change of collision model is local, and it barely impacts the ratio between the wall-clock times required to run (1) a LB simulation and (2) its NS counterpart. By relying on our uniform framework~\cite{COREIXAS_PRE_100_2019}, this work confirmed that by adopting various collision models and adjusting their free parameters (i.e., relaxation frequencies), the linear and numerical properties of LBMs can be drastically improved with only little impact on their efficiency. Interestingly, it was shown that errors flowing from the velocity discretization, and which have a direct impact on the stability (the critical Mach number of the DVBE), could not be canceled out by simply changing the collision model.   

Going into more details, we presented a comprehensive comparison of the linear stability domain of the most common collision models (RM, HM, CM, CHM, K, RR) for two standard configurations encountered in the literature, namely, single relaxation time (SRT) and equilibration of high-order moments (REG). Thanks to this, general conclusions were drawn regarding the stabilization properties of each collision model in the linear regime. In addition, the numerical stability analysis, based on the transport of a vortex by a uniform flow, further allowed us to (1) account for non-linear (de)stabilization mechanisms, and to (2) propose a first evaluation of the impact of (the most stable) collision models on the accuracy of the resulting LBM.

As a conclusion, this work showed that, in terms of stability, accuracy, and for the particular D2Q9 velocity discretization, the SRT-RR, REG-CHM and REG-K collision models are the best candidates (based on static relaxation parameters) for the simulation of isothermal flows in the low-viscosity regime and at finite Mach numbers. This is in accordance with results that were previously reported in the literature~\cite{MALASPINAS_ARXIV_2015,GEIER_CMA_70_2015,COREIXAS_PRE_96_2017,BROGI_JASA_142_2017,MATTILA_PF_29_2017,GEHRKE_CF_156_2017,LI_PRE_100_2019,SHAN_PRE_100_2019}. Yet, SRT and REG configurations are only two possible configurations among many others. Hence, it might be possible to further improve the stability properties of these collision models by fine tuning the free parameters that control the relaxation of high-order moments. This was done, for example, in the context of lattice kinetic schemes where it led to very promising results~\cite{HOSSEINI_PRE_99_2019b}. One could also want to dynamically compute these free relaxation frequencies following a predefined paradigm. One possibility would be to make them depend on (shear and ghost) non-equilibrium populations through an approximation to the maximum entropy principle~\cite{BOSCH_ESAIM_52_2015}. Both approaches are currently under investigation. Extension of these comparative studies to more quantitative testcases, as well as, 3D models (D3Q19, D3Q27, etc), will be presented in forthcoming papers.

\appendix

\section{Recalls on linear stability analyses \label{sec:AppLSA}}

\subsection{Perturbed equations \label{sec:PE}}

To derive linear equations related to perturbations of interest, a Taylor expansion of the populations and of the corresponding collision operator is performed about a mean flow, neglecting second- and high-order terms:
\begin{equation}
 f_i \approx \overline{f_i} + f'_i,
 \label{eq:Perturbed_VDF} 
 \end{equation} 
with $\overline{f}_i=f_i^{eq}\vert_{\overline{\rho},\overline{\bm{u}}}$ (uniform flow), and the overline notation stands for mean quantities in space and time.
Applying the same expansion to the general form of the collision term $\Omega_i$, and noticing that it is linked to $f_j$ through the macroscopic quantities used in the definition of the equilibrium state $f_i^{eq}$, we end up with
\begin{equation}
\Omega_i(f_j) \approx \Omega_i\vert_{\overline{f_j}} + J_{ij}f'_j,
\label{eq:Perturbed_Collision}
\end{equation}
where $J_{ij} = {\left. \frac{\partial\Omega_i}{\partial f_j}\right\vert_{\overline{f_j}}}$ is the Jacobian matrix of the collision operator evaluated at $f_j = \overline{f_j}$. By injecting Eqs.~\retxt{eq:Perturbed_VDF} and~\retxt{eq:Perturbed_Collision} in both the force-free DVBE~\retxt{eq:DVBE} and its space/time discretization~\retxt{eq:StreamCollide}, and noticing that $\overline{f_i}$ are particular solutions of the resulting equations, two perturbed equations are obtained:
\begin{itemize}
\item Perturbed continuous DVBE
\begin{equation}
\partial_t f_i' + \bm{\xi}_i\cdot\bm\nabla f_i' = J_{ij}f'_j.
 \label{eq:ContinuousPerturbedEq}
\end{equation}
\item Perturbed space/time discrete DVBE
\begin{equation}
 f'_i(\bm{x}+\bm{\xi}_i, t + 1) =  \left[\delta_{ij} + J_{ij}\right] f'_j(\bm{x}, t).
 \label{eq:DiscretePerturbedEq}
\end{equation}
\end{itemize}
These two equations are of particular interest. \re{eq:ContinuousPerturbedEq} allows the evaluation of the velocity discretization impact, the equilibrium, as well as the collision model. \re{eq:DiscretePerturbedEq} further highlights the influence of the space/time discretization on the numerical stability and accuracy of the related LBM. 

\subsection{Eigenvalue problem \label{sec:EigPb}}

Solutions of the set of perturbed equations~\retxt{eq:ContinuousPerturbedEq} and~\retxt{eq:DiscretePerturbedEq} are obtained using a Fourier transform. This is equivalent to seeking solutions in the form of monochromatic plane waves evolving in a periodic domain, i.e, $\forall\left(\bm{k},\omega\right) \in \mathbb{R}^{D}\times\mathbb{C}$,
\begin{align}
f'_i &= A_i \exp[\text{i}(\bm{k}\cdot\bm{x}-\omega t)]\nonumber\\[0.1cm]
      &= A_i \:\underset{\text{(a)}}{\underbrace{\exp[\text{Im}(\omega)t]}} \: \underset{\text{(b)}}{\underbrace{\exp[\text{i}(\bm{k}\cdot\bm{x}-\text{Re}(\omega)t)]}}, \label{eq:MonochromPlaneWaves}
\end{align}
where $\bm{k}$ is the (real) wave number, while $\omega$ is the (complex) time frequency of the monochromatic wave. `` $\mathrm{i}$ '' is the imaginary part unit which must not be confused with $i$ (the discrete velocity index). Simply put, (a) is linked to the growth (or dissipation) rate of waves, while (b) gives information about the propagation speed (or dispersion) of waves. 

Injecting \re{eq:MonochromPlaneWaves} into perturbed equations~\retxt{eq:ContinuousPerturbedEq} and~\retxt{eq:DiscretePerturbedEq} leads to the following eigenvalue problem of size $V$:
\begin{align}
\bm{M_{\Omega}^{\mathrm{C}} A} &= \omega \bm{A},\label{eq:EigenvaluePbContinuous}\\
\bm{M_{\Omega}^{\mathrm{D}} A} &= \exp{(-\mathrm{i}\omega)} \bm{A},\label{eq:EigenvaluePbDiscrete}
\end{align}
with $\omega$ (resp. $\exp{(-\mathrm{i}\omega)}$) the eigenvalue, $\bm{M_{\Omega}^{\mathrm{C}}}$ (resp. $\bm{M_{\Omega}^{\mathrm{D}}}$) the matrix associated to the continuous (resp. discrete) perturbed equation, and $\bm{A}$ the eigenvector composed of the perturbations' amplitude. By solving Eqs.~(\ref{eq:EigenvaluePbContinuous}) and~(\ref{eq:EigenvaluePbDiscrete}), eigenvalues are then obtained, and plotting their imaginary and real parts eventually makes it possible to: (i) understand how numerical waves behave in the simulation domain, (ii) confirm if they follow analytic formulas derived from the linearized NS equations, (iii) distinguish the impact of velocity and space/time discretizations of the Boltzmann equation.

Let us continue with the definition of the matrix $\bm{M_{\Omega}}$. It directly depends on the chosen lattice, collision model $\Omega$, wavenumber $\bm{k}$ and mean flow $\left(\overline{\rho},\overline{\bm{u}}\right)$. If we consider the general collision model $\Omega$, its mathematical expression in the continuous case reads as 
\begin{equation}
\bm{M_{\Omega}^{\mathrm{D}}} =\bm{E^{\mathrm{C}}} +\mathrm{i}\bm{J_{\Omega}^{\mathrm{C}}}\quad \mathrm{with}\quad 
 E^{\mathrm{C}}_{ij} = (\bm{\xi}_i\cdot\bm{k})\delta_{ij},
\end{equation}
whereas its discrete counterpart is
\begin{equation}
\bm{M_{\Omega}^{\mathrm{D}}} =\bm{E^{\mathrm{D}}}\left[ \bm{\delta} +\bm{J_{\Omega}^{\mathrm{D}}}\right]\quad \mathrm{with}\quad 
 E^{\mathrm{D}}_{ij} = \exp[-\mathrm{i}(\bm{\xi}_i\cdot\bm{k})]\delta_{ij}.
\end{equation}
$\bm{J_{\Omega}^{\mathrm{C}}}$ and $\bm{J_{\Omega}^{\mathrm{D}}}$ are the Jacobian matrices associated to the collision model $\Omega$ in the continuous and discrete case respectively.
To derive these Jacobian matrices, one can either do it analytically~\cite{CHAVEZMODENA_CF_172_2018,HOSSEINI_PRE_99_2019b,HOSSEINI_PRE_100_2019,WISSOCQ_JCP_380_2019,WISSOCQ_PhD_2019,HOSSEINI_RSTA_378_2020}, or numerically by using a dedicated library. As a first study, the latter option has been adopted, even though analytical formulas for the computation of each Jacobian will be provided in a future investigation.

\section{Collision models and related post-collision populations\label{sec:CollModels}}

Hereafter, the uniform framework~\cite{COREIXAS_PRE_100_2019} used for the derivation of all collision models is briefly recalled. Let us start with the definition of post-collision populations $f_i^*$ based on post-collision raw moments $M_{pq}^*$, with $p,q\leq 2$~\cite{KARLIN_PA_389_2010,LYCETT_BROWN_PF_26_2014,BOSCH_ESAIM_52_2015}: 
\begin{subequations}
\begin{align}
f^{*,\mathrm{RM}}_{(0,0)} &= M_{00}^*-(M_{20}^*+M_{02}^*) + M_{22}^*, \\
f^{*,\mathrm{RM}}_{(\sigma,0)} &= \dfrac{1}{2}\left(\sigma M_{10}^* + M_{20}^* - \sigma M_{12}^* - M_{22}^*\right), \\
f^{*,\mathrm{RM}}_{(0,\lambda)} &= \dfrac{1}{2}\left(\lambda M_{01}^* + M_{02}^* - \lambda M_{21}^* - M_{22}^*\right), \\
f^{*,\mathrm{RM}}_{(\sigma,\lambda)} &= \dfrac{1}{4}\left(\sigma\lambda M_{11}^* + \sigma M_{12}^* + \lambda M_{21}^* + M_{22}^*\right),
\end{align}
\label{eq:Q9postCollVDF_RM}
\end{subequations}
with $(\sigma,\lambda)\in\{\pm 1\}^2$. Post-collision RMs $M_{pq}^*$ are defined as
\begin{subequations}
\begin{align}
M_{00}^*&=M_{00}=M_{00}^{eq}=\rho,\\
M_{10}^*&=M_{10}=M_{10}^{eq}=\rho u_x,\\
M_{01}^*&=M_{01}=M_{01}^{eq}=\rho u_y,\\
M_{20}^*&=(1-\omega_{\nu})M_{20} + \omega_{\nu} M_{20}^{eq},\\
M_{02}^*&=(1-\omega_{\nu})M_{02} + \omega_{\nu} M_{02}^{eq},\\
M_{11}^*&=(1-\omega_{\nu})M_{11} + \omega_{\nu} M_{11}^{eq},\\
M_{21}^*&=(1-\omega_3)M_{21} + \omega_3 M_{21}^{eq},\\
M_{12}^*&=(1-\omega_3)M_{12} + \omega_3 M_{12}^{eq},\\
M_{22}^*&=(1-\omega_4)M_{22} + \omega_4 M_{22}^{eq},
\end{align}
\label{eq:RMcoll}
\end{subequations}
with $M_{pq}=\sum_i \xi_{i,x}^p \xi_{i,x}^q f_i$, $M_{20}^{eq}=\rho (u_x^2 + c_s^2)$, $M_{02}^{eq}=\rho (u_y^2 + c_s^2)$, $M_{11}^{eq}=\rho u_x u_y$, $M_{21}^{eq}=M_{20}^{eq} u_y$, $M_{12}^{eq}=u_x M_{02}^{eq}$, $M_{22}^{eq}=M_{20}^{eq} M_{02}^{eq}/\rho$, and where the conservation of mass and momentum is enforced by imposing zeroth- and first-order moments to their equilibrium values.
$\omega_{\nu}$ is the relaxation frequency related to viscous phenomena, whereas the free parameters $\omega_3$ and $\omega_4$ are those associated to third- and fourth-order moments respectively.

To compute post-collision populations for a given collision model, one just need to relax statistical quantities of interest (HM, CM, CHM and K) in their corresponding moment space. Then, one come back to post-collision RMs using relationships compiled in Appendix D of Ref.~\cite{COREIXAS_PRE_100_2019}, and eventually, one compute post-collision populations $f_i^*$ thanks to \re{eq:Q9postCollVDF_RM}. Let us illustrate this methodology with the example of the collision model based on the CHM space. First, one computes CHMs for $p,q\leq 2$
\begin{equation}
\widetilde{A}_{pq}=\sum_i \widetilde{\mathcal{H}}_{i,pq} f_i,
\end{equation}
where central Hermite polynomials $\widetilde{\mathcal{H}}_{i,pq}$ are defined as~\cite{GRADb_CPAM_2_1949}
\begin{subequations}
\begin{align}
\widetilde{\mathcal{H}}_{i,00}&=1,\\
\widetilde{\mathcal{H}}_{i,10}&=(\xi_{i,x}-u_x),\\
\widetilde{\mathcal{H}}_{i,01}&=(\xi_{i,y}-u_y),\\
\widetilde{\mathcal{H}}_{i,20}&=(\xi_{i,x}-u_x)^2-c_s^2,\\
\widetilde{\mathcal{H}}_{i,02}&=(\xi_{i,y}-u_y)^2-c_s^2,\\
\widetilde{\mathcal{H}}_{i,11}&=\widetilde{\mathcal{H}}_{i,10}\widetilde{\mathcal{H}}_{i,01}=(\xi_{i,x}-u_x)(\xi_{i,y}-u_y),\\
\widetilde{\mathcal{H}}_{i,21}&=\widetilde{\mathcal{H}}_{i,20}\widetilde{\mathcal{H}}_{i,01}=[(\xi_{i,x}-u_x)^2-c_s^2](\xi_{i,y}-u_y),\\
\widetilde{\mathcal{H}}_{i,12}&=\widetilde{\mathcal{H}}_{i,10}\widetilde{\mathcal{H}}_{i,02}=(\xi_{i,y}-u_y)[(\xi_{i,y}-u_y)^2-c_s^2],\\
\widetilde{\mathcal{H}}_{i,22}&=\widetilde{\mathcal{H}}_{i,20}\widetilde{\mathcal{H}}_{i,02}=[(\xi_{i,x}-u_x)^2-c_s^2][(\xi_{i,y}-u_y)^2-c_s^2].
\end{align}
\label{eq:CHpoly}
\end{subequations}
$A_{pq}$ and $\mathcal{H}_{pq}$ stands respectively for Hermite moment and polynomial of order $n=p+q$. The tilde notation extend their definition to the central moment space where discrete velocities $\bm{\xi}_i$ are replaced by their co-moving (or peculiar) counterpart $\widetilde{\bm{\xi}}_i=\bm{\xi}-\bm{u}$.

Knowing that equilibrium CHMs $\widetilde{A}_{pq}^{eq}$ are all zero with the exception of the zeroth-order one, i.e.,
\begin{equation}
\widetilde{A}^{eq}_{00} = \rho, \,\widetilde{A}^{eq}_{10}=\widetilde{A}^{eq}_{01}=\widetilde{A}^{eq}_{11}=\widetilde{A}^{eq}_{20}=\widetilde{A}^{eq}_{02}=\widetilde{A}^{eq}_{21}=\widetilde{A}^{eq}_{12}=\widetilde{A}^{eq}_{22}=0,
\end{equation}
post-collision CHMs $\widetilde{A}_{pq}^*$ are then computed as
\begin{subequations}
\begin{align}
\widetilde{A}_{00}^*&=\widetilde{A}_{00}=\widetilde{A}_{00}^{eq}=\rho,\\
\widetilde{A}_{10}^*&=\widetilde{A}_{10}=\widetilde{A}_{10}^{eq}=0,\\
\widetilde{A}_{01}^*&=\widetilde{A}_{01}=\widetilde{A}_{01}^{eq}=0,\\
\widetilde{A}_{20}^*&=(1-\omega_{\nu})\widetilde{A}_{20} + \omega_{\nu} \widetilde{A}_{20}^{eq}=(1-\omega_{\nu})\widetilde{A}_{20},\\
\widetilde{A}_{02}^*&=(1-\omega_{\nu})\widetilde{A}_{02} + \omega_{\nu} \widetilde{A}_{02}^{eq}=(1-\omega_{\nu})\widetilde{A}_{02},\\
\widetilde{A}_{11}^*&=(1-\omega_{\nu})\widetilde{A}_{11} + \omega_{\nu} \widetilde{A}_{11}^{eq}=(1-\omega_{\nu})\widetilde{A}_{11},\\
\widetilde{A}_{21}^*&=(1-\omega_3)\widetilde{A}_{21} + \omega_3 \widetilde{A}_{21}^{eq}=(1-\omega_3)\widetilde{A}_{21},\\
\widetilde{A}_{12}^*&=(1-\omega_3)\widetilde{A}_{12} + \omega_3 \widetilde{A}_{12}^{eq}=(1-\omega_3)\widetilde{A}_{12},\\
\widetilde{A}_{22}^*&=(1-\omega_4)\widetilde{A}_{22} + \omega_4 \widetilde{A}_{22}^{eq}=(1-\omega_4)\widetilde{A}_{22}.
\end{align}
\label{eq:CHMcoll}
\end{subequations}
Now that post-collision CHMs are known, one needs to express them in terms of post-collision HMs $A^*_{pq}$ using
\begin{subequations}
\begin{align}
A^*_{00}= & \widetilde{A}^*_{00}=\rho,\\
A^*_{10}= & \widetilde{A}^*_{10}+\rho u_x=\rho u_x,\\
A^*_{01}= & \widetilde{A}^*_{01}+\rho u_y=\rho u_y,\\
A^*_{20}= & \widetilde{A}^*_{20}+\rho u_x^2,\\
A^*_{02}= & \widetilde{A}^*_{02}+\rho u_y^2,\\
A^*_{11}= & \widetilde{A}^*_{11}+\rho u_xu_y ,\\
A^*_{21}= & \widetilde{A}^*_{21}+u_y \widetilde{A}^*_{20}+2 u_x \widetilde{A}^*_{11}+\rho u_x^2 u_y,\\
A^*_{12}= & \widetilde{A}^*_{12}+u_x \widetilde{A}^*_{02}+2 u_y \widetilde{A}^*_{11}+\rho u_x u_y^2,\\
A^*_{22}= & \widetilde{A}^*_{22}+2 u_y \widetilde{A}^*_{21}+2 u_x \widetilde{A}^*_{12} +u_y^2 \widetilde{A}^*_{20} +u_x^2 \widetilde{A}^*_{02}+4 u_x u_y \widetilde{A}^*_{11}+\rho u_x^2 u_y^2,
\end{align}
\label{eq:CHM2HM_Q9}
\end{subequations}
before eventually coming back to post-collision RMs $M^*_{pq}$ via 
\begin{subequations}
\begin{align}
M^*_{00} =& A^*_{00}=\rho,\\
M^*_{10} =& A^*_{10}=\rho u_x,\\
M^*_{01} =& A^*_{01}=\rho u_y,\\
M^*_{11} =& A^*_{11},\\
M^*_{20} =& A^*_{20}+\rho c_s^2,\\
M^*_{02} =& A^*_{02}+\rho c_s^2,\\
M^*_{21} =& A^*_{21}+\rho c_s^2 u_y,\\
M^*_{12} =& A^*_{12}+\rho c_s^2 u_x,\\
M^*_{22} =& A^*_{22}+c_s^2(A^*_{20}+A^*_{02}) + \rho c_s^4.
\end{align}
\label{eq:HM2RM_Q9}
\end{subequations}

The above methodology can be applied to any kind of collision model used in this work, with the exception of the RR approach. For the latter, non-equilibrium populations are computed recursively from formulas derived using the Chapman-Enskog expansion at the NS level, i.e., $f^{neq}_i=f^{(1)}_i$ with $f^{(1)}_i \neq (f_i-f_i^{eq})$. It is usually done using the Gauss-Hermite formalism~\cite{MALASPINAS_ARXIV_2015,BROGI_JASA_142_2017}, but for lattices build through tensor products of low-order ones one can rely on the Hermite moment space (see Appendix G in Ref.~\cite{COREIXAS_PRE_100_2019}). Hence, the aforementioned methodology can be reused in the context of the D2Q9-RR with only minor modifications. Starting from Hermite moments $A_{pq}$ and their equilibrium counterpart $A_{pq}^{eq}$, post-collision Hermite moments $A_{pq}^*$ are
computed this time through
\begin{subequations}
\begin{align}
A^*_{00} =& \rho,\\
A^*_{10} =& \rho u_x,\\
A^*_{01} =& \rho u_y,\\
A^*_{11} =& A_{11} - \omega_{\nu} A_{11}^{neq},\\
A^*_{20} =& A_{20} - \omega_{\nu} A_{20}^{neq},\\
A^*_{02} =& A_{02} - \omega_{\nu} A_{02}^{neq},\\
A^*_{21} =& A_{21} - \omega_{3}A_{21}^{neq},\\
A^*_{12} =& A_{12} - \omega_{3}A_{12}^{neq},\\
A^*_{22} =& A_{22} - \omega_{4}A_{22}^{neq}.
\end{align}
\label{eq:HMColl_Q9}
\end{subequations}
with non-equilibrium Hermite moments $A_{pq}^{neq}$ that are derived from the following recursive formulas~\cite{MALASPINAS_ARXIV_2015,BROGI_JASA_142_2017}
\begin{subequations}
\begin{align}
A^{neq}_{21} &= u_y A^{neq}_{20}+ 2 u_x A^{neq}_{11},\\
A^{neq}_{12} &= u_x A^{neq}_{02}+ 2 u_y A^{neq}_{11},\\
A^{neq}_{22} &= u_y^2 A^{neq}_{20} + u_x^2 A^{neq}_{02} + 4 u_x u_y A^{neq}_{11}.
\end{align}
\label{eq:RRformulas_D2Q9}
\end{subequations}
Second-order non-equilibrium moments ($A^{neq}_{20}$, $A^{neq}_{02}$ and $A^{neq}_{11}$) can be computed in several ways. In the present context, the standard approach is adopted, i.e., $A^{neq}_{pq}=\sum_i \mathcal{H}_{pq} (f_i-f_i^{eq})$ for $p+q=2$~\cite{MALASPINAS_ARXIV_2015,BROGI_JASA_142_2017}, but the interested reader may also refer to Ref.~\cite{WISSOCQ_PhD_2019} for an in-depth study on the impact of the initialization of the recursive computation of $A^{neq}_{pq}$. Eventually, formulas \retxt{eq:HMColl_Q9} are used to compute post-collision RMs \retxt{eq:HM2RM_Q9}, and post-collision populations are obtained via Eq.~(\ref{eq:Q9postCollVDF_RM}).

\section{Spectral sampling induced by under-resolved grid meshes \label{sec:filtering}}

\begin{figure}[tbp!]
\centering
\begin{tabular}{c @{\,} c @{\,} c}
\includegraphics[trim=0.3cm 0cm 2cm 0cm, clip=true, totalheight=0.21\textheight]{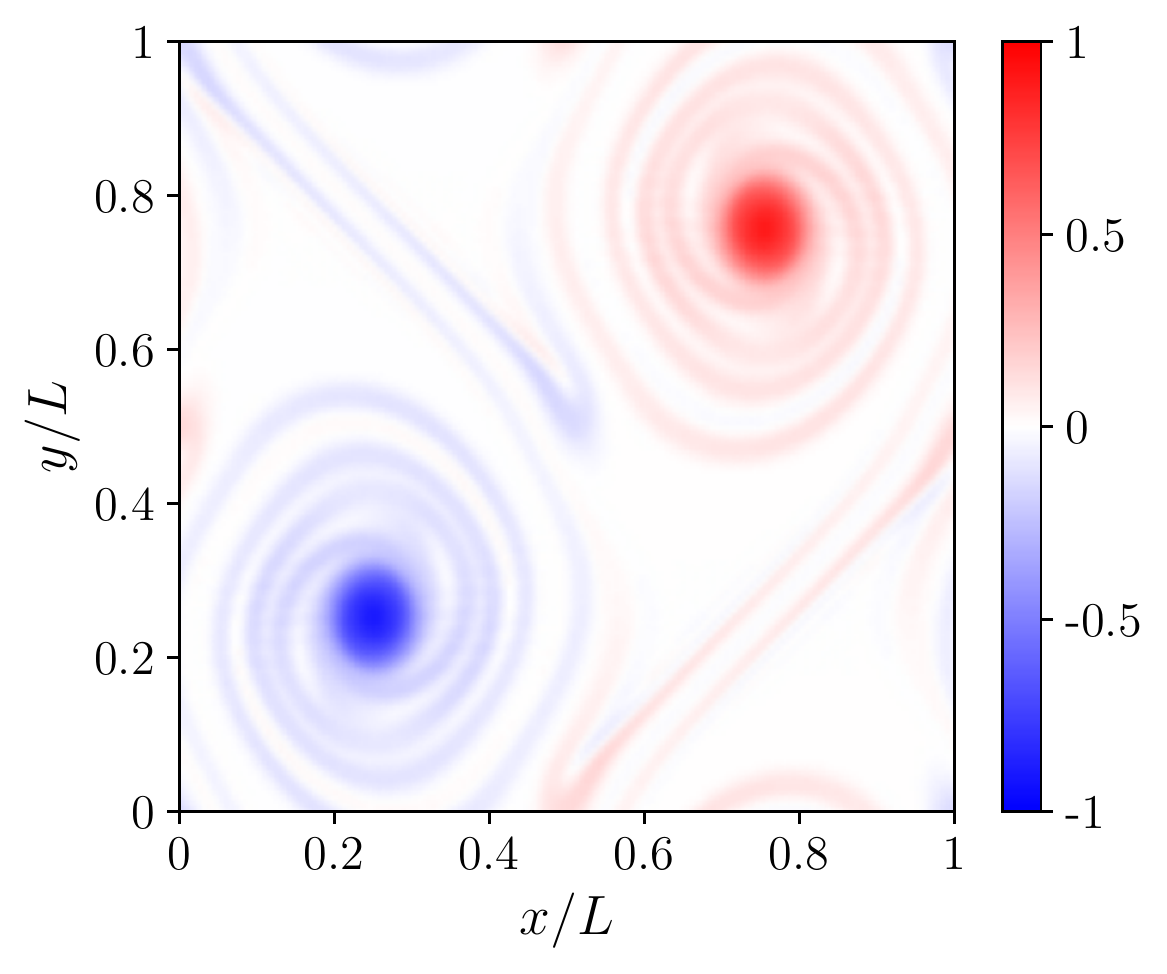} &
\includegraphics[trim=1.7cm 0cm 2cm 0cm, clip=true, totalheight=0.21\textheight]{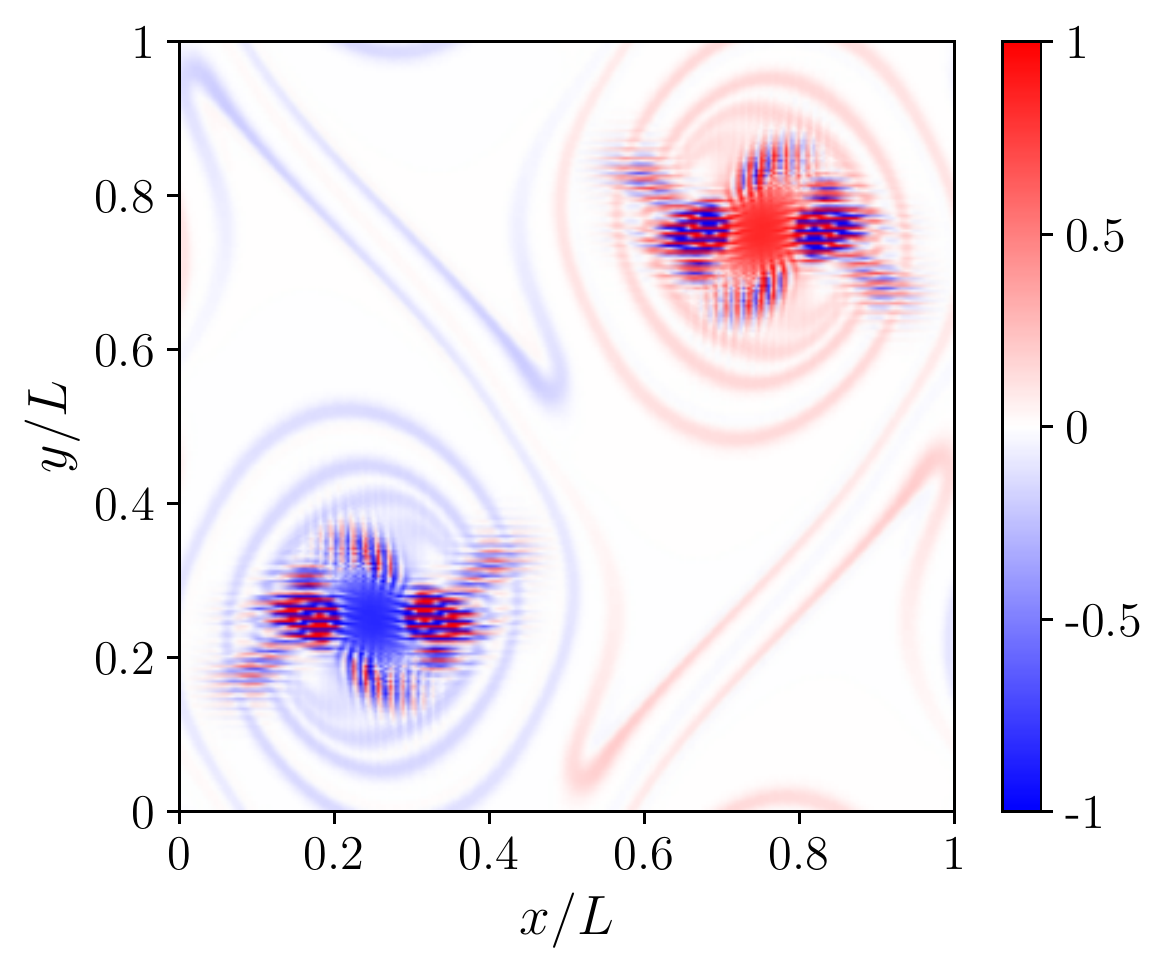} &
\includegraphics[trim=1.7cm 0cm 0.3cm 0cm, clip=true, totalheight=0.21\textheight]{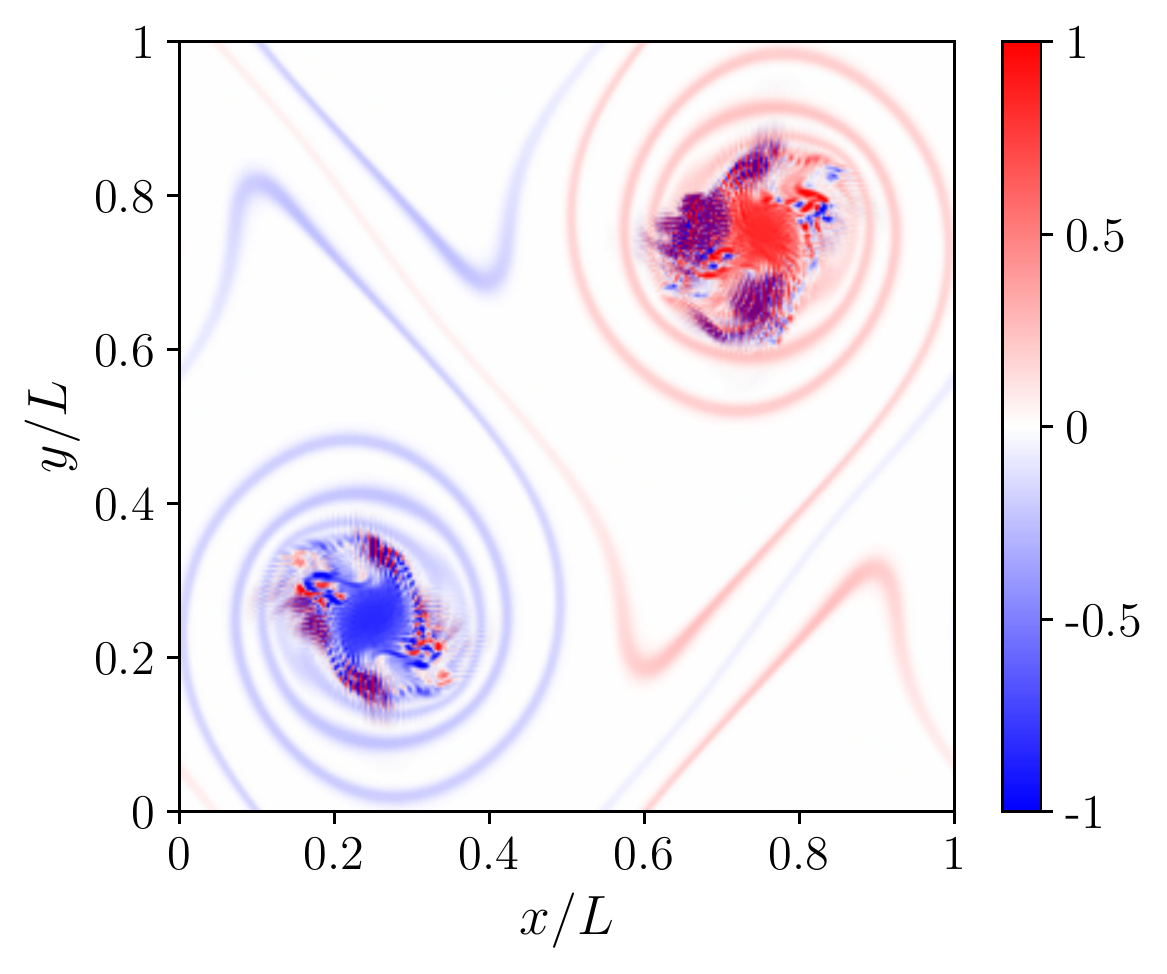}
\end{tabular}
\caption{Rollup of the double shear layer at $\mathrm{Ma}=u_0/c_s=0.12$ and $\mathrm{Re} = u_0L/\nu=3\times 10^4$. Visualization of the $z$-component of the dimensionless vorticity field using (from left to right) $L=128$, $256$ and $512$ grid points in each directions, with the D2Q9 and the PR collision model. Instantaneous snapshots are taken at time $t/t_c =  2$, $1.8$ and $1.7$ respectively, with the characteristic time $t_c=L/ u_0$. From this, it is clear that the finer the mesh grid, the sooner instabilities appear in the simulation domain.}
\label{fig:DSL}
\end{figure}

In a previous work~\cite{COREIXAS_PRE_96_2017}, the numerical stability of the standard (PR~\cite{LATT_MCS_72_2006a}) and recursive (RR~\cite{MALASPINAS_ARXIV_2015,BROGI_JASA_142_2017}) regularization procedures has been investigated using the double shear layer testcase~\cite{BROWN_JCP_122_1995,MINION_JCP_138_1997} --where these collision models correspond (in the present context) to the REG-HM and SRT-RR models respectively. A more in-depth study showed that stability curves (maximal Mach number for a given Reynolds number) depend on the spatial resolution~\cite{COREIXAS_PhD_2018}. For $\mathrm{Re=3\times 10^4}$, and assuming an acoustic scaling to recover the weakly compressible macroscopic behavior~\cite{KRUGER_Book_2017}, it was found that the PR (RR) approach should be stable up to $\mathrm{Ma}=0.14$ ($0.52$) for a resolution based on $L=128$ points per direction, while it drops to $\mathrm{Ma}=0.11$ ($0.49$) when the resolution is increased to $L=256$. Starting from a stable configuration for a given mesh, one would certainly not expect that a better resolution of the simulation domain would deteriorate the stability of the numerical scheme.

This unexpected behavior, which was observed for \emph{all collision models} considered in this work, is illustrated for the D2Q9-PR-LBM in \rf{fig:DSL}, where the simulation was initialized using the same methodology described in our previous work~\cite{COREIXAS_PRE_96_2017}. While the simulation remains stable up to $t=2t_c$ (with $t_c=L/u_0$ the characteristic time) for a rather coarse grid mesh ($L=128$), it becomes unstable near $t=1.8t_c$ for $L=256$. By further refining the grid mesh ($L=512$), stability issues arise even sooner ($t=1.7t_c$). These results highlights the spectral sampling induced by under-resolved grid meshes, that may, for particular configurations, increase the stability of LBMs \emph{whatever the collision model considered}.

\enlargethispage{20pt}

\ack{Fruitful discussions with SA Hosseini are gratefully acknowledged. The authors would also like to thank the reviewers for their pertinent remarks and insightful questions.}

\aucontribute{CC performed the research, carried out the simulations and wrote the paper. GW provided the linear stability code that was further extended for this research, and helped with the interpretation of both linear and numerical results. BC and JL supervised  the  research and revised the manuscript. All authors read and approved the manuscript.}


\dataccess{This article has no additional data.}

\competing{We declare we have no competing interests.}



\begin{thebibliography}{100}
\expandafter\ifx\csname urlstyle\endcsname\relax
  \providecommand{\doi}[1]{doi:\discretionary{}{}{}#1}\else
  \providecommand{\doi}{doi:\discretionary{}{}{}\begingroup
  \urlstyle{rm}\Url}\fi

\bibitem{SUCCI_Book_2018}
Succi S. 2018 \emph{The Lattice {B}oltzmann Equation: For Complex States of
  Flowing Matter}. Oxford University Press.

\bibitem{KRUGER_Book_2017}
Kr\"{u}ger T, Kusumaatmaja H, Kuzmin A, Shardt O, Silva G, Viggen EM. 2017
  \emph{The Lattice {B}oltzmann Method: Principles and Practice}. Springer International Publishing. (\textcolor{blue}{\doi{10.1007/978-3-319-44649-3}})

\bibitem{BHATNAGAR_PR_94_1954}
Bhatnagar P, Gross E, Krook M. 1954 A model for collision processes in gases.
  {I}. {S}mall amplitude processes in charged and neutral one-component
  systems. \emph{Phys. Rev.} \textbf{94}, 511--525. (\textcolor{blue}{\doi{10.1103/PhysRev.94.511}})

\bibitem{GUO_Book_2013}
Guo Z, Shu C. 2013 \emph{Lattice {B}oltzmann Method and Its Applications in
  Engineering}. World Scientific. (\textcolor{blue}{\doi{10.1142/9789814508308_0001}})

\bibitem{COREIXAS_PRE_100_2019}
Coreixas C, Chopard B, Latt J. 2019 Comprehensive comparison of collision
  models in the lattice {B}oltzmann framework: Theoretical investigations. \emph{Phys. Rev. E} \textbf{100}, 033305. (\textcolor{blue}{\doi{10.1103/PhysRevE.100.033305}})

\bibitem{WANG_PRE_94_2016}
Wang P, Wang LP, Guo Z. 2016 Comparison of the lattice {B}oltzmann equation and
  discrete unified gas-kinetic scheme methods for direct numerical simulation
  of decaying turbulent flows. \emph{Phys. Rev. E} \textbf{94}, 043304. (\textcolor{blue}{\doi{10.1103/PhysRevE.94.043304}})

\bibitem{MASSET_Accepted_2019}
Masset PA, Wissocq G. 2019 Linear hydrodynamics and stability of the discrete
  velocity {B}oltzmann equations .\doi{10.1017/jfm.2020.374}. (accepted)

\bibitem{LALLEMAND_PRE_61_2000}
Lallemand P, Luo LS. 2000 Theory of the lattice {B}oltzmann method: Dispersion,
  dissipation, isotropy, {G}alilean invariance, and stability. \emph{Phys. Rev. E} \textbf{61}, 6546--6562. (\textcolor{blue}{\doi{10.1103/PhysRevE.61.6546}})

\bibitem{DELLAR_PRE_65_2002}
Dellar PJ. 2002 Nonhydrodynamic modes and \textit{a priori} construction of
  shallow water lattice {B}oltzmann equations. \emph{Phys. Rev. E} \textbf{65}, 036309. (\textcolor{blue}{\doi{10.1103/PhysRevE.65.036309}})

\bibitem{BOSCH_ESAIM_52_2015}
B{\"o}sch F, Chikatamarla SS, Karlin I. 2015 Entropic multi-relaxation models
  for simulation of fluid turbulence. \emph{ESAIM Proc.} \textbf{52}, 1--24. (\textcolor{blue}{\doi{10.1051/proc/201552001}})

\bibitem{HOSSEINI_PRE_99_2019b}
Hosseini SA, Coreixas C, Darabiha N, Th\'evenin D. 2019 Stability of the
  lattice kinetic scheme and choice of the free relaxation parameter. \emph{Phys. Rev. E} \textbf{99}, 063305. (\textcolor{blue}{\doi{10.1103/PhysRevE.99.063305}})

\bibitem{WISSOCQ_PhD_2019}
Wissocq G. 2019 \emph{Investigating lattice {B}oltzmann methods for
  turbomachinery secondary air system simulations}. Ph.D. thesis, Aix-Marseille Universit\'{e}.

\bibitem{DHUMIERE_PAA_159_1992}
d'Humi\`{e}res D. 1992 Generalized lattice-{B}oltzmann equations. \emph{Prog. Astronaut. Aeronaut.} \textbf{159}, 450--458. (\textcolor{blue}{\doi{10.2514/5.9781600866319.0450.0458}})

\bibitem{DHUMIERES_TRS_360_2002}
d'Humi\`{e}res D. 2002 Multiple-relaxation-time lattice {B}oltzmann models in
  three dimensions. \emph{Philos. Trans. R. Soc. London, Ser. A} \textbf{360}, 437--451. (\textcolor{blue}{\doi{10.1098/rsta.2001.0955}})

\bibitem{SHAN_IJMPC_18_2007}
Shan X, Chen H. 2007 A general multiple-relaxation-time {b}oltzmann collision
  model. \emph{Int. J. Mod. Phys. C} \textbf{18}, 635--643. (\textcolor{blue}{\doi{10.1142/S0129183107010887}})

\bibitem{ADHIKARI_PRE_78_2008}
Adhikari R, Succi S. 2008 Duality in matrix lattice {B}oltzmann models. \emph{Phys. Rev. E} \textbf{78}, 066701. (\textcolor{blue}{\doi{10.1103/PhysRevE.78.066701}})

\bibitem{CHEN_IJMPC_25_2014}
Chen H, Gopalakrishnan P, Zhang R. 2014 Recovery of {G}alilean invariance in
  thermal lattice {B}oltzmann models for arbitrary {P}randtl number. \emph{Int. J. Mod. Phys. C} \textbf{25}, 1450046. (\textcolor{blue}{\doi{10.1142/S0129183114500466}})

\bibitem{GEIER_PRE_73_2006}
Geier M, Greiner A, Korvink JG. 2006 Cascaded digital lattice {B}oltzmann
  automata for high {R}eynolds number flow. \emph{Phys. Rev. E} \textbf{73}, 066705. (\textcolor{blue}{\doi{10.1103/PhysRevE.73.066705}})

\bibitem{GEIER_EPJST_171_2009}
Geier M, Greiner A, Korvink JG. 2009 A factorized central moment lattice
  {B}oltzmann method. \emph{Eur. Phys. J. Spec. Top.} \textbf{171}, 55--61. (\textcolor{blue}{\doi{10.1140/epjst/e2009-01011-1}})

\bibitem{LYCETTBROWN_CMWA_67_2014}
Lycett-Brown D, Luo KH. 2014 Multiphase cascaded lattice {B}oltzmann method. \emph{Comput. Math. Appl.} \textbf{67}, 350 -- 362. (\textcolor{blue}{\doi{10.1016/j.camwa.2013.08.033}})

\bibitem{DUBOIS_CCP_17_2015}
Dubois F, F\'{e}vrier T, Graille B. 2015 Lattice {B}oltzmann schemes with
  relative velocities. \emph{Commun. Comput. Phys.} \textbf{17}, 1088-1112. (\textcolor{blue}{\doi{10.4208/cicp.2014.m394}})

\bibitem{CHEN_PATENT_Collision_2015}
Chen H, Zhang R, Gopalakrishnan P. 2015. Lattice {B}oltzmann collision operators enforcing isotropy and
  {G}alilean invariance. {CA} {P}atent {A}pp. {CA} 2,919,062

\bibitem{MATTILA_PF_29_2017}
Mattila KK, Philippi PC, Hegele~Jr LA. 2017 High-order regularization in
  lattice-{B}oltzmann equations. \emph{Phys. Fluids} \textbf{29}, 046103. (\textcolor{blue}{\doi{10.1063/1.4981227}})

\bibitem{LI_PRE_100_2019}
Li X, Shi Y, Shan X. 2019 Temperature-scaled collision process for the
  high-order lattice {B}oltzmann model. \emph{Phys. Rev. E} \textbf{100}, 013301. (\textcolor{blue}{\doi{10.1103/PhysRevE.100.013301}})

\bibitem{GEIER_CMA_70_2015}
Geier M, Sch\"{o}nherr M, Pasquali A, Krafczyk M. 2015 The cumulant lattice
  {B}oltzmann equation in three dimensions: Theory and validation. \emph{Comput. Math. Appl.} \textbf{70}, 507 -- 547. (\textcolor{blue}{\doi{10.1016/j.camwa.2015.05.001}})

\bibitem{GEIER_JCP_348_2017a}
Geier M, Pasquali A, Sch\"{o}nherr M. 2017 Parametrization of the cumulant
  lattice {B}oltzmann method for fourth order accurate diffusion part {I}:
  Derivation and validation. \emph{J. Comput. Phys.} \textbf{348}, 862 -- 888. (\textcolor{blue}{\doi{10.1016/j.jcp.2017.05.040}})

\bibitem{GEIER_JCP_348_2017b}
Geier M, Pasquali A, Sch\"{o}nherr M. 2017 Parametrization of the cumulant
  lattice {B}oltzmann method for fourth order accurate diffusion part {II}:
  Application to flow around a sphere at drag crisis. \emph{J. Comput. Phys.} \textbf{348}, 889 -- 898. (\textcolor{blue}{\doi{10.1016/j.jcp.2017.07.004}})

\bibitem{SKORDOS_PRE_49_1993}
Skordos PA. 1993 Initial and boundary conditions for the lattice {B}oltzmann
  method. \emph{Phys. Rev. E} \textbf{48}, 4823--4842. (\textcolor{blue}{\doi{10.1103/PhysRevE.48.4823}})

\bibitem{LADD_JSP_104_2001}
Ladd AJC, Verberg R. 2001 Lattice-{B}oltzmann simulations of particle-fluid
  suspensions. \emph{J. Stat. Phys.} \textbf{104}, 1191--1251. (\textcolor{blue}{\doi{10.1023/A:1010414013942}})

\bibitem{LATT_MCS_72_2006a}
Latt J, Chopard B. 2006 Lattice {B}oltzmann method with regularized
  pre-collision distribution functions. \emph{Math. Comput. Simul.} \textbf{72}, 165--168. (\textcolor{blue}{\doi{10.1016/j.matcom.2006.05.017}})

\bibitem{CHEN_PA_362_2006}
Chen H, Zhang R, Staroselsky I, Jhon M. 2006 Recovery of full rotational
  invariance in lattice {B}oltzmann formulations for high {K}nudsen number
  flows. \emph{Physica A} \textbf{362}, 125--131. (\textcolor{blue}{\doi{10.1016/j.physa.2005.09.008}})

\bibitem{MALASPINAS_ARXIV_2015}
{Malaspinas} O. 2015 {Increasing stability and accuracy of the lattice
  {B}oltzmann scheme: recursivity and regularization}. \emph{arXiv preprint arXiv:1505.06900} .

\bibitem{COREIXAS_PRE_96_2017}
Coreixas C, Wissocq G, Puigt G, Boussuge JF, Sagaut P. 2017 Recursive
  regularization step for high-order lattice {B}oltzmann methods. \emph{Phys. Rev. E} \textbf{96}, 033306. (\textcolor{blue}{\doi{10.1103/PhysRevE.96.033306}})

\bibitem{BROGI_JASA_142_2017}
Brogi F, Malaspinas O, Chopard B, Bonadonna C. 2017 {H}ermite regularization of
  the lattice {B}oltzmann method for open source computational aeroacoustics. \emph{J. Acoust. Soc. Am} \textbf{142}, 2332--2345. (\textcolor{blue}{\doi{10.1121/1.5006900}})

\bibitem{JACOB_JT_19_2018}
Jacob J, Malaspinas O, Sagaut P. 2018 A new hybrid recursive regularised
  {B}hatnagar–{G}ross–{K}rook collision model for lattice {B}oltzmann
  method-based {L}arge {E}ddy {S}imulation. \emph{J. Turb.} \textbf{19}, 1051--1076. (\textcolor{blue}{\doi{10.1080/14685248.2018.1540879}})

\bibitem{MANOHA_AIAA_2846_2015}
Manoha E, Caruelle B. 2015 Summary of the {LAGOON} solutions from the benchmark
  problems for airframe noise computations-{III} Workshop. In \emph{21st AIAA/CEAS Aeroacoustics Conference}, p. 2846. (\textcolor{blue}{\doi{10.2514/6.2015-2846}})

\bibitem{PREMNATH_JSP_143_2011}
Premnath KN, Banerjee S. 2011 On the three-dimensional central moment lattice
  {B}oltzmann method. \emph{J. Stat. Phys.} \textbf{143}, 747--794. (\textcolor{blue}{\doi{10.1007/s10955-011-0208-9}})

\bibitem{NING_IJNMF_82_2016}
Ning Y, Premnath KN, Patil DV. 2016 Numerical study of the properties of the
  central moment lattice {B}oltzmann method. \emph{Int. J. Numer. Meth. Fluids} \textbf{82}, 59--90. (\textcolor{blue}{\doi{10.1002/fld.4208}})

\bibitem{GEIER_CF_166_2018}
Geier M, Pasquali A. 2018 Fourth order {G}alilean invariance for the lattice
  {B}oltzmann method. \emph{Comput. Fluids} \textbf{166}, 139 -- 151. (\textcolor{blue}{\doi{10.1016/j.compfluid.2018.01.015}})

\bibitem{GEHRKE_Chapter_2020}
Gehrke M, Banari A, Rung T. 2020 Performance of under-resolved, model-free
  {LBM} simulations in turbulent shear flows. In \emph{Progress in Hybrid {RANS-LES} Modelling} (ed. Y~Hoarau,
  SH~Peng, D~Schwamborn, A~Revell, C~Mockett), pp. 3--18. Cham: Springer
  International Publishing. (\textcolor{blue}{\doi{10.1007/978-3-030-27607-2_1}})

\bibitem{QIAN_EPL_21_1993}
Qian YH, Orszag SA. 1993 Lattice {BGK} models for the {N}avier-{S}tokes
  equation: Nonlinear deviation in compressible regimes. \emph{Europhys. Lett.} \textbf{21}, 255. (\textcolor{blue}{\doi{10.1209/0295-5075/21/3/001}})

\bibitem{CHEN_PRE_50_1994}
Chen Y, Ohashi H, Akiyama M. 1994 Thermal lattice {B}hatnagar-{G}ross-{K}rook
  model without nonlinear deviations in macrodynamic equations. \emph{Phys. Rev. E} \textbf{50}, 2776--2783. (\textcolor{blue}{\doi{10.1103/PhysRevE.50.2776}})

\bibitem{QIAN_EPL_42_1998}
Qian YH, Zhou Y. 1998 Complete {G}alilean-invariant lattice {BGK} models for
  the {N}avier-{S}tokes equation. \emph{Europhys. Lett.} \textbf{42}, 359.

\bibitem{NIE_EPL_81_2008}
Nie XB, Shan X, Chen H. 2008 {G}alilean invariance of lattice {B}oltzmann
  models. \emph{Europhys. Lett.} \textbf{81}, 34005. (\textcolor{blue}{\doi{10.1209/0295-5075/81/34005}})

\bibitem{PRASIANAKIS_PRE_79_2009}
Prasianakis NI, Karlin IV, Mantzaras J, Boulouchos KB. 2009 Lattice {B}oltzmann
  method with restored {G}alilean invariance. \emph{Phys. Rev. E} \textbf{79}, 066702. (\textcolor{blue}{\doi{10.1103/PhysRevE.79.066702}})

\bibitem{DELLAR_JCP_259_2014}
Dellar PJ. 2014 Lattice {B}oltzmann algorithms without cubic defects in
  {G}alilean invariance on standard lattices. \emph{J. Comput. Phys.} \textbf{259}, 270 -- 283. (\textcolor{blue}{\doi{10.1016/j.jcp.2013.11.021}})

\bibitem{SHAN_PRE_100_2019}
Shan X. 2019 Central-moment-based {G}alilean-invariant multiple-relaxation-time
  collision model. \emph{Phys. Rev. E} \textbf{100}, 043308. (\textcolor{blue}{\doi{10.1103/PhysRevE.100.043308}})

\bibitem{KATAOKA_PRE_69_2004a}
Kataoka T, Tsutahara M. 2004 Lattice {B}oltzmann model for the compressible
  {N}avier-{S}tokes equations with flexible specific-heat ratio. \emph{Phys. Rev. E} \textbf{69}, 035701. (\textcolor{blue}{\doi{10.1103/PhysRevE.69.035701}})

\bibitem{DELLAR_PCFD_8_2008}
Dellar PJ. 2008 Two routes from the {B}oltzmann equation to compressible flow
  of polyatomic gases. \emph{Prog. Comput. Fluid Dy.} \textbf{8}. (\textcolor{blue}{\doi{10.1504/PCFD.2008.018081}})

\bibitem{DELLAR_EPL_90_2010}
Dellar PJ. 2010 Electromagnetic waves in lattice {B}oltzmann
  magnetohydrodynamics. \emph{Europhys. Lett.} \textbf{90}, 50002. (\textcolor{blue}{\doi{10.1209/0295-5075/90/50002}})

\bibitem{COELHO_PRE_89_2014}
Coelho RCV, Ilha A, Doria MM, Pereira RM, Aibe VY. 2014 Lattice {B}oltzmann
  method for bosons and fermions and the fourth-order {H}ermite polynomial
  expansion. \emph{Phys. Rev. E} \textbf{89}, 043302. (\textcolor{blue}{\doi{10.1103/PhysRevE.89.043302}})

\bibitem{COELHO_PhD_2017}
Coelho RCV. 2017 \emph{Lattice {B}oltzmann method for semiclassical fluids and
  its application for electron hydrodynamics}. Ph.D. thesis, Federal University of Rio de Janeiro (UFRJ).

\bibitem{GABANNA_PRE_95_2017}
Gabbana A, Mendoza M, Succi S, Tripiccione R. 2017 Towards a unified lattice
  kinetic scheme for relativistic hydrodynamics. \emph{Phys. Rev. E} \textbf{95}, 053304. (\textcolor{blue}{\doi{10.1103/PhysRevE.95.053304}})

\bibitem{AXELSSON_Book_1996}
Axelsson O. 1996 \emph{Iterative solution methods}. Cambridge university press.

\bibitem{GOLUB_Book_3_2012}
Golub GH, Van~Loan CF. 2012 \emph{Matrix computations}, volume~3. JHU Press.

\bibitem{HIRSCH_Book_2007}
Hirsch C. 2007 \emph{Numerical computation of internal and external flows:
  {T}he fundamentals of {C}omputational {F}luid {D}ynamics}. Elsevier.

\bibitem{BLAZEK_Book_2015}
Blazek J. 2015 \emph{Computational {F}luid {D}ynamics: principles and
  applications}. Butterworth-Heinemann, 3 edition.

\bibitem{STERLING_JCP_123_1996}
Sterling JD, Chen S. 1996 Stability analysis of lattice {B}oltzmann methods. \emph{J. Comput. Phys.} \textbf{123}, 196 -- 206. (\textcolor{blue}{\doi{10.1006/jcph.1996.0016}})

\bibitem{WORTHING_PRE_56_1997}
Worthing RA, Mozer J, Seeley G. 1997 Stability of lattice {B}oltzmann methods
  in hydrodynamic regimes. \emph{Phys. Rev. E} \textbf{56}, 2243--2253. (\textcolor{blue}{\doi{10.1103/PhysRevE.56.2243}})

\bibitem{LALLEMAND_PRE_68_2003}
Lallemand P, Luo LS. 2003 Theory of the lattice {B}oltzmann method: Acoustic
  and thermal properties in two and three dimensions. \emph{Phys. Rev. E} \textbf{68}, 036706. (\textcolor{blue}{\doi{10.1103/PhysRevE.68.036706}})

\bibitem{WOLF_GLADROW_Book_2004}
Wolf-Gladrow DA. 2004 \emph{Lattice-gas cellular automata and lattice
  {B}oltzmann models: an introduction}. Springer.

\bibitem{NIU_JSP_117_2004}
Niu XD, Shu C, Chew YT, Wang TG. 2004 Investigation of stability and
  hydrodynamics of different lattice {B}oltzmann models. \emph{J. Stat. Phys.} \textbf{117}, 665--680. (\textcolor{blue}{\doi{10.1007/s10955-004-2264-x}})

\bibitem{DELLAR_PA_362_2006}
Dellar PJ. 2006 Non-hydrodynamic modes and general equations of state in
  lattice {B}oltzmann equations. \emph{Physica A} \textbf{362}, 132 -- 138. (\textcolor{blue}{\doi{10.1016/j.physa.2005.09.012}.})

\bibitem{SIEBERT_PRE_77_2008}
Siebert DN, Hegele LA, Philippi PC. 2008 Lattice {B}oltzmann equation linear
  stability analysis: Thermal and athermal models. \emph{Phys. Rev. E} \textbf{77}, 026707. (\textcolor{blue}{\doi{10.1103/PhysRevE.77.026707}})

\bibitem{RICOT_JCP_228_2009}
Ricot D, Mari\'{e} S, Sagaut P, Bailly C. 2009 Lattice {B}oltzmann method with
  selective viscosity filter. \emph{J. Comput. Phys.} \textbf{228}, 4478 -- 4490. (\textcolor{blue}{\doi{10.1016/j.jcp.2009.03.030}})

\bibitem{MARIE_JCP_228_2009}
Mari{\'e} S, Ricot D, Sagaut P. 2009 Comparison between lattice {B}oltzmann
  method and {N}avier–{S}tokes high order schemes for computational
  aeroacoustics. \emph{J. Comput. Phys.} \textbf{228}, 1056 -- 1070. (\textcolor{blue}{\doi{10.1016/j.jcp.2008.10.021}})

\bibitem{GINZBURG_JSP_139_2010}
Ginzburg I, d'Humi\`{e}res D, Kuzmin A. 2010 Optimal stability of
  advection-diffusion lattice {B}oltzmann models with two relaxation times for
  positive/negative equilibrium. \emph{J. Stat. Phys.} \textbf{139}, 1090--1143. (\textcolor{blue}{\doi{10.1007/s10955-010-9969-9}})

\bibitem{XU_JCP_231_2012}
Xu H, Malaspinas O, Sagaut P. 2012 Sensitivity analysis and determination of
  free relaxation parameters for the weakly-compressible {MRT}–{LBM} schemes. \emph{J. Comput. Phys.} \textbf{231}, 7335 -- 7367. (\textcolor{blue}{\doi{10.1016/j.jcp.2012.07.005}})

\bibitem{DUBOIS_CRM_343_2015}
Dubois F, F\'{e}vrier T, Graille B. 2015 On the stability of a relative
  velocity lattice {B}oltzmann scheme for compressible {N}avier–{S}tokes
  equations. \emph{C. R. M\'{e}canique} \textbf{343}, 599 -- 610. (\textcolor{blue}{\doi{10.1016/j.crme.2015.07.010}})

\bibitem{HOSSEINI_IJMPC_28_2017}
Hosseini SA, Darabiha N, Th{\'e}venin D, Eshghinejadfard A. 2017 Stability
  limits of the single relaxation-time advection--diffusion lattice {B}oltzmann
  scheme. \emph{Int. J. Mod. Phys. C} \textbf{28}, 1750141. (\textcolor{blue}{\doi{10.1142/S0129183117501418}})

\bibitem{CHAVEZMODENA_CF_172_2018}
Ch\'{a}vez-Modena M, Ferrer E, Rubio G. 2018 Improving the stability of
  multiple-relaxation lattice {B}oltzmann methods with central moments. \emph{Comput. Fluids} \textbf{172}, 397 -- 409. (\textcolor{blue}{\doi{10.1016/j.compfluid.2018.03.084}})

\bibitem{KRIVOVICHEV_IJCM__2018}
Krivovichev GV, Mikheev SA. 2018 On the stability of multi-step
  finite-difference-based lattice {B}oltzmann schemes. \emph{Int. J. Comput. Methods} \textbf{16}, 1850087. (\textcolor{blue}{\doi{10.1142/S0219876218500871}})

\bibitem{COREIXAS_PhD_2018}
Coreixas C. 2018 \emph{High-order extension of the recursive regularized
  lattice {B}oltzmann method}. Ph.D. thesis, INP Toulouse.

\bibitem{WILDE_IJNMF_90_2019}
Wilde D, Kr\"amer A, K\"ullmer K, Foysi H, Reith D. 2019 Multistep lattice
  {B}oltzmann methods: Theory and applications. \emph{Int. J. Numer. Meth. Fluids} \textbf{90}, 156--169. (\textcolor{blue}{\doi{10.1002/fld.4716}})

\bibitem{WISSOCQ_JCP_380_2019}
Wissocq G, Sagaut P, Boussuge JF. 2019 An extended spectral analysis of the
  lattice {B}oltzmann method: modal interactions and stability issues. \emph{J. Comput. Phys.} \textbf{380}, 311 -- 333. (\textcolor{blue}{\doi{10.1016/j.jcp.2018.12.015}})

\bibitem{HOSSEINI_PRE_100_2019}
Hosseini SA, Coreixas C, Darabiha N, Th\'evenin D. 2019 Extensive analysis of
  the lattice {B}oltzmann method on shifted stencils. \emph{Phys. Rev. E} \textbf{100}, 063301. (\textcolor{blue}{\doi{10.1103/PhysRevE.100.063301}})

\bibitem{HOSSEINI_RSTA_378_2020}
Hosseini SA, Darabiha N, Th\'{e}venin D. 2020 Compressibility in lattice
  {B}oltzmann on standard stencils: {E}ffects of deviation from reference
  temperature. \emph{Phil. Trans. R. Soc. A} \textbf{378}. (\textcolor{blue}{\doi{10.1098/rsta.2019.0399}})

\bibitem{HUANG_Book_2nd_1987}
Huang K. 1987 \emph{Statistical Mechanics, 2nd Edition}. Wiley, 2 edition.

\bibitem{SEEGER_PhD_2003}
Seeger S. 2003 \emph{The Cumulant Method}. Ph.D. thesis, {TU}-{C}hemnitz.

\bibitem{FOX_QBMM_2014}
Fox R. 2014 Quadrature-based moment methods for kinetic models. In \emph{Workshop on Moment Methods in Kinetic Theory II}. Fields
  Institute, University of Toronto.

\bibitem{SHAN_JFM_550_2006}
Shan X, Yuan XF, Chen H. 2006 Kinetic theory representation of hydrodynamics:
  {A} way beyond the {N}avier–{S}tokes equation. \emph{J. Fluid Mech.} \textbf{550}, 413--441. (\textcolor{blue}{\doi{10.1017/S0022112005008153}})

\bibitem{CHAPMAN_Book_3rd_1970}
Chapman S, Cowling T. 1970 \emph{The Mathematical Theory of Non-uniform Gases:
  An Account of the Kinetic Theory of Viscosity, Thermal Conduction and
  Diffusion in Gases}. Cambridge University Press.

\bibitem{WAGNER_PRE_59_1999}
Wagner AJ, Yeomans JM. 1999 Phase separation under shear in two-dimensional
  binary fluids. \emph{Phys. Rev. E} \textbf{59}, 4366--4373. (\textcolor{blue}{\doi{10.1103/PhysRevE.59.4366}})

\bibitem{PRASIANAKIS_PRE_76_2007}
Prasianakis NI, Karlin IV. 2007 Lattice {B}oltzmann method for thermal flow
  simulation on standard lattices. \emph{Phys. Rev. E} \textbf{76}, 016702. (\textcolor{blue}{\doi{10.1103/PhysRevE.76.016702}})

\bibitem{KARLIN_PA_389_2010}
Karlin I, Asinari P. 2010 Factorization symmetry in the lattice {B}oltzmann
  method. \emph{Physica A} \textbf{389}, 1530 -- 1548. (\textcolor{blue}{\doi{10.1016/j.physa.2009.12.032}})

\bibitem{KARLIN_PRL_81_1998}
Karlin IV, Gorban AN, Succi S, Boffi V. 1998 Maximum entropy principle for
  lattice kinetic equations. \emph{Phys. Rev. Lett.} \textbf{81}, 6--9. (\textcolor{blue}{\doi{10.1103/PhysRevLett.81.6}})

\bibitem{BOGHOSIAN_PRSLA_457_2001a}
Boghosian BM, Yepez J, Coveney PV, Wager A. 2001 Entropic lattice {B}oltzmann
  methods. \emph{Proc. Royal Soc. A} \textbf{457}, 717--766. (\textcolor{blue}{\doi{10.1098/rspa.2000.0689}})

\bibitem{ATIF_PRL_119_2017}
Atif M, Kolluru PK, Thantanapally C, Ansumali S. 2017 Essentially entropic
  lattice {B}oltzmann model. \emph{Phys. Rev. Lett.} \textbf{119}, 240602. (\textcolor{blue}{\doi{10.1103/PhysRevLett.119.240602}})

\bibitem{SAGAUT_CMA_59_2010}
Sagaut P. 2010 Toward advanced subgrid models for lattice-{B}oltzmann-based
  large-eddy simulation: Theoretical formulations. \emph{Comput. Math. Appl.} \textbf{59}, 2194 -- 2199. (\textcolor{blue}{\doi{10.1016/j.camwa.2009.08.051}})

\bibitem{MARIE_JCP_333_2017}
Mari\'{e} S, Gloerfelt X. 2017 Adaptive filtering for the lattice {B}oltzmann
  method. \emph{J. Comput. Phys.} \textbf{333}, 212 -- 226. (\textcolor{blue}{\doi{10.1016/j.jcp.2016.12.017}})

\bibitem{PRASIANAKIS_PRE_78_2008}
Prasianakis NI, Karlin IV. 2008 Lattice {B}oltzmann method for simulation of
  compressible flows on standard lattices. \emph{Phys. Rev. E} \textbf{78}, 016704. (\textcolor{blue}{\doi{10.1103/PhysRevE.78.016704}})

\bibitem{FENG_JCP_394_2019}
Feng Y, Boivin P, Jacob J, Sagaut P. 2019 Hybrid recursive regularized thermal
  lattice {B}oltzmann model for high subsonic compressible flows. \emph{J. Comput. Phys.} \textbf{394}, 82 -- 99. (\textcolor{blue}{\doi{10.1016/j.jcp.2019.05.031}})

\bibitem{RENARD_ARXIV_2020_03644}
Renard F, Feng Y, Boussuge JF, Sagaut P. 2020 Improved compressible hybrid
  lattice {B}oltzmann method on standard lattice for subsonic and supersonic
  flows. \emph{arXiv preprint arXiv:2002.03644} .

\bibitem{TAYYAB_CF_211_2020}
Tayyab M, Zhao S, Feng Y, Boivin P. 2020 Hybrid regularized lattice-{B}oltzmann
  modelling of premixed and non-premixed combustion processes. \emph{Combust. Flame} \textbf{211}, 173 -- 184. (\textcolor{blue}{\doi{10.1016/j.combustflame.2019.09.029}})

\bibitem{GEHRKE_CF_156_2017}
Gehrke M, Jan{\ss}en C, Rung T. 2017 Scrutinizing lattice {B}oltzmann methods
  for direct numerical simulations of turbulent channel flows. \emph{Comput. Fluids} \textbf{156}, 247 -- 263. (\textcolor{blue}{\doi{10.1016/j.compfluid.2017.07.005}})

\bibitem{HAUSSMANN_IJMPC_30_2019}
Haussmann M, Simonis S, Nirschl H, Krause MJ. 2019 Direct numerical simulation
  of decaying homogeneous isotropic turbulence—numerical experiments on
  stability, consistency and accuracy of distinct lattice {B}oltzmann methods. \emph{Int. J. Mod. Phys. C} \textbf{30}, 1--29. (\textcolor{blue}{\doi{10.1142/S0129183119500748}})

\bibitem{STRUCHTRUP_PoF_15_2003}
Struchtrup H, Torrilhon M. 2003 Regularization of {G}rad’s 13 moment
  equations: Derivation and linear analysis. \emph{Phys. Fluids} \textbf{15}, 2668--2680. (\textcolor{blue}{\doi{10.1063/1.1597472}})

\bibitem{TORRILHON_ARFM_48_2016}
Torrilhon M. 2016 Modeling nonequilibrium gas flow based on moment equations. \emph{Annu. Rev. Fluid Mech} \textbf{48}, 429--458. (\textcolor{blue}{\doi{10.1146/annurev-fluid-122414-034259}})

\bibitem{VONNEUMANN_Book_1966}
Von~Neumann J, Burks AW. 1996 \emph{Theory of self-reproducing automata}. University of Illinois Press Urbana.

\bibitem{VANHAREN_JCP_337_2019}
Vanharen J, Puigt G, Vasseur X, Boussuge JF, Sagaut P. 2017 Revisiting the
  spectral analysis for high-order spectral discontinuous methods. \emph{J. Comput. Phys.} \textbf{337}, 379 -- 402. (\textcolor{blue}{\doi{10.1016/j.jcp.2017.02.043}})

\bibitem{SIEBERT_IJMPC_18_2007}
Siebert DN, Hegele LA, Surmas R, Dos~Santos LOE, Philippi PC. 2007 Thermal
  lattice {B}oltzmann in two dimensions. \emph{Int. J. Mod. Phys. C} \textbf{18}, 546--555. (\textcolor{blue}{\doi{10.1142/S0129183107010784}})

\bibitem{LEKKERKERKER_PRA_10_1974}
Lekkerkerker HNW, Boon JP. 1974 Hydrodynamic modes and light scattering near
  the convective instability. \emph{Phys. Rev. A} \textbf{10}, 1355--1360. (\textcolor{blue}{\doi{10.1103/PhysRevA.10.1355}})

\bibitem{DELLAR_PRE_64_2001}
Dellar PJ. 2001 Bulk and shear viscosities in lattice {B}oltzmann equations. \emph{Phys. Rev. E} \textbf{64}, 031203. (\textcolor{blue}{\doi{10.1103/PhysRevE.64.031203}})

\bibitem{ASTOUL_ARXIV_2020_11863}
Astoul T, Wissocq G, Boussuge JF, Sengissen A, Sagaut P. 2020 Analysis and
  reduction of spurious noise generated at grid refinement interfaces with the
  lattice boltzmann method. \emph{arXiv preprint arXiv:2004.11863} .

\bibitem{CAO_PRE_55_1997}
Cao N, Chen S, Jin S, Mart\'{i}nez D. 1997 Physical symmetry and lattice
  symmetry in the lattice {B}oltzmann method. \emph{Phys. Rev. E} \textbf{55}, R21--R24. (\textcolor{blue}{\doi{10.1103/PhysRevE.55.R21}})

\bibitem{GUO_PRE_67_2003}
Guo Z, Zhao TS. 2003 Explicit finite-difference lattice {B}oltzmann method for
  curvilinear coordinates. \emph{Phys. Rev. E} \textbf{67}, 066709. (\textcolor{blue}{\doi{10.1103/PhysRevE.67.066709}})

\bibitem{WATARI_PA_382_2007}
Watari M. 2007 Finite difference lattice {B}oltzmann method with arbitrary
  specific heat ratio applicable to supersonic flow simulations. \emph{Physica A} \textbf{382}, 502 -- 522. (\textcolor{blue}{\doi{10.1016/j.physa.2007.03.037}})

\bibitem{KEFAYATI_JFS_83_2018}
Kefayati GR, Tang H, Chan A. 2018 Immersed boundary-finite difference lattice
  {B}oltzmann method through fluid–structure interaction for viscoplastic
  fluids. \emph{J. Fluids Struct.} \textbf{83}, 238 -- 258. (\textcolor{blue}{\doi{10.1016/j.jfluidstructs.2018.09.007}})

\bibitem{NANNELLI_JSP_68_1992}
Nannelli F, Succi S. 1992 The lattice {B}oltzmann equation on irregular
  lattices. \emph{J. Stat. Phys.} \textbf{68}, 401--407. (\textcolor{blue}{\doi{10.1007/BF01341755}})

\bibitem{XI_PRE_60_1999}
Xi H, Peng G, Chou SH. 1999 Finite-volume lattice {B}oltzmann schemes in two
  and three dimensions. \emph{Phys. Rev. E} \textbf{60}, 3380--3388. (\textcolor{blue}{\doi{10.1103/PhysRevE.60.3380}})

\bibitem{UBERTINI_PRE_68_2003}
Ubertini S, Bella G, Succi S. 2003 Lattice {B}oltzmann method on unstructured
  grids: Further developments. \emph{Phys. Rev. E} \textbf{68}, 016701. (\textcolor{blue}{\doi{10.1103/PhysRevE.68.016701}})

\bibitem{SBRAGAGLIA_PRE_82_2010}
Sbragaglia M, Sugiyama K. 2010 Volumetric formulation for a class of kinetic
  models with energy conservation. \emph{Phys. Rev. E} \textbf{82}, 046709. (\textcolor{blue}{\doi{10.1103/PhysRevE.82.046709}})

\bibitem{GUO_PRE_88_2013}
Guo Z, Xu K, Wang R. 2013 Discrete unified gas kinetic scheme for all {K}nudsen
  number flows: Low-speed isothermal case. \emph{Phys. Rev. E} \textbf{88}, 033305. (\textcolor{blue}{\doi{10.1103/PhysRevE.88.033305}})

\bibitem{GUO_PRE_91_2015}
Guo Z, Wang R, Xu K. 2015 Discrete unified gas kinetic scheme for all {K}nudsen
  number flows. {II}. {T}hermal compressible case. \emph{Phys. Rev. E} \textbf{91}, 033313. (\textcolor{blue}{\doi{10.1103/PhysRevE.91.033313}})

\bibitem{FENG_CF_131_2016}
Feng Y, Sagaut P, Tao WQ. 2016 A compressible lattice {B}oltzmann finite volume
  model for high subsonic and transonic flows on regular lattices. \emph{Comput. Fluids} \textbf{131}, 45 -- 55. (\textcolor{blue}{\doi{10.1016/j.compfluid.2016.03.009}})

\bibitem{YANG_PF_30_2018}
Yang L, Shu C, Yang W, Chen Z, Dong H. 2018 An improved discrete velocity
  method ({DVM}) for efficient simulation of flows in all flow regimes. \emph{Phys. Fluids} \textbf{30}, 062005. (\textcolor{blue}{\doi{10.1063/1.5039479}})

\bibitem{SHI_IJNMF_42_2003}
Shi X, Lin J, Yu Z. 2003 Discontinuous {G}alerkin spectral element lattice
  {B}oltzmann method on triangular element. \emph{Int. J. Numer. Meth. Fluids} \textbf{42}, 1249--1261. (\textcolor{blue}{\doi{10.1002/fld.594}})

\bibitem{MIN_JCP_230_2011}
Min M, Lee T. 2011 A spectral-element discontinuous {G}alerkin lattice
  {B}oltzmann method for nearly incompressible flows. \emph{J. Comput. Phys.} \textbf{230}, 245 -- 259. (\textcolor{blue}{\doi{10.1016/j.jcp.2010.09.024}})

\bibitem{PATEL_IJNMF_82_2016}
Patel S, Min M, Lee T. 2016 A spectral-element discontinuous {G}alerkin thermal
  lattice {B}oltzmann method for conjugate heat transfer applications. \emph{Int. J. Numer. Meth. Fluids} \textbf{82}, 932--952. (\textcolor{blue}{\doi{10.1002/fld.4250}})

\bibitem{SHAO_CMWA_75_2018}
Shao W, Li J. 2018 Three time integration methods for incompressible flows with
  discontinuous {G}alerkin {B}oltzmann method. \emph{Comput. Math. Appl.} \textbf{75}, 4091 -- 4106. (\textcolor{blue}{\doi{10.1016/j.camwa.2018.03.015}})

\bibitem{HE_JCP_146_1998}
He X, Chen S, Doolen GD. 1998 A novel thermal model for the lattice {B}oltzmann
  method in incompressible limit. \emph{J. Comput. Phys.} \textbf{146}, 282 -- 300. (\textcolor{blue}{\doi{10.1006/jcph.1998.6057}})

\bibitem{DELLAR_CMA_65_2013}
Dellar PJ. 2013 An interpretation and derivation of the lattice {B}oltzmann
  method using {S}trang splitting. \emph{Comput. Math. Appl.} \textbf{65}, 129 -- 141. (\textcolor{blue}{\doi{10.1016/j.camwa.2011.08.047}})

\bibitem{DHUMIERE_CMA_58_2009}
d'Humi\`{e}res D, Ginzburg I. 2009 Viscosity independent numerical errors for
  lattice {B}oltzmann models: From recurrence equations to “magic”
  collision numbers. \emph{Comput. Math. Appl.} \textbf{58}, 823 -- 840. (\textcolor{blue}{\doi{10.1016/j.camwa.2009.02.008}})

\bibitem{SHAN_PRE_73_2006}
Shan X. 2006 Analysis and reduction of the spurious current in a class of
  multiphase lattice {B}oltzmann models. \emph{Phys. Rev. E} \textbf{73}, 047701. (\textcolor{blue}{\doi{10.1103/PhysRevE.73.047701}})

\bibitem{LEE_PRE_74_2006}
Lee T, Fischer PF. 2006 Eliminating parasitic currents in the lattice
  {B}oltzmann equation method for nonideal gases. \emph{Phys. Rev. E} \textbf{74}, 046709. (\textcolor{blue}{\doi{10.1103/PhysRevE.74.046709}})

\bibitem{CHEN_IJHMT_76_2014}
Chen L, Kang Q, Mu Y, He YL, Tao WQ. 2014 A critical review of the
  pseudopotential multiphase lattice {B}oltzmann model: Methods and
  applications. \emph{Int. J. Heat Mass Transfer} \textbf{76}, 210 -- 236. (\textcolor{blue}{\doi{10.1016/j.ijheatmasstransfer.2014.04.032}})

\bibitem{CRANK_MPCPS_43_1947}
Crank J, Nicolson P. 1947 A practical method for numerical evaluation of
  solutions of partial differential equations of the heat-conduction type. In \emph{Mathematical Proceedings of the Cambridge Philosophical
  Society}, volume~43, pp. 50--67. Cambridge University Press. (\textcolor{blue}{\doi{10.1017/S0305004100023197}})

\bibitem{LATT_CMA_CorrectedProof_2020}
Latt J, Malaspinas O, Kontaxakis D, Parmigiani A, Lagrava D, Brogi F, Belgacem
  MB, Thorimbert Y, Leclaire S, Li S, Marson F, Lemus J, Kotsalos C, Conradin
  R, Coreixas C, Petkantchin R, Raynaud F, Beny J, Chopard B. 2020 Palabos:
  {P}arallel lattice {B}oltzmann solver. \emph{Comput. Math. Appl.} (in press) (\textcolor{blue}{\doi{10.1016/j.camwa.2020.03.022}})

\bibitem{WANG_IJNMF_72_2013}
Wang Z, Fidkowski K, Abgrall R, Bassi F, Caraeni D, Cary A, Deconinck H,
  Hartmann R, Hillewaert K, Huynh H, Kroll N, May G, Persson PO, van Leer B,
  Visbal M. 2013 High-order {CFD} methods: {C}urrent status and perspective. \emph{Int. J. Numer. Meth. Fluids} \textbf{72}, 811--845. (\textcolor{blue}{\doi{10.1002/fld.3767}})

\bibitem{GENDRE_PRE_96_2017}
Gendre F, Ricot D, Fritz G, Sagaut P. 2017 Grid refinement for aeroacoustics in
  the lattice {B}oltzmann method: A directional splitting approach. \emph{Phys. Rev. E} \textbf{96}, 023311. (\textcolor{blue}{\doi{10.1103/PhysRevE.96.023311}})

\bibitem{BROWN_JCP_122_1995}
Brown DL, Minion ML. 1995 Performance of under-resolved two-dimensional
  incompressible flow simulations. \emph{J. Comput. Phys.} \textbf{122}, 165 -- 183. (\textcolor{blue}{\doi{10.1006/jcph.1995.1205}})

\bibitem{MINION_JCP_138_1997}
Minion ML, Brown DL. 1997 Performance of under-resolved two-dimensional
  incompressible flow simulations, {II}. \emph{J. Comput. Phys.} \textbf{138}, 734 -- 765. (\textcolor{blue}{\doi{10.1006/jcph.1997.5843}})

\bibitem{ANSUMALI_EPL_63_2003}
{S Ansumali}, {I V Karlin}, {H C \"{O}ttinger}. 2003 Minimal entropic kinetic
  models for hydrodynamics. \emph{Europhys. Lett.} \textbf{63}, 798--804. (\textcolor{blue}{\doi{10.1209/epl/i2003-00496-6}})

\bibitem{LYCETT_BROWN_PF_26_2014}
Lycett-Brown D, Luo KH, Liu R, Lv P. 2014 Binary droplet collision simulations
  by a multiphase cascaded lattice {B}oltzmann method. \emph{Phys. Fluids} \textbf{26}, 023303. (\textcolor{blue}{\doi{10.1063/1.4866146}})

\bibitem{GRADb_CPAM_2_1949}
Grad H. 1949 On the kinetic theory of rarefied gases. \emph{Commun. Pure Appl. Math.} \textbf{2}, 331--407. (\textcolor{blue}{\doi{10.1002/cpa.3160020403}})

\end{thebibliography}

\end{document}